\documentclass[aps, amssymb, amsmath, nofootinbib, showpacs, notitlepage, byrevtex, showkeys, prd]{revtex4-1}

\usepackage{bbm}
\usepackage{newtxtext, newtxmath}              

\usepackage{zi4}                               

\usepackage[T1]{fontenc}
\usepackage{booktabs}

\usepackage{graphicx}
\usepackage{bm}
\usepackage{color}

\renewcommand{\vec}[1]{\boldsymbol{#1}}

\newcommand{\ii}{\mathrm{i}}
\newcommand*{\da}[1][]{\mathop{\mathrm{d}\mkern-7mu\mathchar'26\mkern-1mu^{#1}}\mkern-4mu}

\newcommand{\LL}{\mathsf{L}}
\newcommand{\va}{\vec{a}}

\newcommand{\vx}{\vec{x}}
\newcommand{\vy}{\vec{y}}
\newcommand{\vz}{\vec{z}}
\newcommand{\vp}{\vec{p}}
\newcommand{\vq}{\vec{q}}

\newcommand{\vk}{\vec{k}}
\newcommand{\vA}{\vec{A}}
\newcommand{\vB}{\vec{B}}
\newcommand{\vD}{\vec{D}}
\newcommand{\vPi}{\mathbf{\Pi}}
\newcommand{\valpha}{\boldsymbol{\alpha}}

\newcommand{\ve}{\vec{e}}

\newcommand{\nf}{\mathfrak{n}}

\newcommand{\mf}{\mathfrak{m}}

\newcommand{\JJ}{\mathcal{J}}
\newcommand{\II}{\mathcal{I}}

\newcommand{\ab}{\mathsf{a}}
\newcommand{\db}{\mathsf{d}_\mathsf{a}}
\newcommand{\dbh}{\hat{\mathsf{d}}_\mathsf{a}}
\newcommand{\dd}{\mathsf{d}}
\newcommand{\ddh}{\hat{\mathsf{d}}}
\newcommand{\xvec}[1]{\underline{#1}}
\newcommand{\reg}{\mu}

\begin{document}
\title{The effective potential of the Polyakov loop in the Hamiltonian approach to QCD}

\author{Markus Quandt and Hugo Reinhardt}
\affiliation{Institut f\"ur Theoretische Physik\\
Auf der Morgenstelle 14\\
D-72076 T\"ubingen\\
Germany}
\date{\today}

\begin{abstract}
We investigate the effective potential of the Polyakov loop, which is the order 
parameter for the deconfinement phase transition in finite temperature QCD. 
Our work is based on the Hamiltonian approach in Coulomb gauge where finite 
temperature $T$ is introduced by compactifying one space direction. 
We briefly review this approach and extend earlier work in the Yang-Mills sector by 
including dynamical quarks. 
In a first approximation, we follow the usual functional approach and include 
only one-loop contributions to the energy, with the finite temperature 
propagators replaced by their $T=0$ counter parts. It is found that this 
gives a poor description of the phase transition, in particular for the case
of full QCD with $N_f = 3$ light flavours. The physical reasons for this 
unexpected result are discussed, and pinned down to a relative weakness of gluon
confinement compared to the deconfining tendency of the quarks.  
We attempt to overcome this issue by 
including the relevant gluon contributions from the two-loop terms to the 
energy. We find that the two-loop corrections have indeed a tendency to strengthen
the gluon confinement and weaken the unphysical effects in the confining 
phase, while slightly increasing the (pseudo-)critical temperature $T^\ast$ 
at the same time. To fully suppress artifacts in the confining phase, we must 
tune the parameters to rather large values, increasing the critical temperature to
$T^\ast \approx 340\,\mathrm{MeV}$ for $G=SU(2)$. 
\end{abstract}
\maketitle

\section{Introduction} \label{sec:intro}

A detailed understanding of strongly interacting matter under extreme conditions 
(i.e., high temperatures or baryon densities) is among the most challenging and actively 
studied problems in particle physics today. While experimental studies in particular 
at the Large Hadron Collider (LHC) are now starting to probe into the 
physics of the quark gluon plasma, the theoretical description of this topic amounts 
to a detailed computation of the phase diagram of quantum chromodynamics (QCD)
\cite{Karsch2002, Fukushima2011}. Lattice calculations allow for precise ab-initio 
studies at non-zero temperature and vanishing baryon density, while Monte-Carlo 
simulations at non-zero baryon density are hampered by 
the so-called \emph{sign problem} \cite{Karsch2002, Gattringer2016}. Several methods
have been put forward to address this shortcoming, but so far they all seem to be restricted
to rather small chemical potentials. The most promising techniques to overcome this problem 
are, at the time of this writing, the non-perturbative continuum approaches known as 
functional methods. Such tools have therefore become an important part of the theoretical 
study of QCD under extreme conditions, using techniques such as Dyson-Schwinger equations 
(DSE) \cite{Fischer2006, *Alkofer2001, *Binosi2009}, Functional Renormlization group 
(FRG) flow equations \cite{Pawlowski2007, *Gies2012}, covariant variational methods 
\cite{Quandt2014, *Quandt2015} or semi-phenomenological approaches based on a 
massive gluon propagator \cite{*RSTW2016}. One particularly transparent method 
is the so-called \emph{Hamilton Approach} to QCD in Coulomb gauge 
\cite{Feuchter2004, *Feuchter2005, *ERS2007, *Pak2013, *QCDT0}, which is based 
on a variational determination of the QCD ground state wave functional; for a 
recent review see Ref.~\cite{Reinhardt2017}.

QCD has a rich phase structure which can be described by the partition function
depending on temperature and chemical potential (i.e.~baryon density). Virtually
all visible matter in the universe is in the hadronic phase, though different 
phases of QCD may be realized at extremely high densities e.g.~in the core of 
neutron stars \cite{Alford_1998, Alford_2008}. The hadronic phase is characterized by 
permanent colour confinement and the spontaneous breaking of chiral symmetry. 
For the latter, a suitable order parameter is the chiral quark condensate,
while deconfinement is described, at least in the absence of dynamical quarks, 
as a transition from a center symmetric phase at low temperatures to a 
high temperature phase with center symmetry broken \cite{Svetitsky:1985ye}. 
Any quantity that transforms non-trivially under center transformations can 
thus serve as an order parameter for confinement in pure Yang-Mills theory. 
A particularly transparent picture emerges in the imaginary time formalism, 
where finite temperature is introduced by Wick rotating to Euclidean space and 
compactifying the Euclidean time direction to a circle of circumference 
$\beta = 1/T$. Then the Polyakov loop\footnote{Here and in the following, 
$N$ denotes the number of colours in the gauge symmetry group $SU(N)$, and 
$\mathcal{P}$ indicates path ordering.}
\begin{align}
	\LL(\vx) \equiv \frac{1}{N}\,\mathrm{tr}\,\mathcal{P}\exp \left[ - \int_0^\beta dx_0\,
	A_0(x_0,\vx)\right]
	\label{poly}
\end{align}
transforms as $\LL(\vx) \to z \,\LL(\vx)$ under a center transformation with $z \in Z(N)$,
and hence $\langle \LL(\vx) \rangle = 0$ in the center symmetric (confined) phase, 
while $\langle\LL(\vx) \rangle \neq 0$ in the center broken (deconfined) phase. 
The connection between center symmetry and colour confinement comes from the formal
identity
\begin{align}
	\langle \LL(\vx) \rangle = \exp\left[- \beta F_\beta(\vx)\right]\,,
	\label{0}
\end{align}
which relates the Polyakov loop to the free energy $F_\beta(\vx)$ of a single 
quark immersed in the thermal QCD background. It must be emphasized that
this relation is formal, since a single colour charge in the fundamental 
representation cannot be screened by gluons, i.e.~the overlap of states from 
the vacuum sector with single static quark states must vanish. This is a consequence
of Gau\ss' law and has nothing to do with confinement. In particular, it does not 
mean that the eq.~(\ref{0}) cannot be computed or has to vanishes -- it merely 
indicates that $\LL(\vx)$ cannot consistently represent a single static quark
and the usual identification eq.~(\ref{0}) is a short-cut: the true interpretation 
relies on the correlator $\langle \LL(\vx)\cdot \LL(\vy)^\ast\rangle = \exp(- \beta V(r))$
which describes the interaction energy $V(r)$ of a static quark-antiquark pair 
at distance $r=|\vx-\vy|$. Cluster decomposition then leads to $V(\infty)=\infty$
\emph{iff} $\langle \LL\rangle = 0$, i.e. confinement in the center symmetric phase;
a similar reasoning applies to deconfinement in the center broken phase.

The Polyakov loop $\LL(\vx)$ is a rather complicated quantity in continuum approaches, 
mainly because of path ordering. A convenient way to circumvent this problem is to go 
to Polyakov gauge, 
\begin{align}
	\partial_0 A_0^{b_0} = 0\,,\qquad\qquad A_0^{\bar{b}} = 0\,,
\end{align}
where $\{T^{b_0}\}$ are the generators of the Cartan-subgroup $H$, while the remaining 
generators $\{T^{\bar{b}}\}$ span the coset $G/H$ of the colour group $G=SU(N)$. 
In this gauge, the Polyakov loop requires no path ordering,
\begin{align}
	\LL(\vx) = \frac{1}{N}\, \mathrm{tr}\,\exp\left[-\beta A_0^{b_0} \,T^{b_0}\right]\,.
\end{align}
Furthermore, it was argued in Refs.~\cite{MP2008,BGP2010}
that not only the Polyakov loop, but also the simpler quantity $\langle A_0 \rangle $ 
can serve as an order parameter for confinement in this gauge. This statement 
was originally proved for $G=SU(2)$ using Jensen's inequality, but since then 
has also been shown to generalise to $G=SU(3)$ \cite{BH2012} and, using different 
techniques, to an even larger class of compact colour groups  \cite{RSTW2016}. 

For most continuum studies, it is much more convenient to work in \emph{background 
	gauge}, where an external background field $\ab$ is split off the gauge field, 
$A_\mu = \ab_\mu + Q_\mu$, and the quantum field $Q$ is subject to the condition
\begin{align}
	[D_\mu[\ab], Q_\mu] = 0\,,
\end{align}
where $D_\mu[A] \equiv \partial_\mu + A_\mu$ is the covariant derivative.
Since $A_0$ is not a gauge invariant quantity, neither is its effective action, and 
it is not a priori clear that the effective potentials for $\langle A_0 \rangle$ in 
Polyakov and background gauge are identical or even similar. In Refs.~\cite{MP2008, BGP2010} 
is has been argued that gauge invariant features such as the location and order of the 
phase transition can also be extracted from the background gauge potential, 
provided that the background field itself is taken to be in Polyakov gauge, 
\begin{align}
	\partial_0 \,\ab_0^{b_0} = 0\,,\qquad\qquad \ab_0^{\bar{b}} = 0\,.
	\label{Ab}
\end{align} 
The argument is based on the fact that the relevant quantum fluctuations around 
the background field include the ones in Polyakov gauge, if the background itself 
obeys eq.~(\ref{Ab}). This has since been confirmed by lattice simulations and 
numerous continuum approaches, cf.~below. We will also adopt this procedure in the 
following and understand the effective potential of the Polyakov loop as the 
effective potential for a background field obeying eq.~(\ref{Ab}).

\medskip
Once dynamic quarks are introduced, the Polyakov loop is no longer a strict order 
parameter, and lattice calculations indicate that the finite temperature phase 
transition in pure Yang-Mills theory turns into an analytic crossover at a 
significantly lower pseudo-critical temperature, which depends on the exact observable
used in its definition (see \cite{Bazavov_2019} for details and recent numerical
results). This agrees with findings in random matrix models \cite{Dumitru:2012fw}, 
and in various functional approaches 
\cite{BGP2010, MP2008, BH2012, RH2013, Fischer2009, FM2009, FMM2010, RSTW2016, Quandt2016, 
	  Canfora2015, Heffner2012, Reinhardt2016, Reinhardt2012, Heffner2015}, 
all of which compute the effective action 
of the order parameter $\langle A_0 \rangle $ instead of $\langle \LL \rangle $
as in eq.~(\ref{0}) and subsequently reconstruct 
$\langle \LL[A_0] \rangle = \LL[ \langle A_0 \rangle^\ast]$ from the minimizing 
$\langle A_0\rangle^\ast$.

In the Hamilton Approach employed in this paper, the study of the Polyakov loop is 
more involved: while the method is very efficient in computing $T=0$ vacuum properties, 
its generalization to finite temperatures is, at first, hampered by the fact that the 
fields live in 3-dimensional position space, and no Euclidean time is available. 
Finite temperature therefore must be introduced in real time using the
full set of thermal states instead of studying vacuum properties \cite{Heffner2012}.
In addition, the method necessarily works in Weyl gauge, $A_0 = 0$, which also 
prevents us from studying the Polyakov loop directly. These issues were 
overcome in Ref.~\cite{Reinhardt2016}, where the full Euclidean $O(4)$ symmetry 
of the underlying theory was used to introduce a heat bath via compactification of one
\emph{spatial} axis, say the $3$-direction. Finite temperature calculations then involve 
the study of the ground state properties on the semi-compactified spatial manifold 
$\mathbb{R}^2 \times S^1(\beta)$, and the Polyakov loop winds around the 
compactified spatial direction instead of the Euclidean time. 
This setting has been used successfully to compute the deconfinement phase
transition in pure Yang-Mills  theory \cite{Reinhardt2012, RH2013, Heffner2015}. 
In the present study, we extend these calculations to full QCD including 
dynamical quarks. 

\medskip
This paper is organized as follows: In the next section, 
we review the techniques required to formulate the 
Hamiltonian approach in background gauge and at finite temperatures.
The renormalization at the one-loop level is described in 
section \ref{sec:loop1}, which also presents details on our 
numerical methods and the variational kernels used.
In section \ref{sec:results1}, the numerical results for the 
Polyakov loop at one-loop level are presented and discussed. 
We find that the transition region is described well, but the 
numbers do \emph{not} represent the confining phase accurately at 
small temperatures once dynamical quarks are included. 
We attempt to resolve this issue by including the relevant parts of 
the two-loop contribution in section \ref{sec:loop2}. 
There is a residual parameter dependency due to our incomplete 
renormalization at this order, and we discuss the parameter range 
in which the two-loop contribution proves to be benificial. The 
paper is concluded in section \ref{sec:summary} with a 
discussion of our findings and an outlook to further improvements. 

\section{Hamiltonian approach to QCD in background gauge}
\label{sec:ham}

\noindent
For pure Yang-Mills theory, the background field method in the Hamiltonian approach
was discussed in detail in Ref.~\cite{RH2013}. For completeness and to fix our notation, 
we summarize the essential features and then discuss the extension to the 
quark sector.

\subsection{The Hamiltonian in background gauge}
\label{subsec:ham}
\noindent
The canonical quantization of QCD in Weyl gauge $A_0 = 0$ results in the Hamiltonian 
\begin{align}
H = H_{\rm YM} + H_q\,.
\end{align}
Here, the gluon contribution reads
\begin{align}
H_{\rm YM} = \frac{1}{2}\int d^3x\,\left[ g^2 \vPi^2(\vx) + \frac{1}{g^2}\, \vB^2(\vx)\right]
\end{align}
and involves, besides the conjugate momentum operator $\Pi_k = - i \delta / \delta A_k$ 
and the coupling strength $g$, also  the colour magnetic field
$
\vB^a = \nabla \times \vA^a + \frac{1}{2}\,f^{abc}\,\vA^b \times \vA^c\,.
$
The quark sector is simply the usual Dirac Hamiltonian of a massive fermion
(with the standard $4 \times 4$ matrices $\valpha$ and $\beta$) coupled 
covariantly $\big(\vD(\vx) = \nabla + \vA(\vx)\big)$ to the gluons,
\begin{align}
H_q = \int d^3x \,\psi^\dagger(\vx)\,\big[ \valpha\cdot \vD(\vx) + \beta m]\,\psi(\vx)\,.
\label{quarksec}
\end{align}
At this point, the residual gauge symmetry of time-independent gauge transformation
has not yet been fixed. This is reflected by the existence of a time-independent 
constraint (Gau\ss' law),
\begin{align}
\hat{\vD}\cdot \vPi \,\big\vert\,\Phi \rangle = \rho\,\big \vert\,\Phi \rangle\,,
\label{gausslaw}
\end{align}
where the hat denotes the adjoint colour representation,
\[
\hat{D}^{ab}_k = \hat{\partial}_k^{ab} + \hat{A}_k^{ab} = 
\delta^{ab} \partial_k - f^{abc} A_k^c
\]
and 
$
\rho^a(\vx) = - i\,\psi^\dagger(\vx)\,T^a\,\psi(\vx)
$
is the colour charge density of the quarks. We must resolve Gau\ss' law explicitly 
in background gauge. To this end, we first introduce generalised longitudinal 
and transversal projectors,
\begin{align}
\hat{\ell}_{ik} = \ddh_k\,[\ddh\cdot \ddh]^{-1}\,\ddh_k\,,\qquad\qquad
\hat{t}_{ik} = \hat{\delta}_{ik} - \hat{\ell}_{ik}\qquad\qquad\quad
(\hat{\delta}^{ab}_{ik} = \delta^{ab}\delta_{ik})\,,
\label{genproj}
\end{align}
where $\ddh = \hat{\partial} + \hat{\ab}$ is the covariant derivative of the background field. 
For a constant background field $\ab$ in the 
Cartan subalgebra, we have $[\ddh_i, \ddh_k] = 0$ and the generalized projectors 
enjoy the same properties as the ordinary projectors. 

\medskip
We can now follow the resolution of Gau\ss' law in analogy to the standard derivation 
of the gauge fixed Hamiltonian in Coulomb gauge. The result is the 
background gauge-fixed QCD Hamiltonian, 
\begin{align}
H_{\rm fix} = H_{\rm YM}^\perp + H_q + H_C\,, 
\label{Hfix0}
\end{align}
in which all (background) longitudianal fields have been eliminated. 
To simplify the notation, we will therefore drop the superscript '$\perp$' on 
all fields and stipulate that $\vA = \vA^\perp$ and $\vPi = \vPi^\perp$ are 
background transversal in the sense of eq.~(\ref{genproj}), 
unless stated otherwise. The quark section, $H_q$, remains unchanged 
and still reads as in eq.~(\ref{quarksec}). 
In the gluon contribution, however, the gauge fixing has introduced the
Fadeev-Popov-determinant of the background gauge 
\begin{align}
\mathcal{J}_\ab[\vA] = \mathrm{det}\Big[ - \hat{\vD}\cdot \dd\Big]
\label{FPdet}
\end{align}
which enters the kinetic energy,
\begin{align}
H_{\rm YM} = \frac{g^2}{2}\,\int d^3x\, \JJ_\ab^{-1}[\vA]\,\vPi(\vx)\cdot 
\JJ_\ab[\vA]\,\vPi(\vx) + \frac{1}{2g^2} \int d^3x\, \vB(\vx)\cdot \vB(\vx)\,.
\end{align}
In addition, a new non-Abelian colour interaction emerges,
\begin{align}
H_C = \frac{g^2}{2}\int d^3(x,y)\,\JJ_\ab^{-1}[\vA]\,\rho^a_{\rm tot}(\vx) \JJ_\ab[\vA]\,
F^{ab}(\vx, \vy)\,\rho^b_{\rm tot}\,,
\label{coulx}
\end{align}
where
\begin{align}
\hat{F}^{ab}(\vx, \vy) = \int d^3z\,\left[ \big(-\hat{\dd}\cdot \hat{\vD}\big)^{-1}\right]^{ac}(\vx,\vz)
\,\big[- \hat{\dd}\cdot \hat{\dd}]^{-1}(\vz)\,\left[ \big(-\hat{\dd}\cdot
 \hat{\vD}\big)^{-1}\right]^{cb}(\vz,\vy)
 \label{coulkern}
\end{align}
is the background gauge analog of the non-Abelian Coulomb term, which generalizes the 
Coulomb interaction in electrostatics. The total charge $\rho_{\rm tot} = \rho_{\rm YM} + \rho $
receives contributions from both the matter and the gauge fields; in particular, the 
gluon charge $\rho_{\rm YM} = - \hat{\vD} \vPi$ simplifies to 
$\rho_{\rm YM} = - (\hat{\vA}-\hat{\ab}) \vPi$ since $\vPi$ is background transversal.

\subsection{Trial wave functionals for full QCD}
\label{subsec:trial} 
\noindent
In the next step, we have to define trial wave functions $|\Phi\rangle$ which obey the 
constraint $\langle \vA \rangle_\Phi = \ab$. We start with the wave functionals 
of the Hamiltonian approach in Coulomb gauge ($\ab = 0$) \cite{QCDT0, QCDT0Rev},
which employ a product ansatz for the Yang-Mills and quark sector,
\begin{align}
\big\vert\,\Phi[\vA]\big\rangle = \Phi_{\rm YM}[\vA]\, \big\vert\,\Phi_q[\vA]\big\rangle\,.
\end{align}
The gluon part is a modified Gaussian type of functional,
\begin{align}
\Phi_{\rm YM}[\vA] &= \mathcal{N}\cdot \JJ[\vA]^{-\frac{1}{2}}\cdot  \II[A]^{- \frac{1}{2}}\cdot
\widetilde \Phi_{\rm YM}[\vA]\,,
\label{omegaa}
\\[2mm]
\widetilde{\Phi}_{\rm YM}[\vA] &= \exp\left( - \frac{1}{2 g^2} \int d^3(x,y)\,
A_i^a(\vx)\,\omega(\vx-\vy)\,A_i^a(\vy)\right)\,,
\label{omegab}
\end{align}
where $\omega$ is a variation kernel, and $\mathcal{N}$ is a normalization constant
involving $\omega$. The additional normalization $\II[A] = \langle \Phi_q \vert \Phi_q\rangle$
comes from the quark sector, for which a Slater determinant inspired by BCS theory will 
be used  
\cite{QCDT0, QCDT0Rev}\,,
\begin{align}
\vert\,\Phi_q[\vA]\rangle =  
\exp\left[ - \int d^3(x,y)\, \psi_+^{m\dagger}(\vx)\,
K^{mn}(\vx,\vy)\,\psi_-^n(\vy)\right]\,\big \vert 0 \big \rangle\,.
\label{quarkwave}
\end{align} 
Here $\psi_{\pm}$ are the positive/negative energy components of the fermion field, 
$\vert 0 \rangle$ is the bare fermionic vacuum (Dirac sea) and $m,n$ are colour indices 
in the fundamental representation. The variational kernel $K[\vA]$ may depend explicitly
on the gluon field and can be decomposed in Dirac 
structures,
\begin{align}
K^{mn}(\vx, \vy) = \beta \,S^{mn}(\vx, \vy) - i \int d^3 z\,\Big[ V_b^{mn}(\vx,\vy;\vz) 
+ \beta W_b^{mn}(\vx,\vy;\vz)\Big]\,\valpha \cdot \vA^b(\vz)\,,
\label{Kfull}
\end{align}  
where $S$, $V$ and $W$ are variation kernels. Neglecting the coupling to the transversal
gluons in the kernel ($V = W = 0$) leads to a variational equation for $S$ alone, 
which corresponds to the well-known Adler-Davis model \cite{Adler1984}. 
Next, we must shift the trial wave functionals to comply with the constraint
$\langle \vA \rangle = \ab$,
\begin{align}
\big\vert\,\Phi_\ab[\vA] \rangle = \mathcal{N}\, \II[\vA-\ab]^{-\frac{1}{2}}\,
\JJ_\ab[\vA]^{-\frac{1}{2}}\,\widetilde{\Phi}_{\rm YM}[\vA - \ab]\,
\big\vert\,\Phi_q[\vA - \ab] \rangle\,.
\label{trial}
\end{align}
Notice that we have shifted the gauge field argument in all places \emph{except}
the Faddeev-Popov (FP) determinant $\JJ_\ab[\vA]$, which was, however, changed from 
Coulomb gauge to background gauge, cf.~eq.~(\ref{FPdet}). It is then easy to see that 
this wave functional indeed obeys the constratint $\langle \vA \rangle_\Phi = \ab$;
in fact, the expectation value of \emph{any} observable $\Omega$ in the state 
eq.~(\ref{trial}) is 
\begin{align}
\langle \,\Omega[\vA,\vPi,\psi]\rangle_\ab = 
\langle \widetilde{\Omega}[\vA + \ab, \vPi, \psi] \rangle_0
¸\label{Oshift}
\end{align}
with the modified observable $\widetilde{\Omega}$
\begin{align}
\widetilde{\Omega}[\vA, \vPi,\psi] = \JJ_\ab[\vA]^{\frac{1}{2}} \II[\vA-\ab]^{\frac{1}{2}}\,
\Omega[\vA, \vPi, \psi]\, \JJ_\ab[\vA]^{-\frac{1}{2}} \II[\vA-\ab]^{-\frac{1}{2}}\,,
\label{Otilde}
\end{align}
which reduces to $\Omega$ if it contains no functional derivatives, i.e. if it does not 
depend on $\vPi$. The expection value $\langle \cdots \rangle_0$ on the rhs of 
eq.~(\ref{Oshift}) is with the same wave functionals eqs.~(\ref{omegab}) and (\ref{Kfull})
as in Coulomb gauge (hence the subscript '0'), but with the fields $\vA$ being 
background transversal $\ddh \vA = 0$, and with the kernel $\omega$ in eq.~(\ref{omegab})
promoted to a matrix in adjoint colour space.  (This will be discussed in 
the next subsection.) If we take $\Omega[\vA] = \vA$, in particular, 
we have $\widetilde{\Omega}[\vA + \ab] = \vA + \ab$ and thus 
$\langle \vA \rangle_\ab = \langle \vA + \ab \rangle_0 = \ab + \langle \vA \rangle_0 = \ab$,
because $\langle \vA \rangle_0 = 0$ in the original Coulomb gauge wave functional. 

\subsection{QCD propagators in the presence of a background field}
\label{subsec:bg:var}
To carry out the variational approach we have to compute the expectation value 
of the gauge-fixed QCD Hamiltonian in the a trial wave functional (\ref{trial}), 
\begin{align}
\big\langle H_{\rm fix}\big\rangle_\ab = \big\langle \widetilde{H}_{\rm fix}[\vA + \ab]
\big\rangle_0 = \big \langle H_{\rm fix}[\vA + \ab] \big \rangle_0 + \cdots\,,
\label{Hfix}
\end{align}
where the dots indicate higher order terms from moving the determinants in 
eq.~(\ref{Otilde}) past the momentum operators $\vPi$ in $H_{\rm fix}$.  
The gauge field is a connection and thus enters $H_{\rm fix}$ only through the covariant 
derivative; the same holds for the gauge condition $\dbh\,\vA=0$, while the trial wave 
functional used in $\langle \cdots \rangle_0$ are the ones from Coulomb gauge 
eq.~(\ref{omegaa}), which do not depend on the background field explicitly. 
As a consequence, the only effect of the background gauge field in 
$\langle H_{\rm fix}[\vA + \ab]\rangle_0$ as compared to Coulomb gauge $\ab=0$ 
is to replace all ordinary derivatives by covariant background 
derivatives, $\nabla \to \db = \nabla + \ab$. 
This, in turn, has the effect of shifting the momentum argument of Green's 
functions and the variation kernels.

To describe this shift, it is convenient to go to a colour basis in which 
$\ddh$ and $\hat{\ab}$ are diagonal. First we expand the background field 
in the Cartan subalgebra,
\begin{align}
\ab = \sum_{k=1}^r\,\ab_k\,H^k =  
\sum_{\mu} (-i \xvec{\ab}\cdot \xvec{\mu})\, \vert \mu \rangle
\langle \mu \vert\,,
\end{align} 
where $r$ is the rank of the colour group, $H_k$ denotes the Cartan generators,
and the weight vector $\xvec{\mu} = (\mu_1,\ldots,\mu_r)$ contains $r$ 
eigenvalues of the $H_k$. (There are $N$ distinct weight vectors for 
$G=SU(N)$, each of which corresponds to one of the 
$N$ distinct colour vectors $|\mu\rangle$ which diagonalize the Cartan 
generators simultaneously.) 
A similar relation holds in the adjoint representation, 
\begin{align}
\hat{\ab} = \sum_\sigma (-i \xvec{\ab}\cdot \xvec{\sigma})\, \vert \sigma \rangle
\langle \sigma \vert\,,
\end{align}
where the $(N^2-1)$ root vectors $\xvec{\sigma} = (\sigma_1,\ldots,\sigma_r)$
now contain eigenvalues of $\hat{H}_k$ in the adjoint, and the $(N^2-1)$ eigenvectors
$\vert\sigma \rangle$ are adjoint colour vectors  which diagonalize the $\hat{H}_k$
simultaneously. For further details, see appendix \ref{app:cartan}.

\medskip
After expanding the fields 
and kernels in the trial wave functional in the Cartan basis we can Fourier transform them 
in the usual fashion based on translational invariance. The action of the 
covariant background derivative is then 
\begin{align}
\frac{1}{i} \,\dd \exp(i \vp \vx)) &= \sum_\mu \vert \mu \rangle \langle \mu \vert\,
\vp^\mu\,\cdot \exp(i\,\vp \vx)\,,\qquad\qquad
\vp^\mu  = \vp - \xvec{a}\cdot \xvec{\mu}
\\[2mm]
\frac{1}{i} \,\hat{\dd} \exp(i \vp \vx) &= \sum_\sigma \vert \sigma\rangle \langle \sigma \vert\,
\vp^\sigma\,\cdot \exp(i\,\vp\vx)\,,\qquad\qquad
\vp^\sigma  = \vp - \xvec{a}\cdot \xvec{\sigma}\,.
\label{shift}
\end{align}
Since this is the only way in which the background field enters, we can conclude that
the only modification introduced by a constant background field $\ab$ in the Cartan 
algebra is \textbf{(i)} to expand all colour fields in the weight or root vectors, 
as appropriate, and \textbf{(ii)} to shift a momentum variable by 
$(- \xvec{a}\cdot \xvec{\mu})$ if it is associated with a quark field, and by 
$(- \xvec{a}\cdot \xvec{\sigma})$ for gluons and ghosts. 

\subsection{The variational approach in the presence of the background field}
We are now in a position to compute the expectation value of the full gauge fixed QCD 
Hamiltonian in our trial wave functional (\ref{trial}) depending explicitly on 
the background field $\ab$,
\begin{align}
\langle H_{\rm fix}\rangle_\ab &=
\langle H_{\rm YM}^\perp\rangle_\ab + \langle H_q \rangle_\ab + \langle H_C\rangle_\ab \,.
\end{align}
The first (Yang-Mills) piece can be further expanded into three contributions,
\begin{align}
\langle H_{\rm YM}^\perp\rangle_\ab = \langle H_{\rm YM}^A\rangle_\ab + 
\langle H_{\rm YM}^{\rm NA}\rangle_\ab + \langle H_{\rm YM}^q\rangle_\ab\,.
\end{align}
The first term contains the kinetic (electric) energy plus the Abelian part of 
the magnetic energy. It is a 1-loop contribution and involves only the variational 
kernel $\omega$ and the curvature $\chi$ (see below). The second term is 
the contribution from the non-Abelian part of the magnetic energy. 
In the gluon gap equation, it only contributes a (divergent) constant and 
is hence often neglected. Below, we will show that the finite reminder of 
this constant after renormalization may, however, have a significant impact 
on the Polyakov loop; the non-Abelian magnetic energy will be discussed in more 
detail in section \ref{sec:loop2}. The third term comes from the
action of the canonical momentum operator $\vPi$ on the $\vA$-dependent 
part of the quark wave function. It is a  2-loop term that vanishes if
 $V=W=0$, since then the quark wave functional (\ref{quarkwave}) does not 
 couple to $\vA$. 

\medskip\noindent
In the quark sector, $H_q$ contains no momentum operator $\Vec{\Pi}$ and 
the prescription eq.~(\ref{Oshift}) yields two contributions 
\begin{align}
\langle H_q[\vA] \rangle_\ab = \langle H_q[\vA + \ab] \rangle_0 = \langle H_q^\ab\rangle_0 + 
\langle H_q^A\rangle_0\,,
\end{align}
where 
\begin{align}
H_q^\ab &= - i \int d^3x\,\psi^\dagger(\vx)\,\valpha\cdot \dd(\vx)\psi(\vx)
\label{aquark}\\[2mm]
H_q^A &= - i \int d^3x\,\psi^\dagger(\vx)\,\valpha\cdot \vA(\vx)\,\psi(\vx)\,.
\label{fquark}
\end{align}
The second term eq.~(\ref{fquark}) is again a two-loop contribution that vanishes 
if $V=W=0$, because then the expectation value factorizes and 
$\langle \vA \rangle_0 = 0$. The first term eq.~(\ref{aquark}) has both a 
one-loop contribution that only depends on the scalar quark kernel $S$ (see below),
and a 2-loop contribution that vanishes if $V=W=0$.

Finally, the Coulomb term couples the charge densities 
$\rho_{\rm tot} = \rho + \rho_{YM}$ of the gluon and quark. This gives three 
contributions according to the combination of the charges involved,
\begin{align}
\langle H_C \rangle_\ab = \underbrace{\langle H_C^A\rangle_\ab}_{\sim \rho_{\rm YM}\cdot
 \rho_{\rm YM}} 
+ \underbrace{\langle H_C^q\rangle_\ab}_{\sim \rho\cdot \rho} 
+ \underbrace{\langle H_C^{\rm mix}\rangle_\ab}_{\sim \rho\cdot\rho_{\rm YM}}\,.
\end{align}
All three contributions are two-loop terms. The mixed contribution describes 
an interaction between gluons and quarks which is not expected to have a 
significant effect on the Polyakov loop (an inter-quark potential!)  to this order.
We will neglect the mixed contribution in the following. The quark part 
$\langle H_C^q\rangle_\ab$ of the Coulomb interaction is included in the 
quark gap equation, cf. section \ref{sec:loop1:fermi} below, and the gluon part 
$\langle H_C^A\rangle_\ab$ is studied in more detail in section \ref{sec:loop2}.
 
\medskip\noindent
Putting everything together, we can split the full QCD expectation value 
in our background gauge trial wave functional in a bosonic and a 
fermionic part, 
\begin{align}
\langle\, H_{\rm fix}\,\rangle_\ab &= E_B[\ab] + E_F[\ab]\,.
\label{EE}
\end{align}
Both terms have one- and two-loop contributions according to the following chart:
\begin{align}
E_B[\ab] &= 
\underset{\substack{\downarrow\\\text{1-loop\,\,} (\omega,\chi)}}
{\langle H_{\rm YM}^A\rangle_\ab} \quad + \quad 
\underset{\substack{\downarrow\\\text{2-loop\,\,} (\omega,\chi)}}
{\langle H_{\rm YM}^{\rm NA}\rangle_\ab} \quad + \quad 
\underset{\substack{\downarrow\\\text{2-loop\,\,} (\rho_{\rm YM}\,\rho_{\rm YM})
}}{\langle H_C^A\rangle_\ab} \quad + \quad 
\underset{\substack{\downarrow\\\text{2-loop\,\,} (V,W)}}
{\langle H_{\rm YM}^q\rangle_\ab}
\label{EBx}\\[3mm]
E_F[\ab] &= 
\underset{\substack{\downarrow\\\text{1-loop\,\,} (S) \\[1mm] 
\text{2-loop\,\,} (V,W,S) 
}}
{\langle H_q^\ab\rangle_0} \quad + \quad 
\underset{\substack{\downarrow\\\text{2-loop\,\,} (V,W)
}}
{\langle H_q^A\rangle_0} \quad + \quad 
\underset{\substack{\downarrow\\\text{2-loop\,\,} (V,W)}}
{\langle H_C^q\rangle_\ab}\quad  + \quad 
\underset{\substack{\downarrow\\\text{2-loop\,\,} (\rho_{\rm YM}\,\rho)}}
{\langle H_C^{\rm mix}\rangle_\ab}
\label{EFx}
\end{align}
The arrows indicate the variation kernels $\omega, S, V, W$ on which the 
respective contribution depends. 
If we set $V=W=0$ and hence employ a BCS type of wave functional for the quarks, 
and further neglect the mixed 2-loop Coulomb interactions involving both
$\rho$ and $\rho_{\rm YM}$, the result simplifies considerably,
\begin{align}
E_B[\ab] &= 
{\langle H_{\rm YM}^A\rangle_\ab} +
{\langle H_{\rm YM}^{\rm NA}\rangle_\ab} +
{\langle H_C^A\rangle_\ab} 
\label{EBOS}
\\[3mm]
E_F[\ab] &= 
{\langle H_q^\ab\rangle_0} +
{\langle H_C^q\rangle_\ab}
\label{EFERM}
\end{align}
The first term in both the fermionic and bosonic contributions is 1-loop, 
while the remaining terms are all 2-loop contributions.

\subsection{Hamiltonain dynamics at finite temperature}
\label{sec:finiteT}

The developments made so far allow for a computation of the minimal 
energy $\langle H_{\rm fix}\rangle_\Phi$ in all states obeying the 
background field constraint $\langle \vA \rangle_\Phi = \ab$, or at least 
for a subset of states characterized by our ansatz (\ref{omegaa}). This minimal 
energy is the effective potential of the Polyakov loop background $\ab$ at $T=0$. 

As we switch on the temperature, the variational principle still determines the 
minimal energy, when we are actually interested in the \emph{free} energy.
The reason for this shortcoming is that our trial ansatz eq.~(\ref{omegaa}) is 
no longer sufficient at finite temperature: we should instead work with thermal 
states that involve arbitrary excitations above the ground state within a grand 
canonical ensemble. Such an approach has been attempted \cite{Heffner2012}, but
there is a simpler formulation which allows to work with a trial vacuum wave functional 
and the usual minimization of the ground state energy \cite{Reinhardt2012, RH2013, Reinhardt:2016pfe}. 
The finite temperature $T=\beta^{-1}$ is here introduced by a compactification 
of the $x_3$-direction via the boundary conditions
\begin{subequations}
	\begin{align}
		\psi^m(x_1, x_2, x_3 &= \beta/2) = - \psi^m(x_1, x_2, x_3 = -\beta/2)
		\label{bc:dirac}
		\\[2mm]
		\vA^a(x_1, x_2, x_3 &= \beta/2) = \vA^a(x_1, x_2, x_3 = -\beta/2)
		\label{bc:gluon}
	\end{align}
\end{subequations}
for the quark and gluon field, respectively. With these conditions, the original space 
manifold is effectively compactified to a cylinder $\mathbb{R}^2 \times S^1(\beta)$ 
and we use the abbreviation 
\begin{equation}
	\int_{\beta} \dd^3 x \equiv \lim_{\ell \to \infty}\int_{-\ell/2}^{\ell/2} \dd x_1 
	\int_{-\ell/2}^{\ell/2} \dd x_2 \int_{-\beta/2}^{\beta/2} \dd x_3
\end{equation}
for the spatial integration over this manifold. The length $\ell$ of the uncompactified
direction will always be large, and the limit $\ell \to \infty$ projects out  
the grand canonical partition function of QCD at non-zero temperature $T$ and chemical potential
$\mu$ \cite{Reinhardt2016}
\begin{equation}
\mathcal{Z} = \lim_{\ell \to \infty} \, \exp\Big[-\ell E_0(\beta, \mu)\Big] \, , 
\label{eq:partition}
\end{equation}
where $E_0$ is the smallest eigenvalue of the non-hermitean pseudo-Hamiltonian
\begin{equation}
\widetilde{H}(\beta, \mu) \equiv \int_\beta \mathcal{H} + \ii \mu \int_{\beta} \dd^3 x \, 
{\psi}^{\dagger}(\vx) \alpha_3   \psi(\vx) \, . \label{eq:pseudo}
\end{equation}
Here, $\mathcal{H}$ is the usual QCD Hamiltonian density in Coulomb and Weyl gauge 
\cite{Christ1980}, and $\alpha_3$ one of the Dirac matrices.

It should be emphasized that the analysis of Ref.~\cite{Reinhardt2016} exchanges the 
Euclidean time direction $x_0$ with $x_3$ (and likewise for all vector quantities) assuming 
relativistic $O(4)$ invariance of the underlying Euclidean field theory. 
In particular, it does not hold for non-relativistic quark models or 
effective nuclear theories that single out a fixed reference frame. 
For the case of a vanishing chemical potential $\mu = 0$, the Hamiltonian
$\widetilde{H}(\beta,0)$
is hermitean, all energy eigenvalues are real and the limit
$\ell \to \infty$ of the non-compactified directions projects out the ground 
state contribution in eq.~(\ref{eq:partition}). (For $\mu \neq 0$ and real, 
the situation is more complicated and we defer $\mu \neq 0$ to a forthcoming 
investigation.) 

For explicit calculations, it is convenient to switch to momentum 
space. From the general boundary conditions eq.~(\ref{bc:dirac}), continuity of the 
wave functional implies, for instance, for the quark kernel,
\begin{subequations}
\begin{align}
K(\vx + \beta \mathbf{e}_3, \vy) = K(\vx, \vy + \beta \mathbf{e}_3) = - K(\vx, \vy) \,,
\end{align}
\end{subequations}
and a similar relation for the scalar dressing function 
$S(\vx-\vy)$ if we set $V = W = 0$. The Fourier representation thus takes the form
\begin{equation}
S(\vx) = \int_{\beta} \da^3 p \, \exp\Big[\ii\, (\vp_{\perp} + 
\tilde{\Omega}_{\nf} \mathbf{e}_3) \cdot \vx\Big]\, S(\vp_{\perp}, 
\tilde{\Omega}_{\nf}) \, ,
\label{Sfourier}
\end{equation}
where $\vp_{\perp} = p_1\mathbf{e}_1 + p_2 \mathbf{e}_2$ is the planar momentum perpendicular to 
the compactified direction of the heat bath, and 
\begin{equation}
\tilde{\Omega}_{\nf} = \frac{\pi (2\nf + 1)}{\beta}\,,\qquad\qquad \nf \in \mathbb{Z}
\label{mats}
\end{equation}
are the fermionic Matsubara frequencies. Furthermore, we have introduced the short-hand 
notation 
\begin{equation}
\int_{\beta} \da^3 p \cdots \equiv \int \frac{d^2 p_\perp}{(2 \pi)^2} \, \frac{1}{\beta} 
\sum_{\nf = -\infty}^{\infty}\cdots \, .
\label{matsmeasure}
\end{equation}
A similar relation holds for the bosonic kernel $\omega$, with the bosonic 
Matsubara frequencies
\begin{equation}
\Omega_{\nf} = \frac{2 \pi\nf}{\beta}\,,\qquad\qquad \nf \in \mathbb{Z}\,.
\label{matsB}
\end{equation}

\section{The effective potential of the Polyakov loop at one-loop} \label{sec:loop1}
\noindent
We are now in the position to compute the effective potential of 
the background field representing the Polyakov loop. We start with the 
1-loop contributions in eq.~(\ref{EBOS}) and (\ref{EFERM}).
Since the background field $\ab$ in $3$-direction is constant, 
the space volume always factorizes and we really compute the energy 
\emph{density}
$e[\ab] = E_B[\ab] / V_3 = E_B[\ab] / (V_\perp\,\beta)$.

\subsection{Boson contribution}
\label{sec:boson:loop1}
\noindent
We begin with the bosonic 1-loop contribution in eq.~(\ref{EBOS}). A straightforward 
calculation at $T=0$ gives \cite{RH2013}
\begin{align}
e_B[\ab] &= \sum_{\sigma} \int \frac{d^3p}{(2\pi)^3}\,\big[\omega_\sigma(\vp) - 
\chi_\sigma(\vp)\big]\,,
\label{eb1}
\end{align}
where the sum is over all roots $\sigma$ of $SU(N)$. The curvature $\chi(\vk)$ is the 
contribution from the ghost loop which can be related to the ghost form factor 
and the gluon kernel $\omega$ through a separate DSE \cite{Heffner2015}.
As explained before, the background field $\ab$ lives in the Cartan subalgebra 
and is taken to be constant in the $3$-direction. It enters $\langle H_{\rm YM}^A\rangle_\ab$ 
only through the covariant derivative when using eq.~(\ref{Oshift}), and the 
kernels $\omega_\sigma(\vp)$ and $\chi_\sigma(\vp)$ are hence obtained from their 
$T=0$ counterparts through a shift $\vp \to \vp^\sigma$ in the momentum argument, 
$\omega_\sigma(\vp) = \omega(\vp^\sigma)$, cf.~eq.~(\ref{shift}).
This prescription can also be seen explicitly in the DSE for the curvature $\chi_\sigma$ 
and the gap equation obtained by minimizing the bosonic energy, 
 \begin{align}
 \omega_\sigma(\vp)^2 = \vp_\sigma^2 + \chi_\sigma(\vp)^2 + I_0^{NA} + I_C^A(\vp, \sigma)\,.
 \label{gap0}
 \end{align}
 Here, $I_0^{\rm NA}$ is a (colourless) tadpole term from the non-Abelian 
 part of the magnetic energy in eq.~(\ref{EBOS}), while $I_C^A$ comes from
 the 2-loop Coulomb-term in eq.~(\ref{EBOS}). Both expressions are divergent and 
 require renormalization, cf.~section \ref{sec:renorm1} below. Let us 
 mention at this point already that the renormalization of the tadpole term 
 $I_0^{\rm NA}$ requires a gluon mass counterterm, cf.~eq.~(\ref{Ect}) below. 
 If the coefficient is adjusted to cancel the 
 divergences in $I_0^{\rm NA}$ at zero temperature, we expect no further 
 divergences to appear (from this term) at finite temperature and non-vanishing
 background field. This is indeed the case, though the explicit verification is 
 rather tricky, cf.~section \ref{sec:loop2NA}. For the logarithmic divergence 
 in the Coulomb two-loop term $I_C^A$, we do not as yet have a full 
 renormalization at finite temperature. In the present paper, we will take 
 a more pragamtic approach and identify, isolate and then subtract the 
 divergence as usual. The remaining free parts are not fixed by a 
 renormalization condition at $T=0$, but we treat the corresponding 
 counterterm coefficient as a free parameter.   
 
 It should be emphasized that 
 eq.~(\ref{eb1}) is the \emph{self-consistent} energy obtained after 
 inserting the gap equation into the full energy and truncating 
 at one-loop level. Since the gap equation mixes loop orders, it will thus 
 effectively contain 2-loop contributions. We must hence ensure that 
 eq.~(\ref{gap0}) holds -- maybe in renormalized form --  when using 
 eq.~(\ref{eb1}). The quark sector does not couple directly to the 
 gluon sector at this level, provided that we also waive the direct coupling of the 
 quarks to the gluon field $\vA$ in the trial wave function, i.e.~we set the 
 kernels $V = W = 0$. In this case, the gluon sector is identical to the Yang-Mills 
 case, and we refer to Ref.~\cite{Heffner2015} for a detailed discussion of the 
 boson kernels $\omega$ and $\chi$ at $T=0$. The two-loop terms missing in
 eq.~(\ref{eb1}) will be discussed in section \ref{sec:loop2} below.
 
 As we switch on the temperature, the initial $O(3)$ rotation symmetry is 
 broken, and the kernel $\omega$ can no longer be transversal.
 Instead, the compactification of the $3$-direction gives rise to 
 two distinct Lorentz structures, $\omega = \omega^\perp\,t^\perp + \omega^\|r\,t^\|$,
 where the projectors are (at $\ab = 0$ for simplicity \cite{Heffner2015}) 
\begin{align}
 t^\perp_{ij}(\vp) = (1 - \delta_{i3})\,\Big(\delta_{ij} - \frac{p_i p_j}{\vp_\perp^2}\Big)
 (1-\delta_{j3})\,,\qquad\qquad\quad
 t^\|_{ij}(\vp) = t_{ij}(\vp) - t_{ij}^\perp(\vp)\,.
 \end{align}
 As indicated, the two distinct Lorentz structures involve two distinct 
 scalar variation kernels $\omega^\perp(\vp)$ and $\omega^\|(\vp)$, and likewise 
 for the curvature $\chi$. The Lorentz trace of the boson kernel, previously
 $\mathrm{tr}\,\omega = 2 \omega(\vp)$, now becomes
 \[
\mathrm{tr}\,\omega = \omega^\perp(\vp) \,\mathrm{tr}\, t^\perp(\vp) + \omega^\|(\vp)\,
\mathrm{tr}\,t^\|(\vp) = \omega^\perp(\vp) + \omega^\|(\vp)\,,
\]
and the colour trace turns into the sum over roots. Finally, the integration in 
the compactified dimension is replaced by a Matsubara sum and eq.~(\ref{eb1})
turns into
\begin{align}
e_B(\ab,\beta) = \frac{1}{2}\sum_{\sigma} \int \frac{d^2 p_\perp}{(2\pi)^2}\,
\frac{1}{\beta}\sum_{\nf \in \mathbb{Z}}
\Big[ \omega^\perp(\vp_\nf^\sigma) - \chi^\perp (\vp_\nf^\sigma)
+ \omega^\|(\vp_\nf^\sigma) - \chi^\| (\vp_\nf^\sigma)\Big]\,,
\label{deriv1}
\end{align}
where we have now applied the shift as in eq.~(\ref{shift}),
\begin{align}
\vec{p} \mapsto \vp^\sigma_\nf \equiv 
\vec{p}_\perp + (\Omega_\nf - \xvec{\sigma}\cdot \xvec{\ab})\,\hat{\vec{e}}_3\,,
\qquad\qquad \Omega_\nf \equiv2 \pi \nf / \beta\,. 
\end{align}
The quantity eq.~(\ref{deriv1}) is still 
infinite because it contains the (free) energy of the vacuum. Since the Polyakov loop 
represents a single static quark immersed in the thermal QCD ground state, its 
effective potential must be understood as the \emph{change} of the free energy
due to the presence of the background,
\begin{align}
\bar{e}(\ab, \beta) \equiv e(\ab, \beta) - e(0, \beta)\,.
\label{deriv2}
\end{align}
The subtraction is most easily performed after Poisson resumming the 
Matsubara sum,\footnote{This technique is based on the simple
distributional identity
\begin{align}
\frac{1}{\beta}\sum_{\nf = - \infty}^\infty u(\Omega_\nf) =
\frac{1}{\beta}\sum_{\nf=-\infty}^\infty
u\big(\frac{2 \pi \nf + \varphi}{\beta}\big) = 
\int\limits_{-\infty}^\infty \frac{dp_3}{2\pi}\sum_{\mf=-\infty}^\infty 
\exp\left[\ii \mf (\beta q_3 - \varphi)\right]\,u(q_3)\,,
\label{poisson}
\end{align}
valid for suitable test functions $u$ and arbitrary $\varphi \in [0, 2\pi]$.  The case 
$\varphi = 0$ corresponds to bosons,  $\varphi = \pi$ to fermions.
}
\begin{align}
\bar{e}_B(\ab,\beta) = \sum_\sigma \int \da^3 p \sum_{\mf \in \mathbb{Z}} 
e^{i \mf \beta p_z} \big[ g_B(\vp^\sigma) - g_B(\vp)\big]\,,
\label{deriv2a}
\end{align}
where $\vp^\sigma = \vp_\perp + \ve_3\,(p_z - \ab \cdot \xvec{\sigma})$, 
and we have defined
\begin{align}
g_B(\vp) \equiv \frac{1}{2}\Big[\omega^\perp(\vp) - \chi^\perp(\vp) + 
\omega^\|(\vp) - \chi^\|(\vp)\Big]\,.
\label{gB}
\end{align}
Next we shift $p_z \to p_z + \ab \cdot \xvec{\sigma}$ 
and introduce the dimensionless background shift
\begin{align}
\Delta^\sigma \equiv \frac{\beta}{2 \pi}\,\big(\ab \cdot \xvec{\sigma}\big)\,.
\end{align}
This gives
\begin{align}
\bar{e}_B(\ab,\beta) = \sum_\sigma\sum_{\mf \in \mathbb{Z}} \int \da^3 p
\,e^{i \mf \beta p_z}\,\Big[e^{2 \pi i \mf \Delta^\sigma} - 1\Big]\,g_B(\vp)\,.
\label{deriv3a}
\end{align}
Due to the subtraction eq.~(\ref{deriv2}), the term with $\mf = 0$, i.e.~the 
vacuum energy at $T=0$, does not contribute to this expression. Furthermore, the 
term in the bracket vanishes for the trivial root $\sigma = (0,0,\ldots)$, 
while the non-trivial roots of $SU(N)$ always come in pairs with opposite sign, 
so that
\[
\sum_\sigma \Big[e^{2 \pi i \mf \Delta^\sigma} - 1\Big] = - \sum_\sigma
\Big[1 - \cos(2 \pi \mf \Delta^\sigma)\Big]\,.
\] 
Eq.~(\ref{deriv3a}) can now be rewritten as 
\begin{align}
\bar{e}_B(\ab,\beta) = -2 \sum_\sigma\sum_{\mf = 1}^\infty 
\Big[1 - \cos(2 \pi \mf \Delta^\sigma)\Big]
\int \da^3 p \,\cos(\mf \beta p_z) \,g_B(\vp)\,.
\label{deriv3}
\end{align}
In the last step, we non-dimensionalize the (free) energy density and rewrite 
it in a form suitable for later numerical evaluation:
\begin{align}
u_B(\ab,\beta) &\equiv \beta^4\,\big[ e_B(\ab,\beta) - e_B(0,\beta)\big]
= \frac{2}{\pi^2} \sum_{\mf=1}^\infty \sum_\sigma 
\frac{1 - \cos(2 \pi \mf \Delta^\sigma)}{\mf^4}\,h_B(\beta \mf)
\label{deriv4}
\\[2mm]
h_B(\lambda) &= - \pi^2 \lambda^4\,\mathsf{Re}\int \da^3 p \,e^{i \lambda p_z}\,g_B(\vp)\,.
\label{deriv5} 
\end{align}
The temperature dependence is completely encoded in the function $h_B(\lambda)$. We compute it
by introducing spherical coordinates $(p,\vartheta,\varphi)$ for $\vp$ and note 
that the polar angle $\varphi$ is cyclic due to the residual $O(2)$ symmetry in 
the $xy$-plane mentioned above. Changing variables 
$\vartheta \to \xi \equiv \cos \vartheta$, we obtain
\begin{align}
h_B(\lambda) = - \frac{\lambda^4}{4}\int_0^\infty dp\,p^2 \int_{-1}^1 d\xi\,
\cos(\lambda p \xi)\, g_B(p,\xi) \,.
\label{deriv6}
\end{align}
Since $\omega(\vp) \sim p$ and $\chi(\vp)\sim 1/p$ at large $p$, we also have 
$g_B(\vp) \sim p$ for $p \to \infty$ and eq.~(\ref{deriv6}) is apparently UV divergent.
We will investigate this issue in more detail in sec.~\ref{sec:renorm1}. 
If the kernels happen to be $O(3)$-symmetric, i.e.~if they do not depend on the angle against
the heat bath, the $\xi$-integration can be performed and we obtain
\begin{align}
h_B(\lambda) = - \frac{\lambda^3}{2}\int_0^\infty dp\,p \sin(\lambda p) g_B(p)\,.
\label{hB2}
\end{align}
Even if we do not have $O(3)$ symmetry, we can still use eq.~(\ref{hB2}) with the 
$g_B(p)$ replaced by the angular average 
\begin{align}
\bar{g}_B(p) \equiv \frac{\lambda}{2} \int_{-1}^{+1} d\xi\,
\frac{\cos(\lambda p  \xi)}{\sin(\lambda p)} \,p g_B(p,\xi)
\to \frac{1}{2} \int_{-1}^1 d\xi \,g_B(p,\xi) \quad \mbox{at }\lambda \to 0\,.
\label{ave}
\end{align}
As indicated, this reduces to the simple integral average when $\lambda \ll 1$, i.e.~at 
very high temperatures. Thus, for any finite temperature kernel, we can do the angular average 
eq.~(\ref{ave}) and then employ the $O(3)$ symmetric relations such as eq.~(\ref{hB2}).

\medskip\noindent
For $G=SU(2)$, the roots are $\{-1,0,+1\}$, so that $\Delta^\sigma = \{-x,0,+x\}$ in 
terms of the fundamental domain (\emph{Weyl alcove})
\begin{align}
x \equiv \frac{\beta \ab^3}{2 \pi} \in [0,1]\,,
\label{weylsu2}
\end{align}
on which center symmetry acts by $x \to 1 - x$. The center symmetric point for 
$G=SU(2)$ is therefore $x=\frac{1}{2}$, while $x\in\{0, 1\}$ are the maximally 
center breaking points. The contribution of the trivial root drops 
out of the Poisson sum in eq.~(\ref{deriv4}) and we find
\begin{align}
u_B(x,\beta) = \frac{4}{\pi^2} \sum_{\mf=1}^\infty \frac{1-\cos(2 \pi \mf x)}{\mf^4}
\,h_B(\mf \beta)\,,\qquad\qquad x \in [0,1]\,.
\label{uBsu2}
\end{align}
To check this formula, recall that the 1-loop effective potential in 
\emph{perturbation theory} requires only tree-level kernels, i.e.~we can set 
$\omega_{\perp}(p) = \omega_{\|}(p) = p$ and $\chi_{\perp}(p) = \chi_{\|}(p) = 0$ 
to this order, so that $g_B(p) = p$ and hence,
from eq.~(\ref{hB2}), $h_B(\lambda) = 1$. (We will discuss the calculation of $h_B$
and the treatment of the UV divergences below.) The corresponding effective 
potential for the Polyakov loop in $G=SU(2)$ is
\begin{align}
u_B(x,\beta) = \frac{4}{\pi^2} \sum_{\mf=1}^\infty \frac{1-\cos(2 \pi \mf x)}{\mf^4}
= \frac{4}{3}\pi^2 x^2(1-|x|)^2\,,\qquad\qquad x \in[-1,1]\,.
\label{weiss223}
\end{align}
This is indeed the Weiss potential \cite{Weiss1981} usually obtained in one-loop 
thermal perturbation theory. 

Let us also generalize eq.~(\ref{uBsu2}) to the colour group $SU(3)$ which has rank 2
so that the Polyakov loop background field has two colour components, $\ab^3$ and $\ab^8$. 
As a parametrization of the Polyakov loop (or the Weyl alcove), we choose 
\begin{align}
	x & \equiv \frac{\beta\ab^3}{2\pi} \in [0,1]
	\nonumber\\[2mm]
	y & \equiv \frac{\beta\ab^8}{2\pi} \in [0,2 / \sqrt{3}]\,.
\end{align}
Since the Weyl alcove for $G=SU(3)$ is triangular, the square region for $(x,y)$
defined above actually covers a single Weyl alcove plus two adjacent 
half-alcoves. The effective potential of the $SU(3)$ Polyakov loop background 
is again given by eq.~(\ref{deriv2a}); the only difference to $G=SU(2)$ is the 
root sum, which now runs over $N^2-1 = 8$ root vectors, of which $N(N-1) = 6$ are 
non-vanishing. The non-zero roots come in pairs with both signs, and the three 
non-vanishing \emph{positive} roots lead to different momentum shifts.
After performing the root sum, it follows that the bosonic $SU(3)$ Polyakov 
loop potential is simply a sum of three SU(2) potentials,
\begin{align}
	u_B(x,y,\beta) = u_B(x,\,\beta) + u_B\Big(\frac{x + y \,\sqrt{3}}{2},\,\beta
	\Big) +  u_B\Big(\frac{x - y \,\sqrt{3}}{2},\, \beta
	\Big)
	\label{uBsu3}
\end{align}

\subsection{Fermion 1-loop contribution}\label{sec:loop1:fermi}

In the present study, we waive the explicit coupling to the gluon sector
in the trial wave functional ($V = W = 0$), so that only a single scalar variation
kernel $S(\vp)$ remains in the quark sector. The corresonding expectation value 
to the Fermi part of the energy in eq.~(\ref{EFERM}) contains a one-loop term 
involving the free quark Hamiltonian, and a two-loop contribution involving the 
Coulomb potential. We proceed as in the boson case and first vary the total 
fermion energy w.r.t.~the kernel $S(\vp)$ to obtain a fermionic gap equation.
The solution, which releates one- and two-loop orders, is then inserted back 
into eq.~(\ref{EFERM}) to obtain the \emph{self-consistent} quark energy up to
and including two-loop order. At this stage (and not earlier), the fermion energy 
is truncated to one-loop order;  through the gap equation, it will actually 
contain parts of the two-loop term in eq.~(\ref{EFERM}), in a self-consistent 
manner.\footnote{If we truncated the energy to one-loop order \emph{prior} to 
the variation, the gap equation (\ref{AD}) would turn into $M(k) = m$, i.e. 
the Coulomb interaction and chrial symmetry breaking would be absent.} 

In more detail, the fermion gap equation is best formulated in terms of the 
\emph{mass function} $M(\vp)$, which follows from the variational kernel 
$S(\vp)$ via
\begin{equation}
	M(\vp) =  2 | \vp| \cdot \frac{S(\vp)}{1 - S^2(\vp)} \,.
	\label{def:mass}
\end{equation}
The value $M(0)$ can be viewed as a dynamically generated quark mass
which breaks chiral symmetry. In the case of a vanishing background field $\ab = 0$,
the fermion gap equation formally agrees with the model proposed by 
Adler and Davis \cite{Adler1984}
\begin{equation}
	M(k) = m + \frac{C_F}{8 \pi^2} \int_0^\infty dq\,\int_{-1}^1 dz\, q^2 V_C(q)\,\frac{M(Q) - 
		\big[1 + q z / k\big]\,M(k)}{\sqrt{Q^2 + M(Q)^2}}\,.
\label{AD}
\end{equation}
Here, $Q = |\vk + \vq| = \sqrt{k^2 + q^2 + 2 k q z}$ and $C_F = \frac{N_c^2 - 1}{2N_c}$ 
is the value of the quadratic Casimir of the colour group $SU(N_c)$. Initially, the 
gap equation also involves the non-Abelian Coulomb kernel, eq.~(\ref{coulkern}), which can 
however be replaced, to this order, by its vacuum expectation value 
$g^2 \langle \hat{F}^{ab}(\vx,\vy)\rangle \approx \delta^{ab}\,V_C(\vx,\vy)$.
The long-ranged part of the variational solution in the Yang-Mills sector is well 
described by a linear rising Coulomb potential, $V_C = - \sigma_C \,| \vx - \vy|$, 
which amounts to the Fourier transform $V_C(p) = 8 \pi \sigma_C / p^4$ to be 
used in eq.~(\ref{AD}). The numerical solution for the mass 
function $M(p)$ at $T=0$ is shown in the left panel of Fig.~\ref{fig1}. 

Next we insert the solution of eq.~(\ref{AD}) back into eq.~(\ref{EFERM}) and 
introduce finite temperature as before. After a straightforward calculation, we can 
employ the Poisson resummation formula (\ref{poisson}) for fermions to obtain
\begin{align}
e_F(\ab, \beta)  = -N_f \sum_\mu \int  \da^3 p \sum_{\mf = -\infty}^\infty
e^{i \beta \mf (p_z - \pi / \beta)}\,\frac{p_\perp^2 + (p_z^\mu)^2}
{\sqrt{p_\perp^2 + (p^\mu_z)^2 + M(\vp^\mu)}}\,.
\end{align}
Here, $N_f$ is the number of (light) quark flavours, the z-component in the momentum
 variable $\vp^\mu$ is shifted according to 
$p_z^\mu = p_z - \ab\cdot \xvec{\mu}$, and the sum is over all weights $\mu$ of $SU(N)$. 
The remaining calculation also follows the bosonic case:
we shift the integration variable 
$p_z \to p_z + \ab\cdot \xvec{\mu}$, Poisson resum the Matsubara series and subtract 
the vacuum contribution ($\ab = 0$). For the result, we introduce the quantities
\begin{align}
\Delta^\mu = \frac{\beta}{2\pi}\,\big(\ab \cdot \xvec{\mu}\big)\,,\qquad\qquad\quad
g_F(\vp) = \frac{\vp^2}{\sqrt{\vp^2 + M(\vp)^2}}\,,
\label{deriv11}
\end{align}
and the self-consistent quark contribution to the effective potential of the Polyakov 
loop becomes, at one-loop level, 
\begin{align}
u_F(\ab,\beta) &\equiv \beta^4\,\big[ e_F(\ab,\beta) - e_F(0,\beta)\big]
 = - \frac{N_f}{\pi^2} \sum_{\mf= - \infty}^\infty (-1)^\mf\,
\frac{\sum_\mu \big[1 - e^{2 \pi i \mf \Delta^\mu}\big]}{\mf^4}\,h_F(\beta \mf)
\label{deriv14}
\\[2mm]
h_F(\lambda) &= - \pi^2 \lambda^4 \,\mathsf{Re} \int \da^3 p\,e^{i \lambda p_z}\,g_F(\vp)
= - \frac{\lambda^3}{2}\int_0^\infty dp\,p \sin(\lambda p) g_F(p)\,.
\label{deriv16}
\end{align}
As indicated, only the real part of the Fourier integral contributes, because 
$g_F(\vp)$ must be even under the flip\footnote{In the original Matsubara formulation, 
this corresponds to the sign change $\Omega_\nf \to \Omega_{-(\nf+1)} = - \Omega_\nf$. 
For $G=SU(2)$, in particular, the invariance under this flip can be seen explicitly 
since the part odd in $p_z$ is also odd in $\mf$ and hence vanishes after the Poisson 
summation over $\mf$.} $p_z \to (- p_z)$.
Upon comparing eq.~(\ref{deriv5}) with eq.~(\ref{deriv16}), we realize that the 
factors $h_B(\lambda)$ and $h_F(\lambda)$ are constructed in the same way, with 
$g_B(\vp)$ in the bosonic case replaced by $g_F(\vp)$ in the fermionic case. 

Next we compute the weight sum in eq.~(\ref{deriv14}). For $G=SU(2)$, the fundamental
representation is 2-dimensional and we have hence two weights 
$\mu = \pm1/2$ so that $\Delta^\mu = 
(\ab\cdot \xvec{\mu})\,\beta/(2\pi) = \pm \beta \ab / (4 \pi) = \pm x/2$.
After a short calculation,
\begin{align}
u_F(x, \beta) = - \frac{4 N_f}{\pi^2}\sum_{\mf=1}^\infty (-1)^\mf\,\frac{1 - \cos(\pi \mf x)}
{\mf^4}\,h_F(\beta \mf)
\label{deriv50}
\end{align}
For $G =SU(3)$, the fundamental representation is 3-dimensional and we have thus 
three weights
\begin{align*}
\xvec{\mu}_1 = \frac{1}{2}\begin{pmatrix} 1 \\  1 / \sqrt{3}\end{pmatrix}\,,\qquad\qquad
\xvec{\mu}_2 = \frac{1}{2}\begin{pmatrix} -1 \\ 1 / \sqrt{3}\end{pmatrix}\,,\qquad\qquad
\xvec{\mu}_3 = \begin{pmatrix} 0 \\ -1 / \sqrt{3}\end{pmatrix}\,,
\end{align*} 
so that $\Delta^{1,2} = (\pm x + y / \sqrt{3}) / 2$ and $\Delta^3 = - y / \sqrt{3}$.
The weight sum now becomes
\begin{align}
\sum_{\mu}  \Big[1 - e^{2 \pi i \mf \Delta^\mu} \Big]
= \sum_\theta \big[1 - \cos(\pi \mf \theta)\big] + \text{terms odd in $\mf$}\,,
 \label{deriv25}
\end{align}
where $\theta$ runs ove the three values $\pm x + y / \sqrt{3}$ and $-2 y / \sqrt{3}$.
The terms odd in $\mf$ drop out in eq.~(\ref{deriv14}), because $h_F(-\lambda) = h_F(\lambda)$
as explained earlier. Furthermore, the term with $\mf = 0$ does not contribute due to the 
subtraction of the trivial background. Combining terms with $(\pm \mf)$, we can 
again express the $SU(3)$ result as a sum over $SU(2)$ potentials, 
\begin{align}
u_F(x, y, \beta) = \frac{1}{2}\,\Bigg[ u_F\Big(x + \frac{y}{\sqrt{3}}, \beta\Big) + 
u_F\Big(- x + \frac{y}{\sqrt{3}},\beta\Big) + 
u_F\Big(-2 \frac{y}{\sqrt{3}},\beta\Big) \Bigg]\,.
\label{deriv80}
\end{align}
To check these equations, consider a free massless fermion, $M(p) = 0$, where 
$g_F(p) = p$ and hence $h_F(\lambda)=1$ as in the bosonic case. The 
\emph{perturbative} quark contribution to the effective potential of the $SU(2)$ 
Polyakov loop is therefore
\begin{align}
u_F(x, \beta) 
= - \frac{4 N_f}{\pi^2} \sum_{\mf=1}^\infty (-1)^\mf\,\frac{1 - \cos(\pi \mf x)}{\mf^4}
=N_f\,\frac{\pi^2}{6}\left(x^2 - \frac{1}{2}\,x^4\right)\,,\qquad\qquad\quad
x \in [-1,1]\,.
\label{deriv127}
\end{align}
This one-loop result  agrees with the quark part of the standard expression \cite{Weiss1981}.

\subsection{Renormalization and 1-loop numerics}
\label{sec:renorm1}
Let us next study the counter terms necessary to render the bosonic gap equation 
(\ref{gap0}) finite. (At $V = W = 0$, the fermionic gap equation (\ref{AD}) 
is UV-finite if only the long-ranged part of the Coulomb potential is retained.) 
The quadratically divergent tadpole contribution $I_0^{NA}$ 
is cancelled by a gluon mass counter term of the form \cite{ERSS2008}
\begin{align}
	H_{\rm ct} = \frac{C_0}{2 g^2}\,\int d^x\, \vec{A}^a(\vx) \vec{A}^a(\vx)\,.
	\label{Ect}
\end{align}
When added to the original Hamiltonian, this contributes the constant $C_0$ 
to the gap equation, which then takes the form
\begin{align}
	\omega_\sigma(\vp)^2 = p_\sigma^2 + \chi_\sigma(\vp)^2 + I_0^{NA} + C_0 + 
	I_C^A(\vp,\sigma)\,.
	\label{gap1}
\end{align}
The Coulomb term is a 2-loop contribution which also requires 
renormalization and an additional counter term \cite{ERSS2008}. This is discussed in 
detail in section \ref{sec:loop2}. For the moment, it is sufficient 
to note that the relevant counter term (with coefficient $C_1$) would result in the modified
gap equation 
\begin{align}
	\omega_\sigma(\vp)^2 = p_\sigma^2 + \chi_\sigma(\vp)^2 +\big[ I_0^{NA} + C_0 \big]
	+ \big[ I_C^A(k,\sigma) + 2 C_1 \,\chi(\vp) \big]\,.
	\label{gap1a}
\end{align}
Numerical investigations \cite{Feuchter2004, *Feuchter2005, *ERS2007, *Pak2013, *QCDT0}
 show that the last term in eq.~(\ref{gap1a}) can safely 
be neglected in the gap equation -- it may, however, contribute in the total energy
and this is investigated in section \ref{sec:loop2} below. The remaining terms yield
a gluon and ghost propagator that agrees very well with the lattice calculations
\cite{BQR2009} and, in particular, the analytical Gribov prediction. Furthermore, 
a perimeter law for the 't Hooft loop is only possible if the finite remainder 
of the last term in eq.~(\ref{gap1a}) vanishes \cite{RE2007}. All these arguments 
strongly suggest that the Coulomb term can be neglected in the gap equation, while 
it may play a significant role in the total energy.

The tadpole term $I_0^{\rm NA}$ depends on both the temperature and the background field 
$\ab$, but not the external momentum. We can therefore adjust the counter term 
coefficient $C_0$  to cancel the quadratic divergence in the tadpole. The finite remainder,
\begin{align}
	 c_0  \equiv I_0^{\rm NA} + C_0
	\label{palx}
\end{align}
is a free renormalization constant that parametrizes the theory. It is important 
to note that the renormalization occurs without a background field and at zero 
temperature. It is generally expected that the divergences (and hence the counter terms)
should be independent of temperature and the background field. This means that the 
sum of the temperature-dependent contributions from $I_0^{NA}$ and the $T=0$ counter term 
must be UV finite. This is indeed the case, though the explicit proof is rather involved,
cf.~section \ref{sec:loop2}.

Neglecting the Coulomb term as discussed above, the renormalized gap equation now takes the form 
(at $\ab = 0$ and $T=0$ for simplicity) 
\begin{align}
	\omega(\vp)^2 = p^2 + \chi(\vp)^2 + c_0\,.
	\label{gapren}
\end{align}
At one-loop level, we require no further counter term. The DSE
for the curvature may require a ghost wave-function renorma\-lization, but this 
is automatically included when we compute the curvature (at given $\omega(p)$)
from the gap equation, rather than through its DSE. 
Before inserting the gap equation, the 1-loop boson energy density 
eq.~(\ref{deriv1}) including the counter term becomes
\begin{equation}
	e_B(\ab, \beta) = \sum_\sigma \int \da^3 p\,\frac{\big[\omega(p_\sigma) - 
		\chi(p_\sigma)\big]^2 + \vp_\sigma^2 + c_0}{\omega(p_\sigma)}
\end{equation}
where we have not distinguished the two Lorentz structures for simplicity.
(We will only use the $T=0$ solutions for $\omega(p)$ in the following.)
Inserting the renormalized gap equation, the counter term contribution 
formally drops out and we are left with
\begin{equation}
	e_B(\ab, \beta) = \sum_\sigma \int \da^3 p\,\big[\omega(p_\sigma) - \chi(p_\sigma)\big]\,,
	\label{compens}
\end{equation}
just as in the unrenormalized case eq.~(\ref{deriv1}). The counter term 
$c_0$ thus enters only indirectly via the modification of the curvature 
through the gap equation (\ref{gapren}).

\bigskip\noindent
Even after renormalizing the gap equation, the profiles $h_B(\lambda)$ and 
$h_F(\lambda)$ entering the effective potential of the Polyakov loop are 
apparently UV divergent. Since these divergences cannot be cancelled by any 
$T=0$ counter term in the theory, they must be spurious. 
To see this, note that the $T=0$ vacuum energy (including all possible 
divergences and counter terms) is already subtracted in eq.~(\ref{deriv4})
and eq.~(\ref{deriv14}), respectively, so that $h_B$ and 
$h_F$ may not contain ($T=0$) divergences. More precisely, the leading  
UV divergence for $h_B(\lambda)$ in eq.~(\ref{hB2}) comes from $g_B \simeq p$
at large $p$, which leads to the expression
\[
h_B(\lambda) \simeq - \frac{\lambda^3}{2} \int_0^\infty dp\,p^2 \sin(\lambda p)
= - \frac{1}{2}\int_0^\infty dq\,q^2 \sin(q)\,.
\]
This is formally divergent but independent of $\lambda$ (and hence temperature). 
Any counter term for it would have to be temperature-independent, too, 
but all available $T=0$ counter term have already been exhausted in the 
renormalization of the gap equation above.

There are at least three ways to deal with the spurious divergences in the functions 
$h(\lambda)$:
\begin{enumerate}
\item introduce a regulator $e^{- \mu p}$ in the momentum integral and perform the 
limit $\mu \to 0$ outside the integral;
\item perform integration by parts and throw away the boundary contribution from 
$p = \infty$;
\item do contour integration and throw away the large circle at complex infinity 
$|p| \to \infty$.
\end{enumerate}
All methods are equivalent and the first two are also suitable for numerical 
evaluation. Let us briefly check the regulator method for the model 
$g(p) = p^\alpha$ with $\alpha > 0$:
\begin{align}
h(\lambda) &= 
- \frac{\lambda^3}{2}\lim_{\reg\to 0} \int_0^\infty dp\,p \sin (\lambda p)\,
p^\alpha\,\,e^{- \reg p} \nonumber \\[2mm]
&= - \frac{\lambda}{2} \lim_{\reg\to 0} \int_0^\infty dq\,q \,(q/\lambda)^\alpha
\,e^{-\reg q / \lambda} \,\sin q\nonumber \\[2mm]
&= \lim_{\reg\to 0} - \frac{\lambda^{1-\alpha}}{2}\,(1 + \reg^2)^{-(1 + \alpha/2)}
\Gamma(\alpha + 2)\,\sin\big[(\alpha + 2) \arctan (\lambda / \reg)\big]
\nonumber \\[2mm]
& = - \frac{1}{2}\,\lambda^{1-\alpha}
\Gamma(\alpha + 2)\sin\left(\frac{\pi}{2}\,(\alpha+2)\right)\,.
\label{qn}
\end{align}
For a free gauge boson, we have $\omega(p) = p$ and $\chi(p) = 0$ which implies
$g_B(p) = p$ and thus $\alpha = 1$. In the free fermionic case, we have $M(\vp)=0$ and 
thus also $g_F(p) = p$ from eq.~(\ref{deriv11}). Free particles are therefore 
always characterized by $\alpha = 1$, which implies $h(\lambda) = 1$.
This was used in the derivation of the Weiss formula above. 

\medskip
As a second example, take a free massive boson, $\omega(p) = g_B(p) = \sqrt{p^2 + m^2}$,
and employ the integration by parts technique. We have to do 4 integrations by 
parts and drop the momentum-independent boundary terms, to arrive at the finite
contribution 
\begin{align}
h(\lambda) &= - \frac{1}{2 \lambda} \int_0^\infty dp\,\sin(\lambda p)\,
\frac{d^4}{dp^4} \Big[p \sqrt{p^2 + m^2}\Big]
= \int_0^\infty dp\,\frac{15 m^4 p\,\sin(\lambda p)}{2 \lambda \,(p^2 + m^2)^{\frac{7}{2}}}
= \frac{(m \lambda)^2}{2}\, K_2(m \lambda)\,,
\end{align} 
where $K_2$ is a modified Bessel function. In the massless limit $m \to 0$, or at high 
temperatures $\lambda \to 0$, we obtain again $h(0) = 1$. This can be interpreted as follows:
the techniques used to derive the effective action of the Polyakov loop, eqs.~(\ref{uBsu2})
and (\ref{deriv50}), can be adapted to evaluate the free energy and the pressure 
of thermal QCD. In that case, the same function $h(\lambda)$ appears as a multiplicative 
factor, and the remaining factors are such that the high temperature limit $h(0)=1$ 
for each degree of freedom saturates the Stefan-Boltzmann law. Thus, the value $h(0)$ 
at high temperatures can be interpreted as counting the perturbative degrees of freedom 
as given by the Stefan-Boltzmann law \cite{Quandt:2017poi}.
 
\bigskip\noindent
In our numerical code, we have always used the regulator method explained above 
to deal with the spurious divergences in $h_B(\lambda)$ and $h_F(\lambda)$. 
Furthermore, we have always used the $T=0$ kernels even at finite temperature.
In the gluon sector, this is a standard procedure in functional methods, 
based on the lattice observations that the gluon propagator is only mildly 
affected by temperatures up to $T=2 T^\ast$. Furthermore, there are 
qualitative arguments \cite{BGP2010} which suggest that the finite temperature 
corrections to the gluon kernel are of higher order in the effective potential 
for the gauge-invariant Polyakov loop, and it is assumed that this
carries over to the present  background gauge calculation. However, a 
stringent proof does not exist and the justification is essentially 
\emph{a posteriori}. 

For the gluon sector, the Coulomb gauge propagator in both lattice 
\cite{BQR2009} and variational \cite{Feuchter2004, *Feuchter2005, *ERS2007, *QCDT0} calculations can be well described by the \emph{Gribov formula},
\begin{align}
\omega(p) = \sqrt{p^2 + \frac{M_G^4}{p^2}}\,,
\label{gribovx}
\end{align}
with the Gribov mass $M_G \simeq 880\,\mathrm{MeV}$ that sets the overall scale 
in the gluon sector. The curvature is then fixed by the gap equation,
\begin{align}
\chi(p) = \sqrt{- c_0 + \frac{M_G^4}{p^2}}\,,
\label{gribovx2}
\end{align}
and we must have $c_0 < 0$ so that the curvature is real for all momenta. 
Lattice calculations indicate that these shapes are only mildly affected by 
finite temperatures up to $T = 2 T^\ast$, while a corresponding calculation 
in the Hamiltonian approach has not yet been carried out.

\begin{figure}
	\centering
	\includegraphics[width=0.48\linewidth]{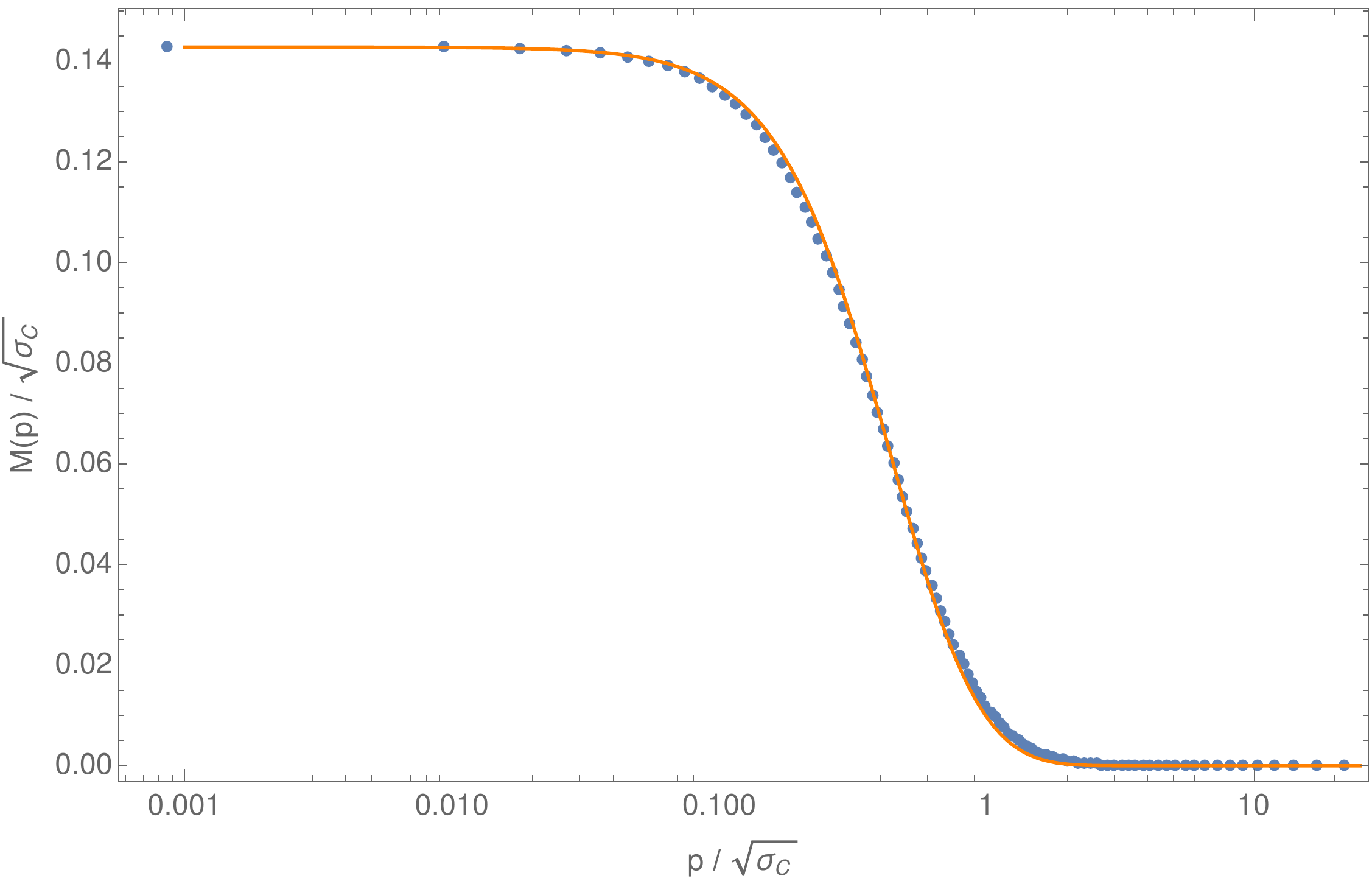}
	\hfill
	\includegraphics[width=0.48\linewidth]{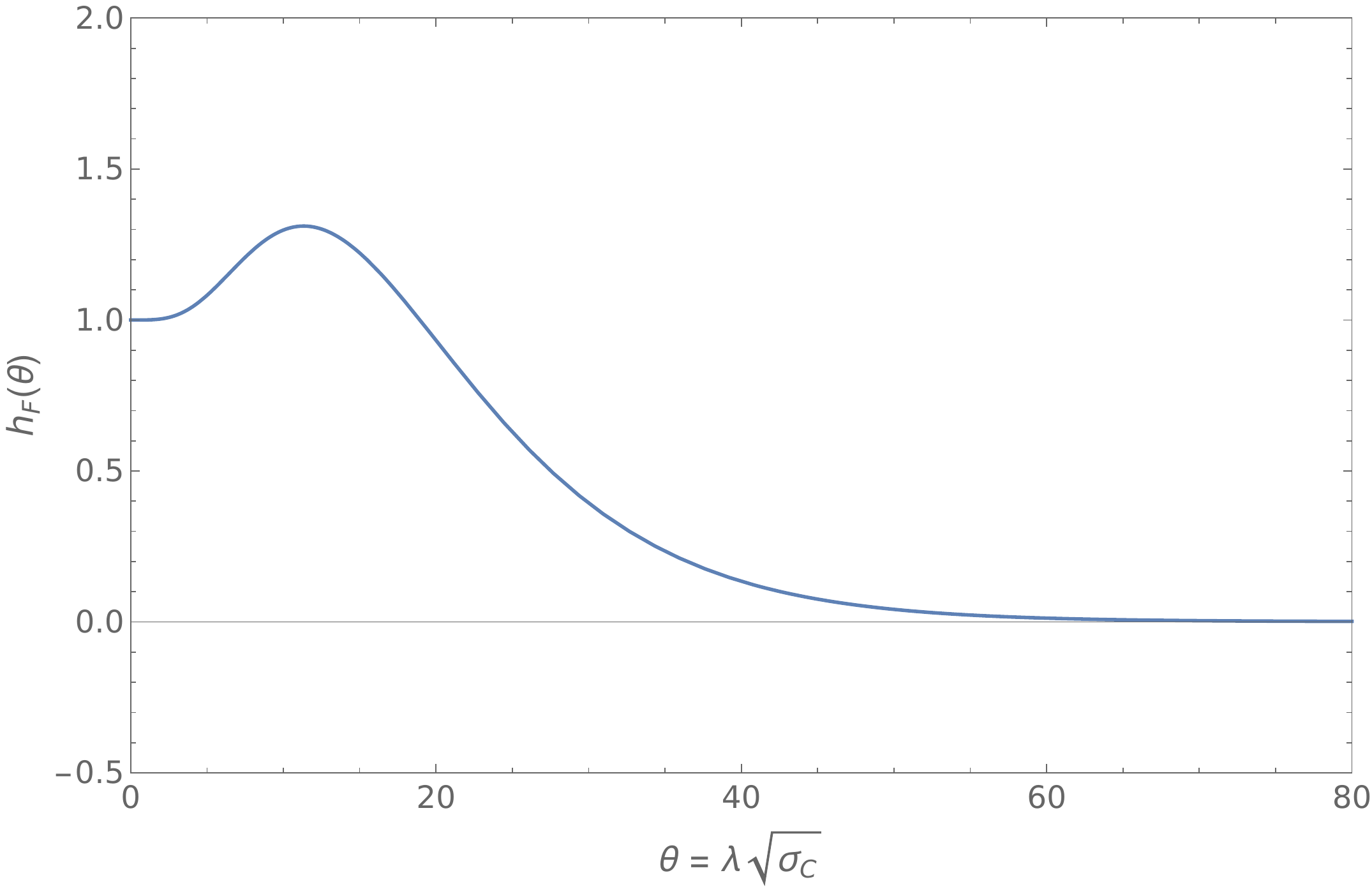}
	\caption{\emph{Left}: Smooth fit to the numerical solution of the Adler-Davis equation. 
		\emph{Right}: The Fourier transform $h_F(\lambda)$ for fermions; the dimensionfull
	argument $\lambda$ is measured in units of $\sqrt{\sigma_c}$.}	
    \label{fig1}
\end{figure}

\bigskip\noindent
In the quark sector, the solution to the gap equation (\ref{AD}) at $T=0$ 
can be parametrized in a veriety of ways, for instance
\begin{align}
M(p) = \frac{M_0}{\big[1 + (p/p_0)^2\big]^2}\,.
\label{adfit}
\end{align}
The mass parameters are naturally measured in units of the Coulomb string tension,
$\sqrt{\sigma_C} = 695\,\mathrm{MeV}$ and \footnote{We use a conservative estimate 
$ \sigma_C = 2.5\,\sigma$ for the Coulomb string tension in terms of the Wilson 
string tension\cite{BQR2009}. Other studies favour values up to 
$\sigma_C = 4 \,\sigma$, which would mean that $\sqrt{\sigma_C} \approx M_G 
\approx 880\,\mathrm{MeV}$.}
\begin{align}
SU(2)\,:&\qquad M(0) / \sqrt{\sigma_C} = 0.143\,,
&  p_0 / \sqrt{\sigma_C} &= 0.610 
\\[2mm]
SU(3)\,:& \qquad M(0) / \sqrt{\sigma_C} = 0.190\,,
& p_0 / \sqrt{\sigma_C}&= 0.813\,.
\end{align}
This is shown in Fig.~\ref{fig1}, together with the numerical solution of 
the Adler-Davis equation. In the right panel, we present the 
resulting Fourier transform $h_F(\lambda)$ according to eq.~(\ref{deriv16}).
Other fits for the mass function $M(\vp)$ may give a slightly better 
$\chi^2/\text{dof}$, but this has virtually zero impact on $h_F(\lambda)$. 

Below the chiral phase transition, the quark mass function shows only a mild 
angular dependence caused by the violation of $O(3)$ invariance due to the 
heat bath \cite{Quandt:2018bbu}. This would have to be averaged over angles 
similarly to eq.~(\ref{ave}) and provides only a minor correction to the 
$T=0$ form eq.~(\ref{adfit}). 
Above the chiral phase transition, however, the mass function quickly 
vanishes for all momenta \cite{Quandt:2018bbu}. From eq.~(\ref{deriv11}),
this means that $g_F(p) = p$ and hence $h_F = 1$. 
Recall that the argument of $h_F$ is $\lambda = \mf \beta \simeq \beta$, 
because the $\mf = 1$ term dominates the Poisson sum in the potential 
eq.~(\ref{deriv50}) at virtually all temperatures. In the right panel 
of Fig.~\ref{fig1}, $\lambda$ is measured in units of $\sqrt{\sigma_C}$, 
so that the vanishing mass function would set $h_F = 1$  for all 
$\lambda \le \sqrt{\sigma_C} / T_\chi  \approx 7$, i.e. it would only slightly
suppress the bump at small $\lambda$. This has, however, only a minor effect 
on the effective potential of the Polyakov loop at high temperatures, and 
none at small temperatures $T < T_\chi$. This justifies the use of 
the $T=0$ solution \emph{a posteriori}.

\section{Numerical results at one-loop level}
\label{sec:results1}
\begin{figure}
	\centering
	\includegraphics[width=0.48\linewidth]{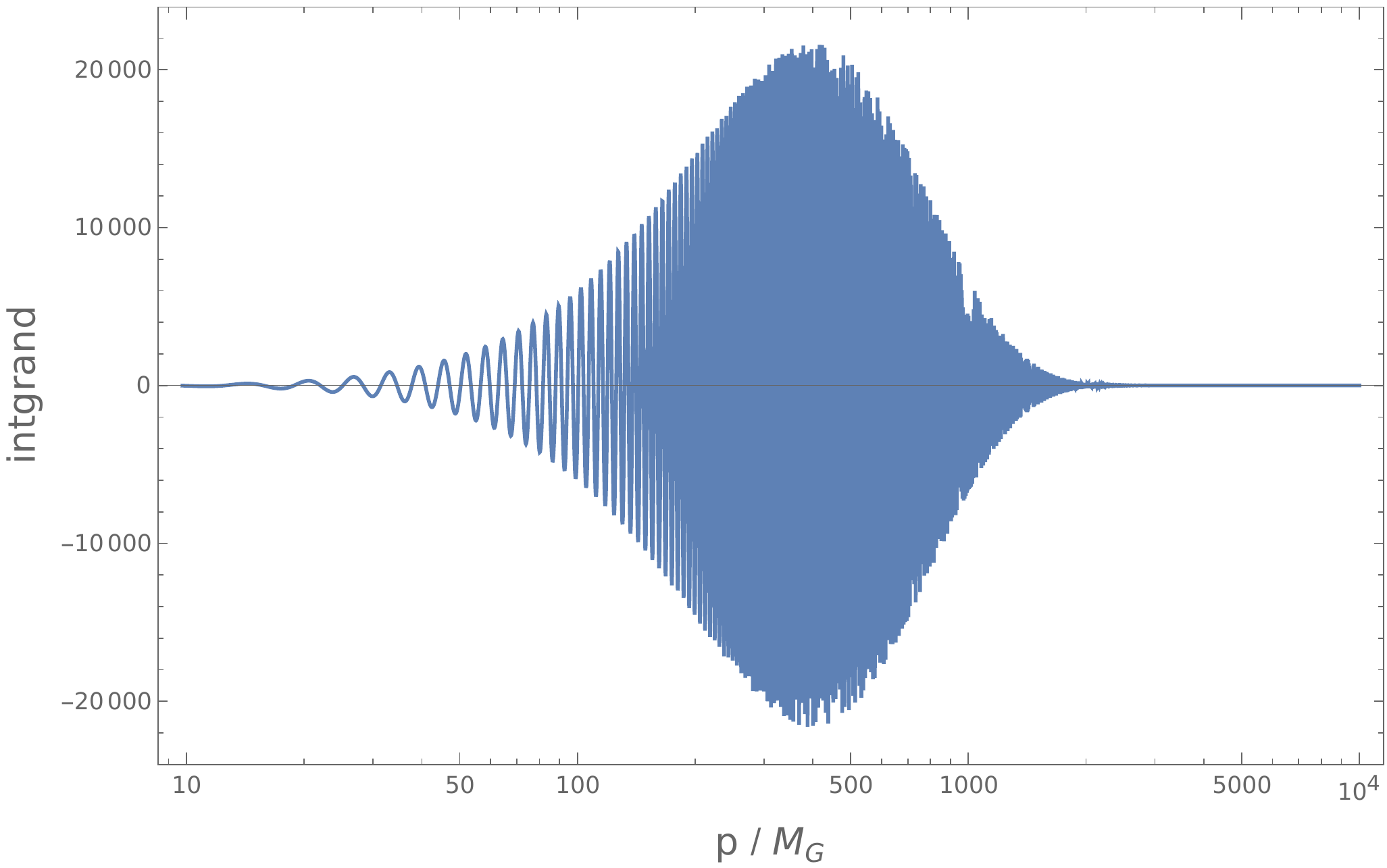}
	\hfill
	\includegraphics[width=0.48\linewidth]{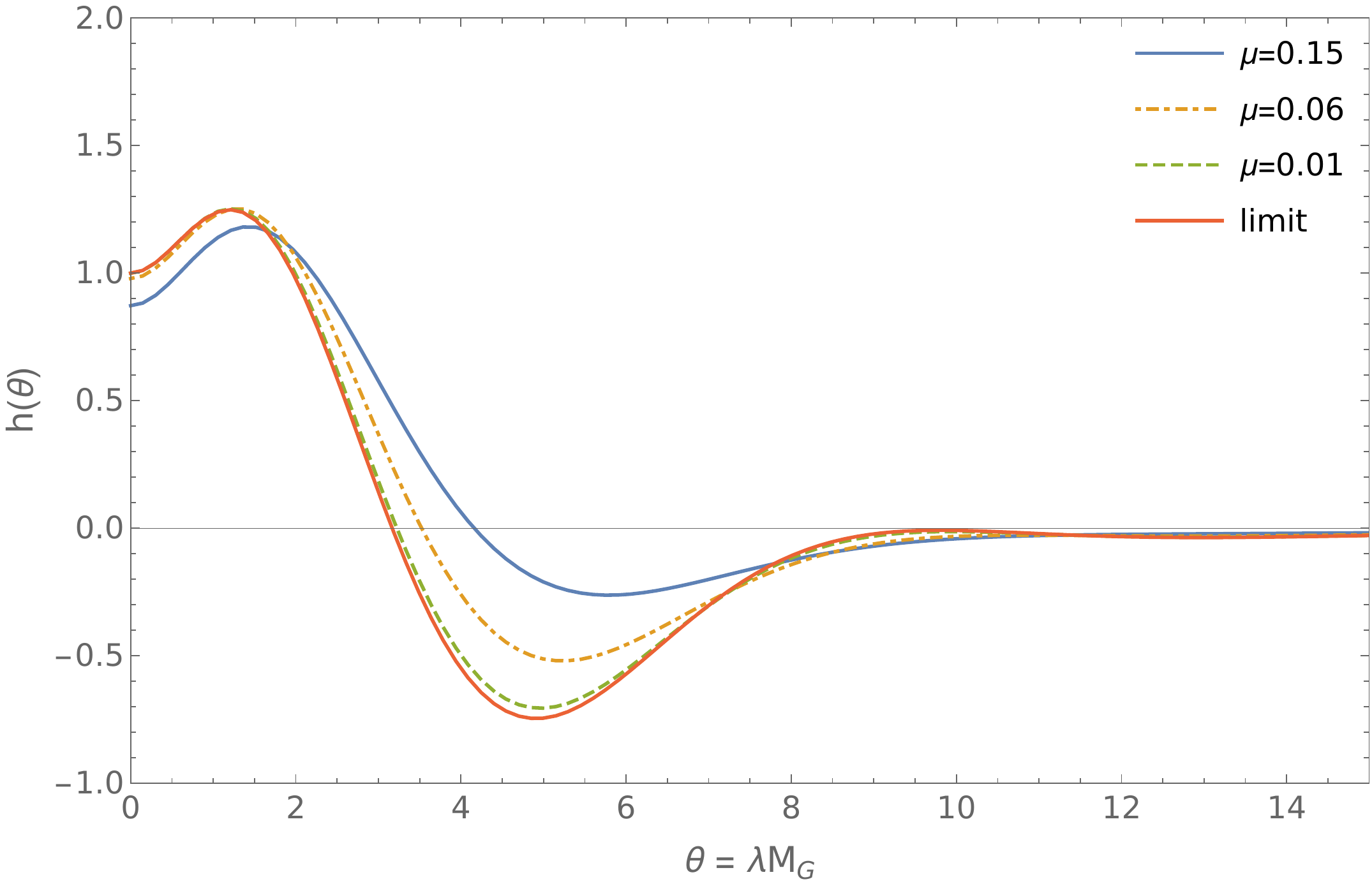}
	\caption{\emph{Left}:The integrand of the Fourier transform eq.~(\ref{hh2})
		for gluons using the Gribov formula. \emph{Right}: Result of the double-exponential 
		Fourier transformation using various regulators.}	
	\label{fig2}
\end{figure}

\noindent
The main numerical challenge at one-loop order is the accuracte computation 
of the Fourier transform 
\begin{align}
h(\lambda) = - \frac{\lambda^3}{2}\int_0^\infty dp\,p \sin(\lambda p) g(p)\,,
\label{hh2}
\end{align}
where $g(p)$ is either $g_B(p)$ eq.~(\ref{gB}) or $g_F(p)$ eq.~(\ref{deriv11}) for gluons and quarks,
respectively. To visualize the problem, the left panel of Fig.~\ref{fig2} shows 
the integrand in eq.~(\ref{hh2}) as a function of the momentum $p$ for the 
Gribov formula eq.~(\ref{gribovx}) at $\lambda = 10$ and 
$c_0 = 0$, for a small regulator $\mu = 0.006 \ll 1$ in the extrapolation
\begin{align}
h(\lambda) = - \frac{\lambda^3}{2} \lim_{\mu \to 0}
\int_0^\infty dp\,p \sin(\lambda p) g(p)\,e^{- \mu p}\,.
\label{hh3}
\end{align}
The numerical issue of the wildly oscillating integrand is clearly visible.
Nonetheless, we have chosen to use the regulator method for our numerical code, 
since integration by parts may involve higher numerical derivatives, which are 
much less reliable. If an analytical expression for $g(p)$ is given, we may actually 
take a combination of first integrating twice by parts (analytically), and then 
applying the regulator method, which yields the best results. The Fourier 
transforms can then be done to high accuracy using double exponential algorithms, 
combined with Richardson extrapolation to the limit $\mu \to 0$. This 
is shown in the right panel of Fig.~\ref{fig2}, where the resulting 
transform $h_B$ for the Gribov formula eq.~(\ref{gribovx}) is plotted as a function 
of the dimensionless variable $\theta = \lambda M_G$ at various regulators 
$\mu \ll 1$. The convergence to the limit is clearly visible but requires 
quite small values for $\mu$. 

Note that the scales on the horizontal axis in Figs.~\ref{fig1} and \ref{fig2} 
are slightly different. Unless stated otherwise, we will non-dimensionalize 
all quantities in the numerical code using appropriate units of the 
Coulomb string tension. The numerical value 
$\sqrt{\sigma_C} \approx 695\,\mathrm{MeV}$ is then used to produce 
absolute numbers in various plots. 

\begin{figure}
	\centering
	\hfill
	\includegraphics[width=0.48\linewidth]{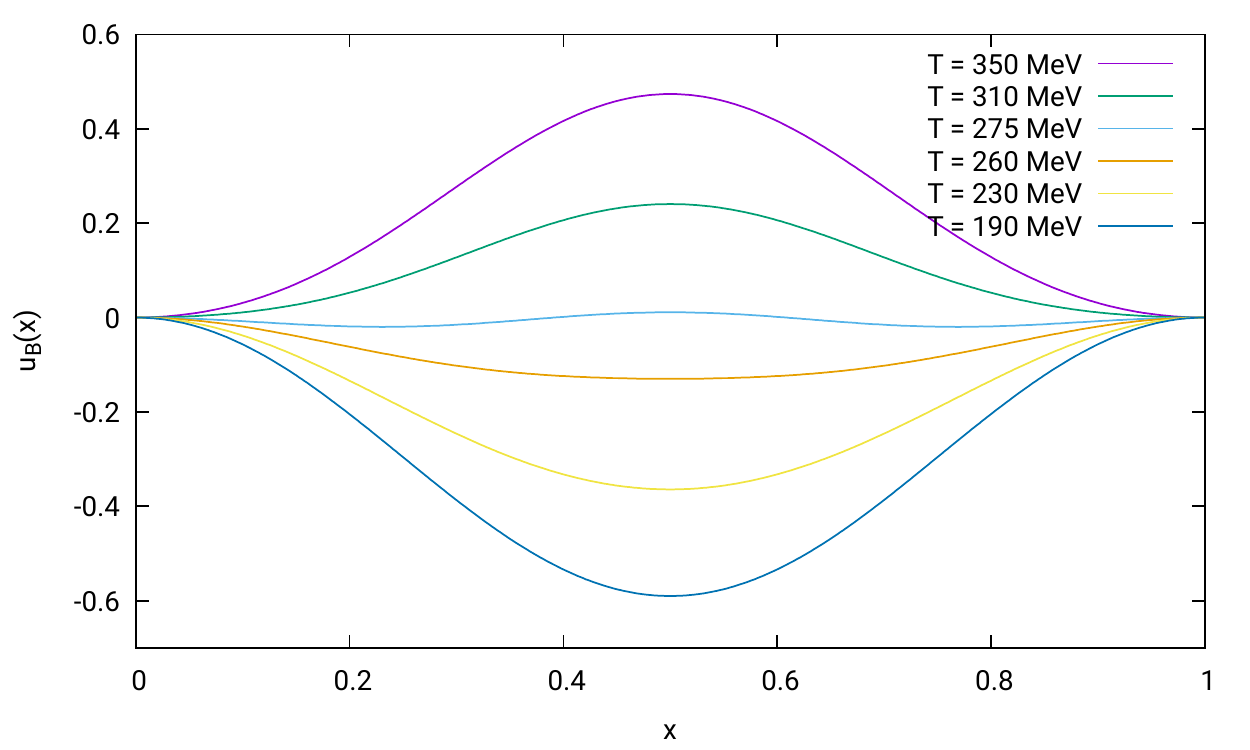}
	\hfill
	\includegraphics[width=0.38\linewidth]{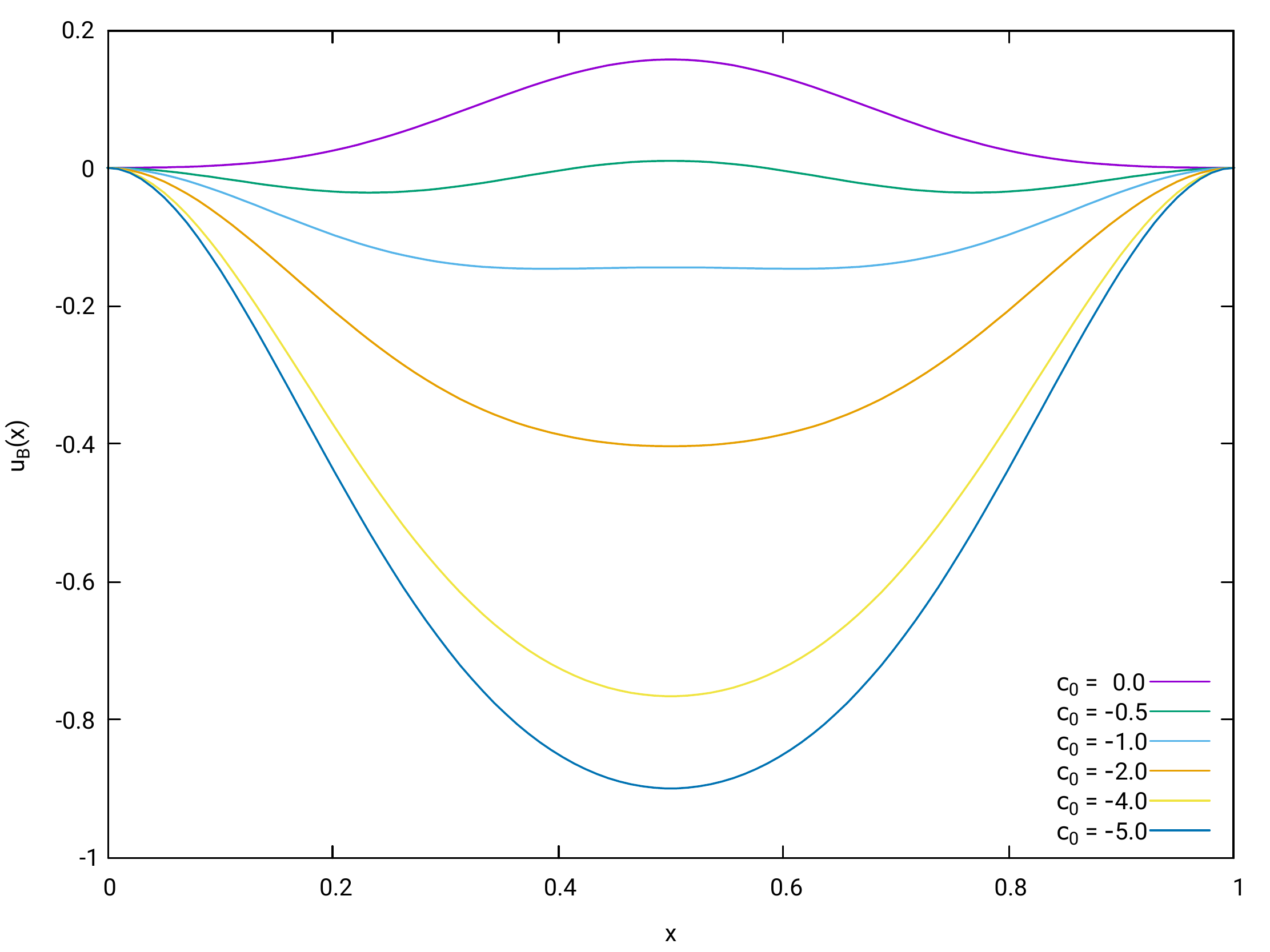}
	\hfill
	\caption{Gluon part of the effective potential of the SU(2) Polyakov loop
		for various temperatues at $c_0 = 0$ (\emph{left}) and for a fixed temperature 
		$T = 302 \,\mathrm{MeV}$ at various renormalization constants $c_0 \le 0$
		(\emph{right}).}	
	\label{fig3}
\end{figure}

\bigskip\noindent
With $h_B(\lambda)$ and $h_F(\lambda)$ at hand, the computation of the effective 
potential of the Polyakov loop is a simple matter of summing the corresponding 
Poisson series, cf.~eq.~(\ref{uBsu2}) and (\ref{deriv14}) for the colour group 
$SU(2)$. Since $h_B(\lambda)$ and $h_F(\lambda)$ are bounded, the Poisson series 
converges at least as $1/\mf^4$ so that very few terms are necessary to saturate
the sum, even at higher temperatures. At lower and intermediate temperatures,
the first term $\mf=1$ in the Poisson series gives already an accurate description,
and we observe e.g.~from eq.~(\ref{uBsu2}) that 
\begin{align}
u_B(x, \beta) \approx \frac{4}{\pi^2}\,\big[1 - \cos(2 \pi x)\big]\,h_B(\beta)\,. 
\label{97}
\end{align}
This has its minimum at the center-breaking points $x \in \{0,1\}$ if $h_B(\beta) > 0$
(deconfinement) and flips over to a minimum at the center symmetric point $x = 1/2 $
(confinement) if $h_B(\beta) < 0$.
The phase transition thus occurs through a sign change in the Fourier transform 
$h_B(\beta)$ and the critical temperature is determined by the zero, 
$h_B(\beta^\ast) = 0$. 
Fig.~\ref{fig3} shows the 
effective potential $u_B(x)$ of the Polyakov loop eq.~(\ref{weylsu2}) for various 
temperatures and $c_0 = 0$. In the right panel of this figure, we have plotted 
the potential at a fixed temperature $T = 302\,\mathrm{MeV}$ for various values of 
$-c_0 \ge 0$. As can be seen, increasing the renormalization parameter $-c_0$ 
makes for a stronger confinement in the boson sector. This also increases the 
critical temperatur $T^\ast$ in the pure Yang-Mills case, because higher temperatures
are necessary to overcome the strong confinement.

\begin{figure}
	\centering
	\includegraphics[width=0.48\linewidth]{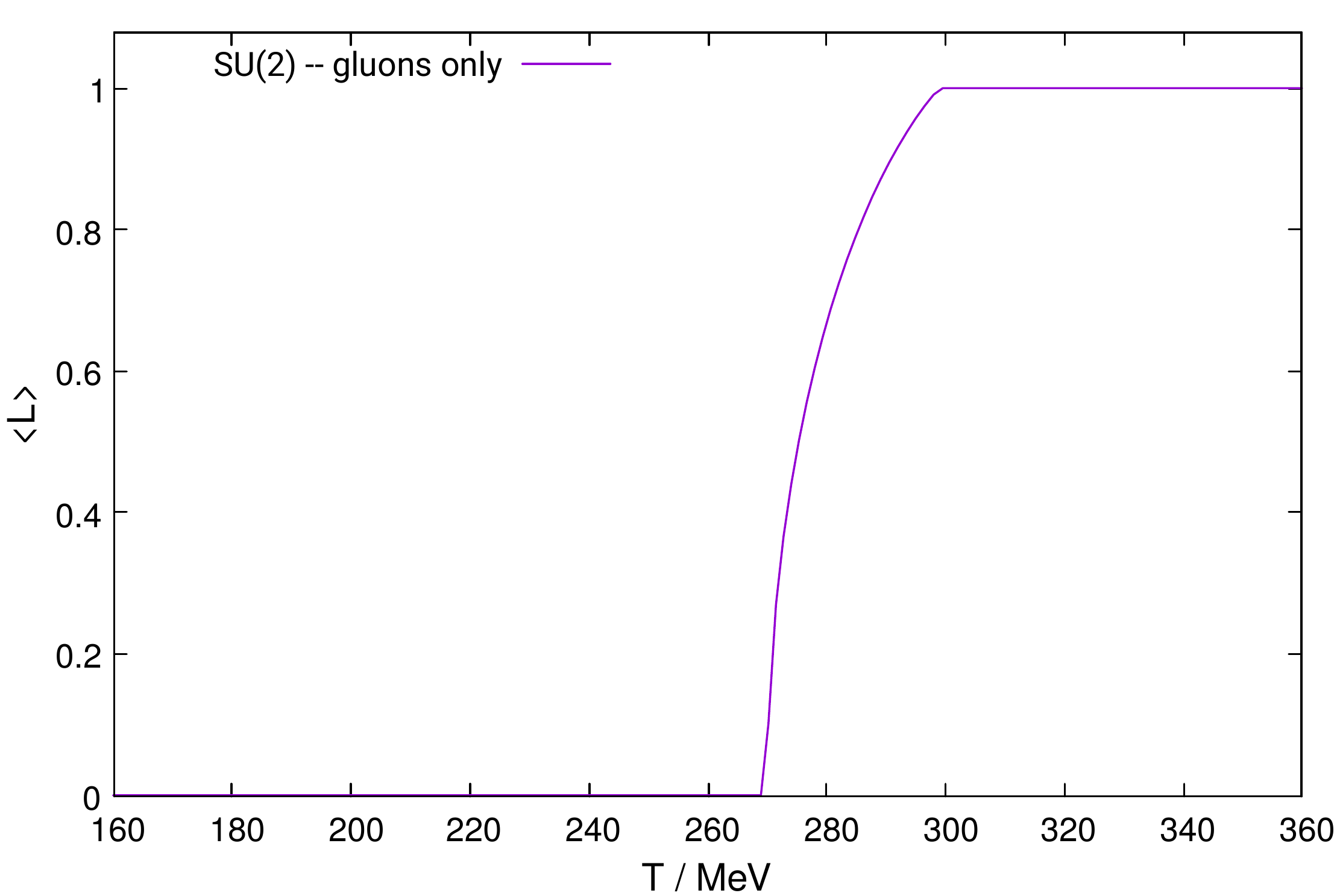}
	\hfill
	\includegraphics[width=0.48\linewidth]{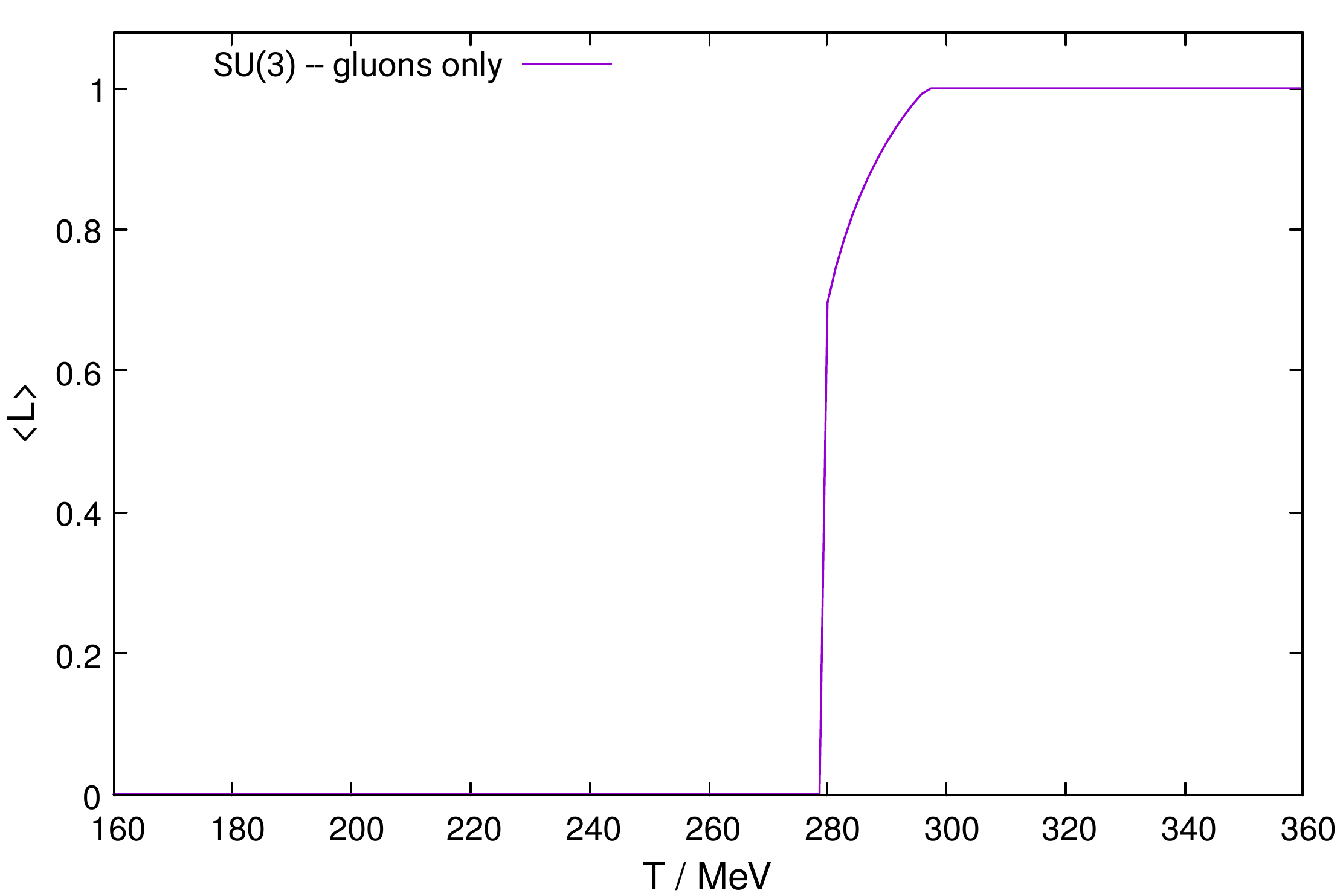}
	\caption{The bosonic contribution to the Polyakov loop at 1-loop level,
		as a function of temperature, for the standard renormalization constant
		$c_0 = 0$.	The left panel is for $SU(2)$ (\emph{left}), 
		and the right panel is for $SU(3)$.}	
	\label{fig4}
\end{figure}

\medskip\noindent
From the minimum $x^\ast$ of the Polyakov loop potential, we can compute the expectation 
value of the Polyakov loop itself through
\begin{align}
SU(2) : \quad & \quad 
\langle L \rangle = \frac{1}{2} \,\mathrm{tr}
\exp\left( - \beta \ab_3^\ast \,\sigma_3 / 2i\right) = \cos(\pi x^\ast)
\\[2mm]
SU(3) : \quad & \quad 
\langle L \rangle = \frac{1}{3}\,\mathrm{tr}
\exp\left( - \beta \ab_3^\ast \,\lambda_3 / 2i - \beta \ab_8^\ast\,\lambda_8 / 2i\right) =
\frac{1}{3}\,\sqrt{1 + 4\,\cos(\pi x^\ast)\,\big[ \cos(\pi x^\ast) + 
\cos\big(2 \pi\,y^\ast / \sqrt{3}\big)\big]}\,.
\end{align}
The result is plotted for $c_0 = 0$ in Fig.~\ref{fig4}. We observe the well-known
second order phase transition for $G=SU(2)$ at a critical temperature of 
$T^\ast \approx 266  \,\mathrm{MeV}$, while the transition is 
first order for $G=SU(3)$ with a critical temperature of 
$T^\ast \approx 278\,\mathrm{MeV}$.

\begin{figure}
	\centering
	\includegraphics[width=0.48\linewidth]{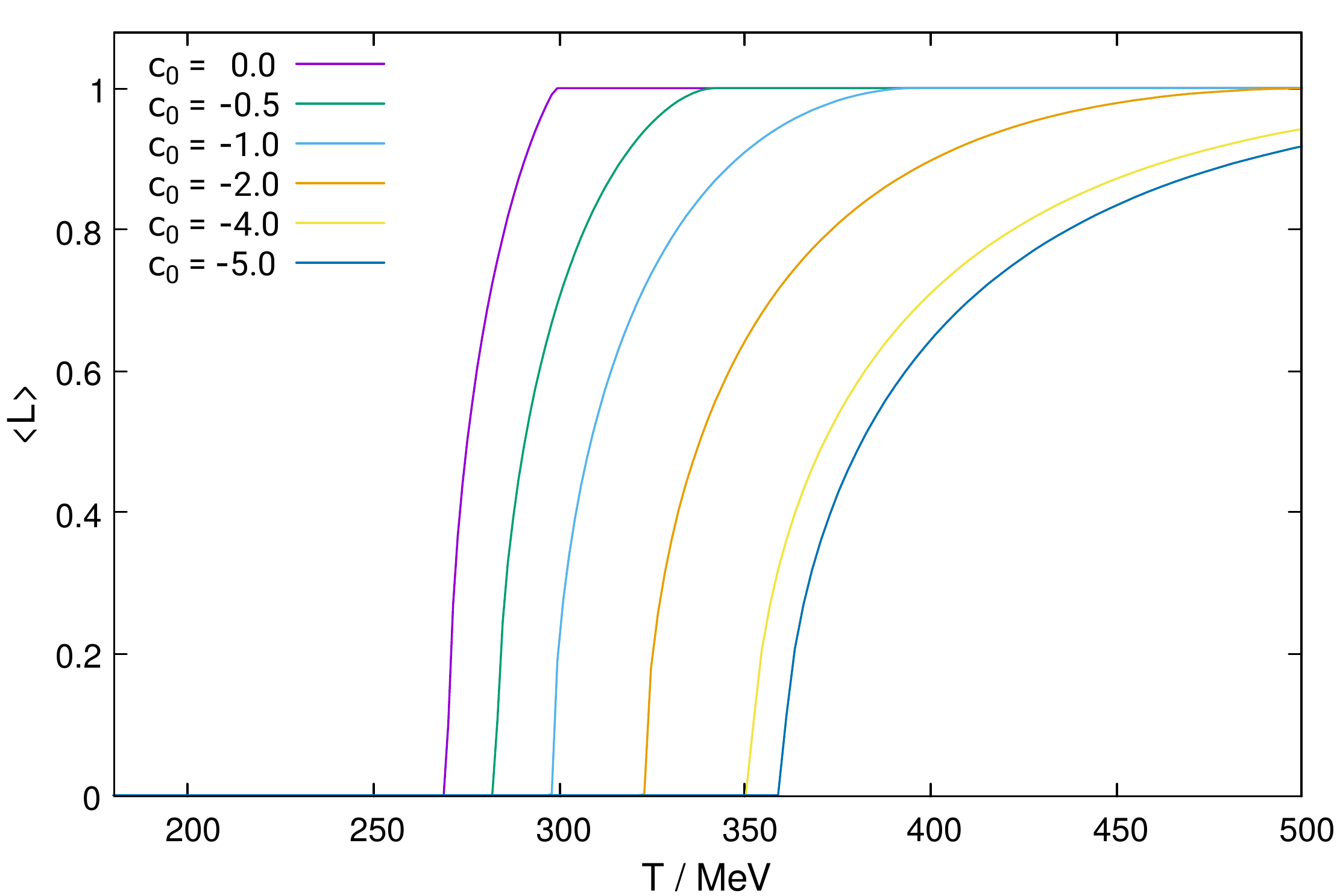}
	\hfill
	\includegraphics[width=0.48\linewidth]{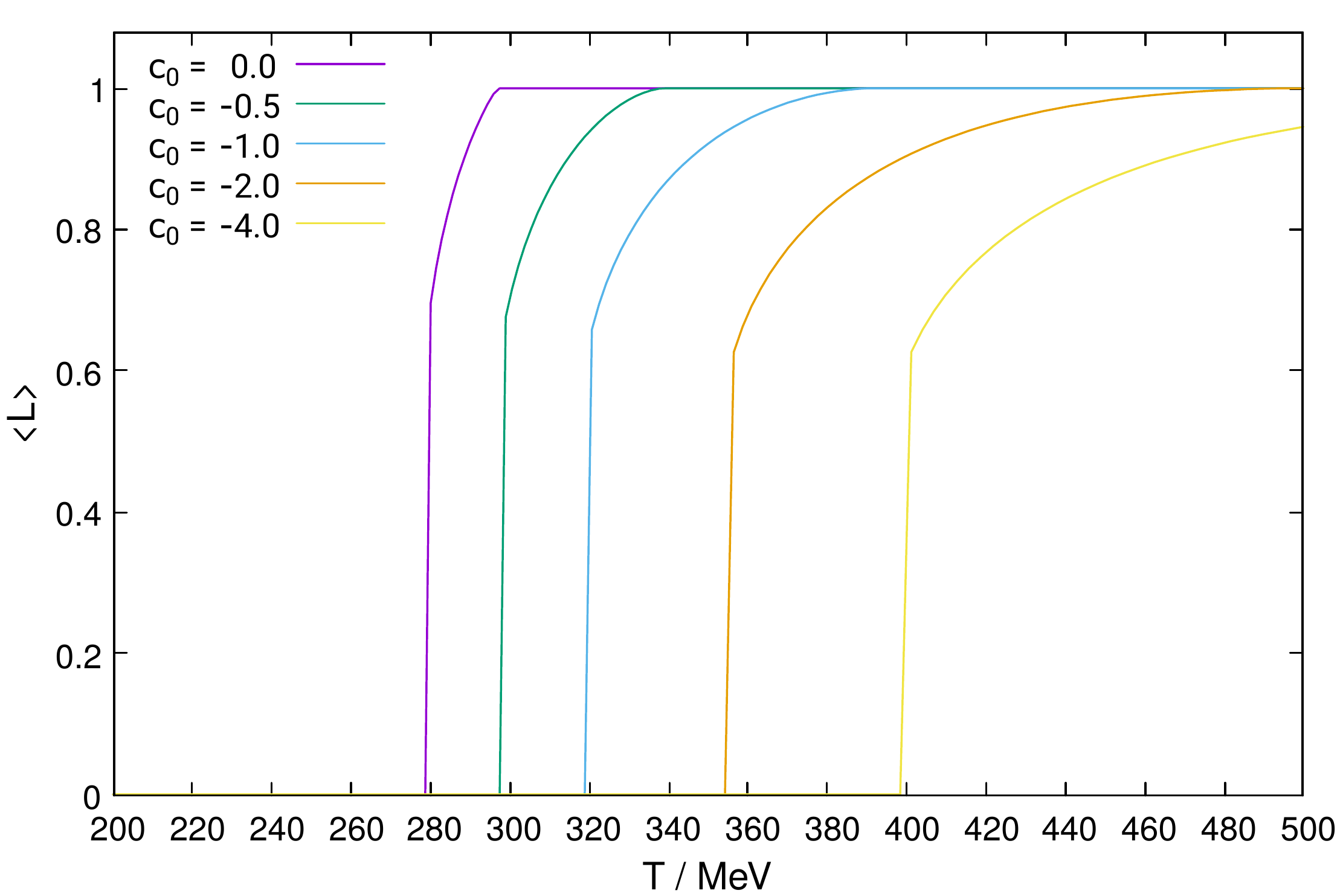}
	\caption{The bosonic contribution to the Polyakov loop at 1-loop level
		as a function of temperature, for various values of the 
		renormalization parameter $c_0$.
		The left panel is for $G=SU(2)$ (\emph{left}), and the right 
		panel is for $G=SU(3)$.}	
	\label{fig5}
\end{figure}

\medskip\noindent
As explained earlier, increasing the renormalization parameter $-c_0$ 
makes for a stronger gluon confinement, i.e.~the critical temperature 
in the pure Yang-Mills case increases. This can be seen in Fig.~\ref{fig5} 
where the Polyakov loop is plotted at various values of $c_0$. 
The critical temperature measured on the lattice for $G=SU(2)$ is 
$T^\ast = 306\,\mathrm{MeV}$, which indicates that a value
in the range $-c_0 \simeq 1.0\ldots 2.0$ is compatible with the lattice 
and describes the transition at least as well as the standard choice 
$c_0= 0$. For $G=SU(3)$ shown in the right panel of Fig.~\ref{fig5},
the agreement with the lattice favours the standard value $c_0 = 0$, 
but the good agreement with the transition temperature $T^\ast \approx 
278\,\mathrm{MeV}$ on the lattice must be considered accidential given 
our approximations. Within the expected accuracy of our calculation, 
values up to $c_0 \simeq -1.5$ are still compatible with the lattice 
findings. The bottom line is that the acceptable range for the 
renormalization parameter $c_0$ is, for both colour groups, 
about $c_0 \in [-1.5,\ldots,0]$, and the value $c_0 = 0$ used in earlier
studies is usually preferred, at least for $SU(3)$.

\bigskip\noindent
Next, let us include the fermion contribution at one-loop level.
Numerically, the calculation of the quark contribution is very 
similar to the gluon case, with the  mass function $M(p)$ entering
first the function $g_F(p)$ in eq.~(\ref{deriv11}), which is then
Fourier transformed, using the techniques described earlier, 
into the profile function $h_F(\lambda)$ in eq.~(\ref{deriv16}).
This function plays a similar role as in the boson case: 
it provides a profile prefactor $h_F(\beta)$ for the dominant 
term in the effective potential of the Polyakov loop,
eqs.~(\ref{deriv50}) and (\ref{deriv80}). 
The positive sign of $h_F$ in Fig.~\ref{fig1} thus indicates 
deconfinement at all temperatures, which is expected on physical
grounds: quarks should turn the phase transition into 
a soft \emph{crossover} while leaving confinement intact at
small temperatures below the dynamical quark mass. 

\begin{figure}
	\centering
	\includegraphics[width=0.48\linewidth]{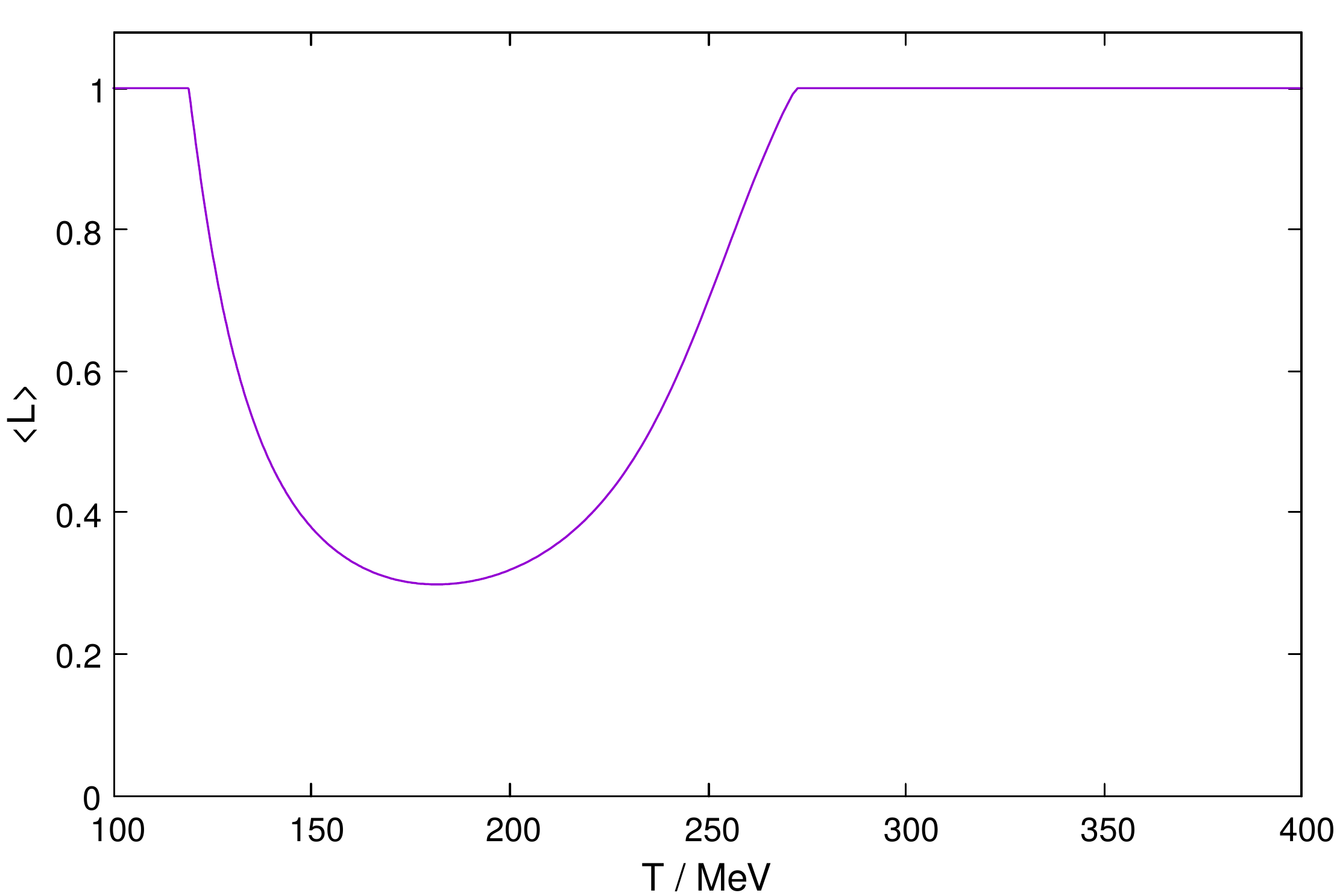}
	\hfill
	\includegraphics[width=0.48\linewidth]{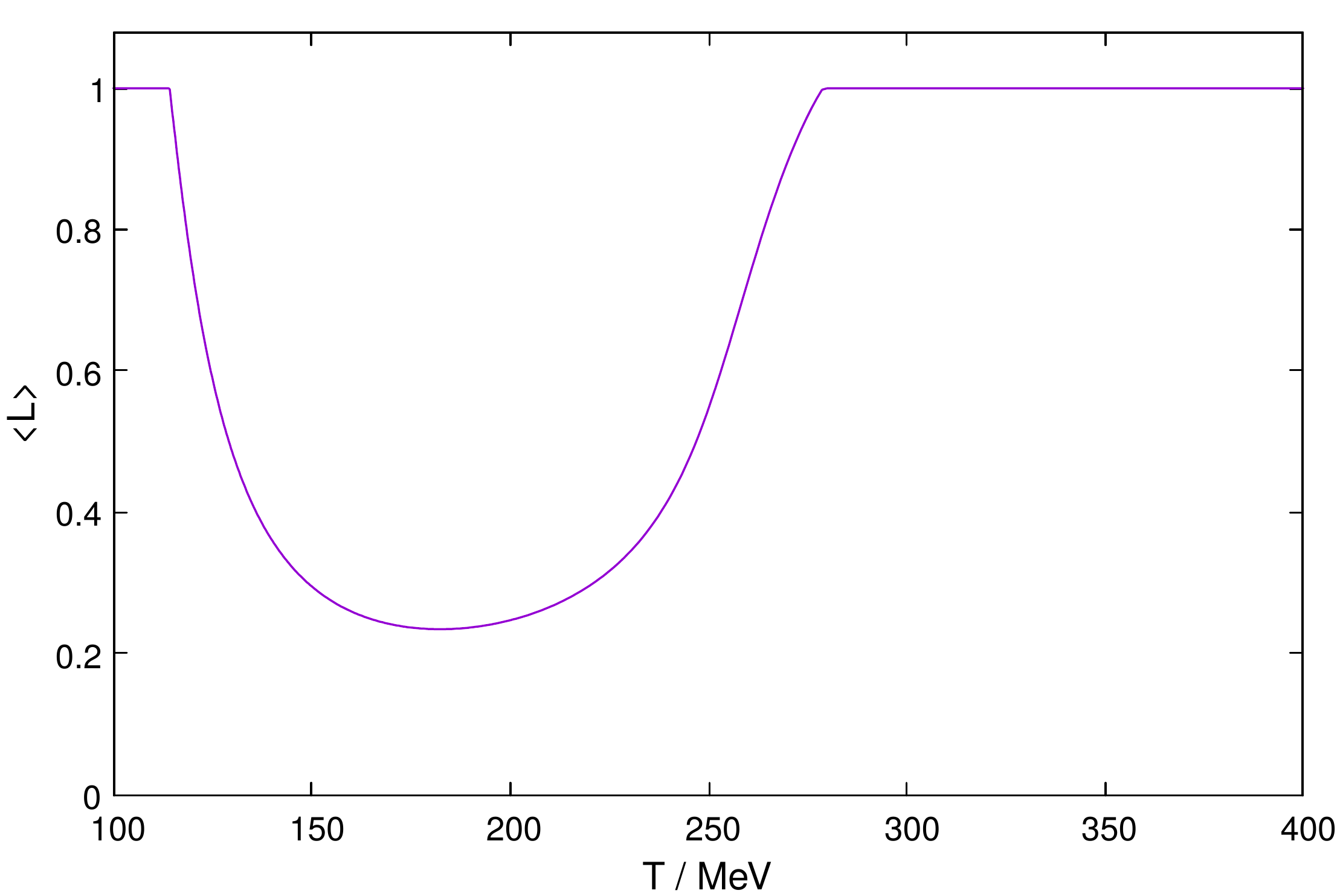}
	\caption{The 1-loop contribution to the Polyakov loop in full QCD
		as a function of temperature for $N_f = 1$ and $c_0 = 0$. 
		The left panel is for $G=SU(2)$ (\emph{left}), and the right 
		panel is for $G=SU(3)$.}	
	\label{fig6}
\end{figure}

Surprisingly, these reasonable expectations are \emph{not}
fully met at one-loop level, as can be seen in Fig.~\ref{fig6}: 
while the transition is indeed softened into a crossover, the 
quarks start to dominate at temperatures below $T^\ast/2$, 
so that the confinement eventually breaks down and the Polyakov 
loop approaches $\langle \LL \rangle = 1$ again. This occurs in
the same way for both colour groups $SU(2)$ and $SU(3)$. 
The situation becomes even worse when the number $N_f$ of 
(light) quark flavours is increased. 
 
\bigskip\noindent
To understand the physics behind these unexpected findings, 
consider the gluon 1-loop contribution to the effective potential, eq.~(\ref{97}).
As explained in the paragraph after eq.~(\ref{97}) confinement is a 
consequence of a negative sign in $h_B(\beta)$, which effectively flips the 
shape of the effective potential. From our discussion of the profile $h_B$ 
in section \ref{sec:renorm1}, we also know that this quantity essentially 
measures the number of active massless particles in the spectrum, 
in the sense that each free massless particle contributes $(+1)$ to $h_B$ 
(ghosts contribute with a negative sign).
This is the basis of the confinement mechanism in covariant functional 
approaches \cite{BGP2010, MP2008, BH2012, RH2013, Fischer2009, FM2009, FMM2010, RSTW2016,Quandt2016,Canfora2015}:  Perturbatively, there are 3 covariantly 
transversal gluon modes, one massless longitudinal mode which decouples 
from the dynamics, and two ghost degrees of freedom, for a total of 
$3+1-2=2 > 0$, which is reflected in $h_B \to 2$ at high temperatures. As we lower 
the temperature, the three transversal modes become massive through 
interactions and eventually decouple, so that the mode count is 
$0+1-2 = -1 < 0$. We now have $h_B \to -1$ and hence confinement.
The salient point here is that confinement is caused predominantly by 
the ghost degrees of freedom and this makes for a very \emph{strong} 
confinement, which cannot be overcome by quarks at low temperature, 
since the quarks become massive at $\beta \to \infty$ and 
hence tend to $h_F \to +0$, cf.~Fig.~\ref{fig1}. 
The quarks thus soften the transition, but cannot overcome the strong 
confinement caused by \emph{ghost dominance}.

By contrast, the Hamiltonian approach predicts a profile $h_B(\beta)$ 
which is strongly negative (confining) at intermediate temperatures, but 
approaches $h_B \to -0$ at low temperatures (large $\lambda$), 
cf.~Fig.~\ref{fig2}. 
This means that the one-loop confinement in the Hamiltonian approach to 
Yang-Mills theory is actually very \emph{weak} or \emph{fragile} at low 
temperatures. This fragility does \emph{not} show up in the Polyakov loop, 
because the effective potential still attains its tiny negative minimum at 
$x = \frac{1}{2}$, so that $\langle \LL \rangle = 1$. However, even the 
smallest deconfining effect, such as $N_f = 1$ flavour of    quarks, can overcome 
the weak confinement. 

The Hamiltonian approach has only physical (transversal) gluon modes, and there
is no trace of ghost dominance at small temperatures. In fact, the ghosts tend 
to nullify the strong gluon confinement induced by the Gribov propagator, 
which can be seen directly from eq.~(\ref{eb1}): without the curvature, 
confinement would be strong, actually \emph{too} strong as the transition 
temperature would increase to unreasonable values. There is hence a 
\emph{ghost compensation} rather than a \emph{ghost dominance} at 
one-loop level in the Hamiltonian approach, and the confining phase is
essentially devoid of light degrees of freedom. This leads to problems 
with confinement, but it may actually be closer to the true physical picture:
in the ghost dominance scenario, the abundance of massless ghost particles 
leads to unphysical results for most thermodynamic quantities such as 
a remanent pressure or a negative energy density below the phase transition 
\cite{Quandt:2017poi}. This happens because the true physical picture is an 
exponentially suppressed partition function and a vanishing pressure, as the 
lightest colourless glueball excitation has a mass of well above $1\,\mathrm{GeV}$.
A vacuum of suppressed (compensated) excitations, as in the Hamiltonian approach, 
is hence not completely without physical merits.

\begin{figure}
	\centering
	\includegraphics[width=0.48\linewidth]{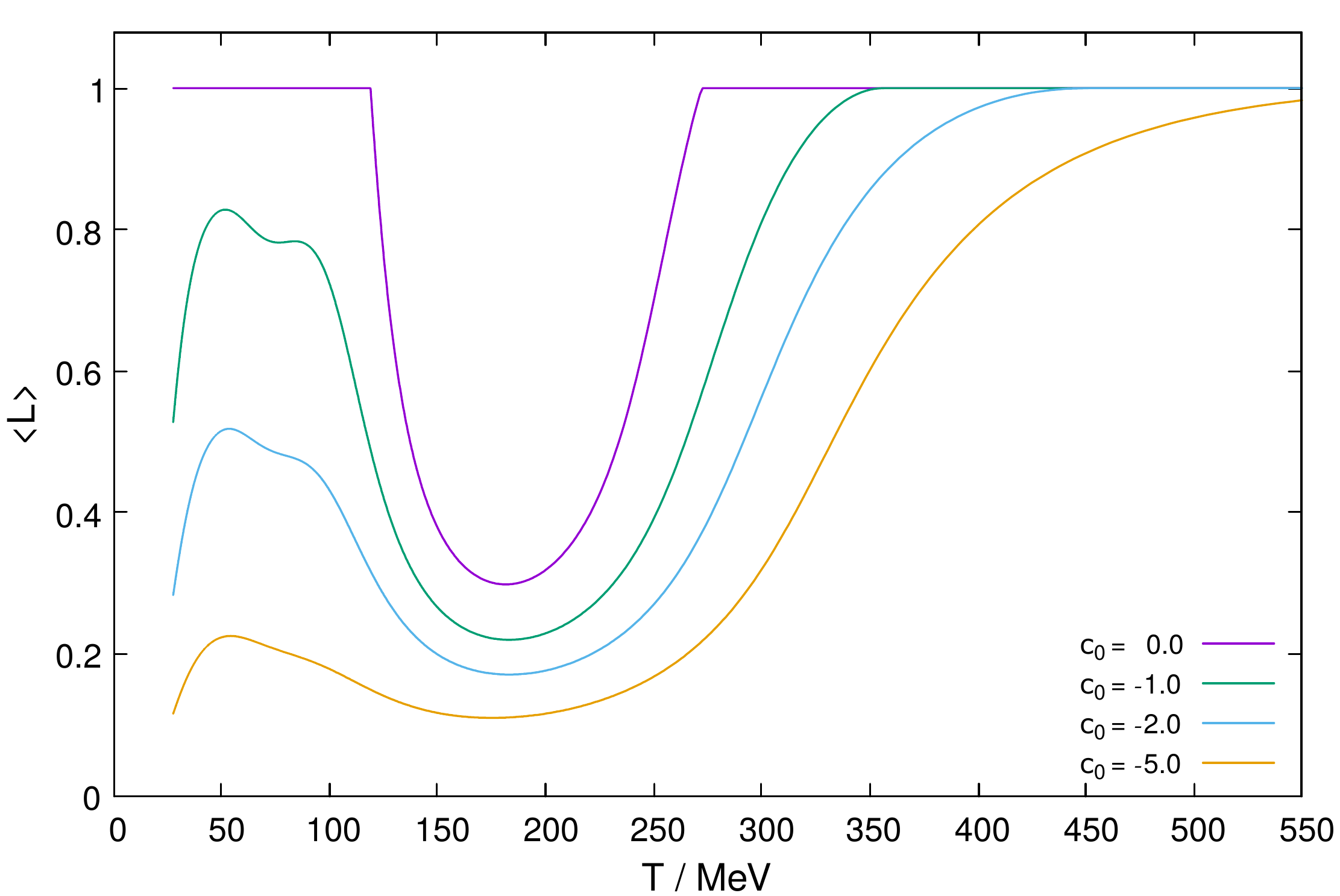}
	\hfill
	\includegraphics[width=0.48\linewidth]{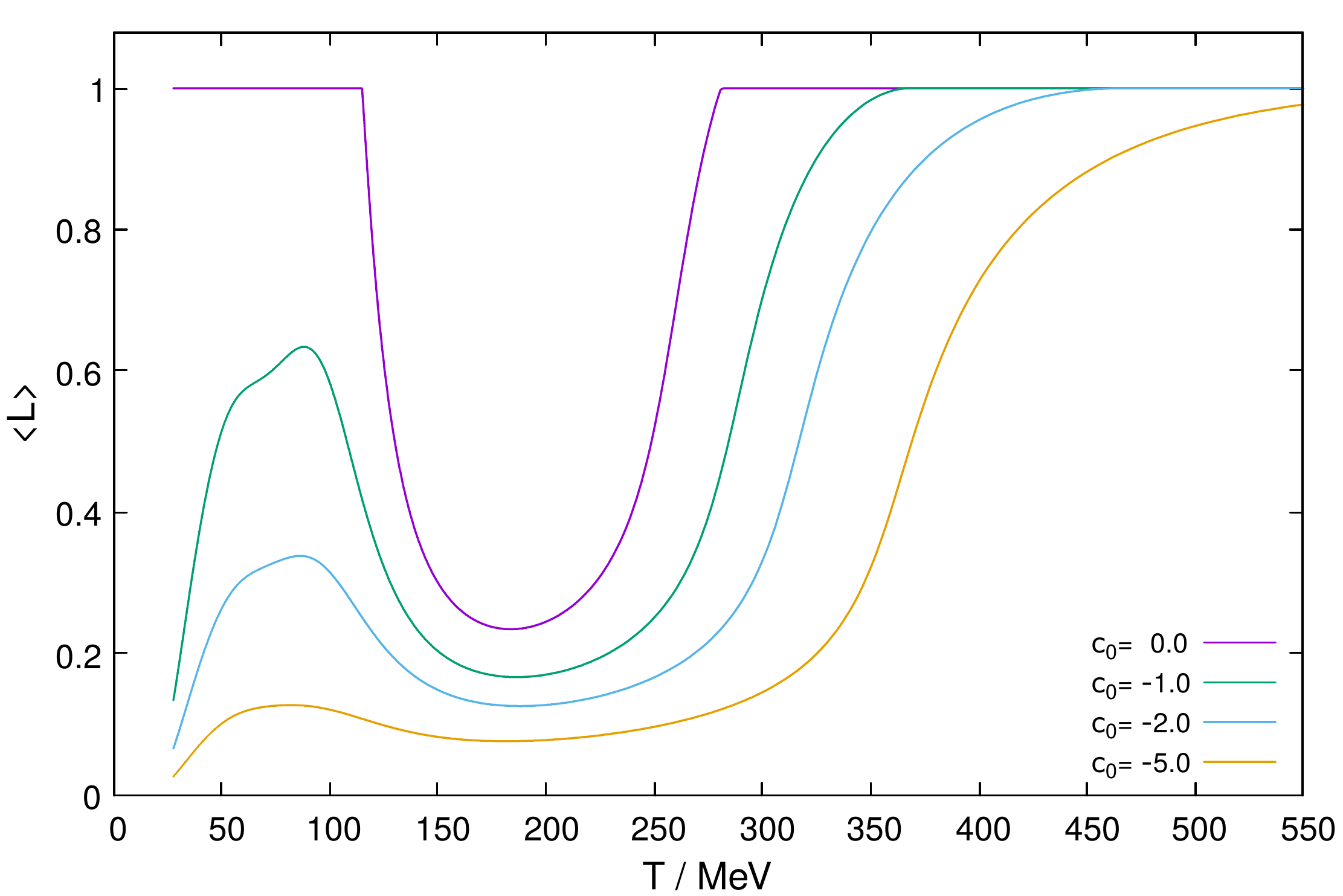}
	\caption{The Polyakov loop in full QCD including all one-loop contributions,
		for $N_f = 1$ flavours and various values of the renormalization parameter 
		$c_0$. The left panel is for  $SU(2)$, and the right panel for $SU(3)$.}   
	\label{fig7}
\end{figure}

Still, we need to fully understand the mechanism of how confinement comes about 
in the Hamiltonian approach. Modifications to the quark sector such as 
explicit coupling to gluons in the variational ansatz will not change the 
physical picture qualitatively
-- quarks will still be deconfining and $h_F \to +0$ at low temperatures. 
The real cause of the problem is not the strength of the quark deconfinement, 
but rather the weakness of the gluon confinement. This can be seen when 
improving the confining strength through the undetermined counter term 
$c_0$, cf.~Fig.~\ref{fig7}. As can be clearly seen, all curves with 
$c_0 < 0$ tend to $\langle \LL \rangle \to 0$ at very small temperatures. 
Thus, even a very small perturbation of the delicate balance between
gluons and ghosts through the counter term $c_0 < 0$ is sufficient to 
eventually restore confinement. This indicates that the inclusion of 
two-loop gluon contributions, though mostly irrelevant for the gap 
equation, may just give sufficient contributions to the gluon energy 
to restore enough gluon confinement for a decent physical picture. 
In the next section, we will therefore study the qualitative effects 
of the gluon 2-loop contributions to the effective action of the Polyakov 
loop.\footnote{Recall that $c_0$ is the finite part of the counter term to the 
non-Abelian magnetic energy, which is a two-loop contribution. There is also a 
two-loop contribution to the quark sector from the Coulomb term in eq.~(\ref{EFERM}). 
It has partially been included in our fermion 1-loop calculation through the 
self-interaction in the gap equation, and is not expected to contribute to the 
solution of the weak gluon confinement. We therefore defer its study to a future 
investigation.}  

\section{Gluonic two-loop contributions} \label{sec:loop2}
In the previous sections, we have repeatedly stressed that the gap equation 
mixes loop orders, and the self-consistent one-loop contribution to the energy actually 
contains parts of the two-loop terms. This rises the question whether the self
consistent two-loop contribution must be corrected to avoid double counting.
As explained in appendix \ref{app:sc}, the two-loop energy $E_2$ (valid for \emph{all} kernels) 
\emph{differs} from the self-consistent two-loop energy $E_2^{\rm sc}$ (valid only for solutions 
of the gap equation) by a subtraction which compensates for the two-loop terms 
moved from $E_2$ into $E_1^{\rm sc}$ via the gap equation. In other words,
$E_1 + E_2 = E_1^{\rm sc} + E_2^{\rm sc}$ for solutions of the gap equation, and 
we must use $E_2^{\rm sc}$ if we also used the self-consistent 1-loop energy, and 
$E_2$ otherwise.

In our present investigation, this subtlety does not matter: the only two-loop term 
included in the gap equation is the $T=0$ contribution $c_0$ from the tadpole term, 
i.e.~the non-Abelian magnetic field. (We do not include the Coulomb term or any 
finite temperature corrections to the gap equation.) This means that the 
self-consistency correction must only be applied to this particular two-loop term. 
As further explained in appendix \ref{app:sc}, the correction is substantial and 
would actually flip the sign of the $T=0$ tadpole in $e_B[\va]$. Since the 
contribution is, however, independent of temperature and the background field, 
it drops out when computing 
the Polyakov loop from the \emph{change} in the energy, $e_B[\va] - e_B[0]$. 
The bottom line is hence that $E_2 = E_2^{\rm sc}$ in the present study, and 
no self-consistency correction should be applied to any two-loop term. 
 
\subsection{The non-Abelian magnetic energy}
\label{sec:loop2NA}
\noindent
From eq.~(\ref{EBOS}), there are two 2-loop contributions to the gluon sector.
In the present subsection, we first study the term arising from the non-Abelian part of 
the colour magnetic field, 
\begin{align}
\langle H^{\rm NA}_{\rm YM}\rangle_\ab = 
\left\langle \frac{1}{4 g^2} \int d^4x\,\big[ A_i(\vx), A_k(\vx) \big]\,
\big[ A_i(\vx), A_k(\vx) \big]\right \rangle_\ab\,.
\end{align}
Since our trial wave functional eq.~(\ref{omegab}) in the bosonic sector is Gaussian,
Wick's theorem entails
\begin{align}
\langle H^{\rm NA}_{\rm YM}\rangle_\ab = \frac{f^{eab} f^{ecd}}{4 g^2} \int d^3 x\,
\Big[ D^{ab}_{ik}(\vx,\vx) D^{cd}_{ik}(\vx,\vx) + D^{ac}_{ii}(\vx) D^{bd}_{kk}(\vx)
+ D^{ad}_{ik}(\vx,\vx) D^{bc}_{ki}(\vx,\vx) \Big]\,,
\label{pal1}
\end{align}
where $D^{ab}_{ik}(\vx,\vy) = \langle A_i^a(\vx)\,A_k^b(\vy)\rangle_\ab$ is the gluon 
propagator in the presence of a background field, and $f^{abc}$ are the structure 
coefficients of the colour algebra.

\medskip
Let us first consider the case $T=0$ without a background field. Global 
colour and Lorentz invariance implies that the gluon propagator is colour diagonal
and transversal in this case, i.e.~we have in momentum space
$D^{ab}_{ik}(\vp) = \delta^{ab}\,t_{ik}(p)\,D(p)$. With our trial 
wave functional (\ref{omegab}), the propagator is $D(p) = g^2 / [2\omega(p)]$ 
in terms of the variational kernel $\omega(p)$. We note in passing  that the first 
term in eq.~(\ref{pal1}) vanishes due to the colour symmetry of the propagator. 
For the remaining terms, it is easy to work out the colour and Lorentz traces 
which results in
\begin{align}
\langle H^{\rm NA}_{\rm YM}\rangle_0 = \frac{N(N^2-1)}{16}\,g^2 V_3 \int 
\frac{d^3p}{(2\pi)^3}\int \frac{d^3 q}{(2\pi)^3}\,\frac{3 - (\hat{p}\,\hat{q})^2}
{\omega(p)\,\omega(q)}\,.
\label{pal5}
\end{align}
The gap equation follows from the variation of the energy functional with respect 
to the gluon propagator. For the non-Abelian magnetic energy, this yields
\begin{align}
\frac{\delta}{\delta \omega(k)^{-1}} \langle H^{\rm NA}_{\rm YM}\rangle_0
= \frac{N^2-1}{2\,(2\pi)^3}\,V_3 \cdot\frac{Ng^2}{4}\int \frac{d^3q}{(2\pi)^3} \,
\frac{3 - (\hat{k}\hat{q})^2}{\omega(q)}\,.
\label{pal2}
\end{align}
Note that we have pulled out a prefactor which is common to all contributions to the gap equation, 
cf.~eq.~(\ref{fracx}). To renormalize eq.~(\ref{pal2}), we must add the mass counterterm
(\ref{Ect}) to the original Hamiltonian. Its contribution to the total energy can also be 
evaluated by Wick's theorem,
\begin{align}
\langle H_{\rm ct} \rangle_0 = 
\frac{C_0}{2 g^2}\cdot 2 (N^2-1) V_3 \int \frac{d^3q}{(2\pi)^3} \,D(q) = 
\frac{N^2-1}{2}\,V_3\,C_0\,\int \frac{d^3q}{(2\pi)^3}\frac{1}{\omega(q)}\,.
\label{pal52}
\end{align}
The variation with respect to the gluon propagator gives a constant
\begin{align}
\frac{\delta}{\delta \omega(k)^{-1}} \langle H_{\rm ct}\rangle_0  =  \frac{N^2-1}{2\,(2\pi)^3}\,V_3 \cdot C_0\,.
\label{pal3}
\end{align}
We note that the same prefactor as in eq.~(\ref{pal2}) has appeared. The condition 
that the counter term cancels the divergence in eq.~(\ref{pal2}) is satisfied if
$C_0$ depends on the cutoff (not the momentum) in such a way that\footnote{After 
the angular integration, the loop integral is independent of the external 
momentum $k$ and depends only on the UV cutoff $\Lambda$.}
\begin{align}
c_0 \equiv C_0 + \frac{Ng^2}{4}\int \frac{d^3q}{(2\pi)^3}\, 
\frac{3 - (\hat{k}\hat{q})^2}{\omega(q)}
\label{pal4}
\end{align}
is finite. The finite coefficient $c_0$ is precisely the 
renormalization parameter of the same name introduced earlier in eq.~(\ref{palx})
of the previous section.

\medskip
We will now show that eq.~(\ref{pal4}) is also sufficient to render the entire 
non-Abelian magnetic energy finite, even in the presence of a background field 
and at finite temperature. The modifications to the non-Abelian magnetic energy 
eq.~(\ref{pal5}) necessary to account for for finite temperature and a constant
background field in the Cartan subalgebra follow the techniques described 
at length in section \ref{sec:loop1}: we introduce finite temperature in the 
loop integrals by compactifying the (spatial) direction of the heat bath.
Furthermore, we replace the colour trace by a sum over all roots and note that 
the constant background in the Cartan subalgebra only enters via the covariant 
derivative. After Fourier transformation, this amounts to (i) replacing the 
integration in the direction of the heat bath by a Matsubara sum and (ii) 
shifting the momentum arguments in the kernels $\vp \to \vp^\sigma$ as in 
eq.~(\ref{shift}). After Poisson resumming the Matsubara sum and shifting 
the loop integration, we end up with
\begin{align}
\langle H^{\rm NA}_{\rm YM}\rangle_\ab = 
\frac{g^2}{16}\,V_2 \beta \int 
\frac{d^3p}{(2\pi)^3}\int \frac{d^3 q}{(2\pi)^3}\,\frac{3 - (\hat{p}\,\hat{q})^2}
{\omega(p)\,\omega(q)}\,\sum_{\rho,\sigma,\tau} | f_{\rho,\sigma,\tau}|^2 
\sum_{\nf = - \infty}^\infty 
\exp\big[ i \nf \beta \,(q_3 +  \xvec{\sigma}\cdot \xvec{\ab})\big]
\sum_{\mf = - \infty}^\infty 
\exp\big[ i \mf \beta \,(q_3 +  \xvec{\tau}\cdot \xvec{\ab})\big]\,.
\label{pal6}
\end{align}
For the simple case of colour group $SU(2)$, the structure constants $f_{\rho,\sigma,\tau}$ 
in the Cartan basis read
\begin{align}
f_{\rho,\sigma,\tau} = \epsilon_{\rho \sigma \tau} = \{ \epsilon_{-1,0,1} = 1\,\,
\text{and cyclic}\}\,,\qquad\qquad \rho,\sigma,\tau \in \{-1,0,1\}\,.	
\label{pal7}
\end{align}
Further details can be found in appendix \ref{app:cartan}. For all colour groups 
$SU(N)$ we have  $|f_{\rho,\sigma,\tau}|^2 = N (N^2-1)$
when summed over all roots. This shows that the $\mf = \nf = 0$  term 
in eq.~(\ref{pal7}) agrees with the $T=0$ vacuum contribution eq.~(\ref{pal5}). 
This term is independent of the background field and will drop out once the 
change of the free energy due to the background is considered, cf.~eq.~(\ref{deriv2}).
Of the remaining terms, only the ones are singular in which one of the two Poisson 
indices vanishes. (This is intuitively clear, but the proof is rather technical and thus 
deferred to appendix \ref{app:nonmag}.)
	
Using eq.~(\ref{schnabel}) for the colour trace, the singular terms in eq.~(\ref{pal6}) read
\begin{align}
\langle H^{\rm NA}_{\rm YM}\rangle_\ab \,\Big\vert_{\rm sing} = 
2 N \,\frac{g^2}{16}\, \beta V_2 \int 
\frac{d^3p}{(2\pi)^3}\int \frac{d^3 q}{(2\pi)^3}\,\frac{3 - (\hat{p}\,\hat{q})^2}
{\omega(p)\,\omega(q)}\,
\sum_{\mf = - \infty \atop \mf \neq 0}^\infty \sum_{\sigma} 
\exp\big[ i \mf \beta \,(q_3 +  \xvec{\sigma}\cdot \xvec{\ab})\big]\,.
\label{pal8}
\end{align}
By the same technique, the counter term $H_{\rm ct}$ eq.~(\ref{Ect}) contributes, at finite 
temperature and in the presence of a background field,
\begin{align}
\langle H_{\rm ct} \rangle_\ab = \frac{C_0}{2}\,\beta V_2\, \int \frac{d^3 q}{(2\pi)^3}\,
\frac{1}{\omega(q)}\,\sum_{\mf = - \infty}^{\infty}\sum_\sigma 
\exp\big[ i \mf \beta \,(q_3 +  \xvec{\sigma}\cdot \xvec{\ab})\big]\,.
\label{pal10}
\end{align}
The $\mf = 0$ vacuum contribution of this expression cancels the divergence 
in the vacuum contribution of the non-Abelian magnetic energy, cf.~eq.~(\ref{pal4}).
These terms would, however, drop out anyhow when computing the effective potential 
of the Polyakov loop. In addition, however, the $T=0$ renormalization (\ref{pal4})
is also sufficient to cancel all UV divergences in the finite temperature corrections,
as can be seen explictily from eqs.~(\ref{pal8}) and (\ref{pal10}). The combination 
of these two expression gives the finite contribution
\begin{align}
\langle H^{\rm NA}_{\rm YM}\rangle_\ab \,\Big\vert_{\rm sing, ren} = 
\frac{c_0}{2}\,\beta V_2\, \int \frac{d^3 q}{(2\pi)^3}\,
\frac{1}{\omega(q)}\,\sum_{\mf = - \infty \atop \mf \neq 0}^{\infty}\sum_\sigma 
\exp\big[ i \mf \beta \,(q_3 +  \xvec{\sigma}\cdot \xvec{\ab})\big]\,.
\label{pal12}	
\end{align}
This is again a one-loop term which now depends \emph{explicitly} on the 
renormalization constant $c_0$. (In section \ref{sec:loop1}, the dependency
on $c_0$ was only indirect via the curvature computed from the gap equation.)
To complete the non-Abelian magnetic field, we must also add the 
finite two-loop contribution from the terms $(\mf\neq 0,\, \nf \neq 0)$ 
in eq.~(\ref{pal6}). The techniques used to treat these terms numerically 
are identical for the Coulomb contribution studied in the next section, and 
we defer the details to section \ref{sec:loop2N} below.

\subsection{The gluon part of the Coulomb potential}
\label{sec:loop2C}
The last term in eq.~(\ref{EBOS}) is the contribution of the Non-Abelian 
Coulomb term to the gluon energy, i.e.~the expectation value of eq.~(\ref{coulx})
in the gluon sector,
\begin{align}
\langle H_C^A \rangle_\ab = \frac{g^2}{2}\int d^3(x,y)\,\Big\langle \JJ_\ab^{-1}[\vA]\,
\rho^a_{\rm YM}(\vx) \JJ_\ab[\vA]\,\hat{F}^{ab}(\vx, \vy)\,\rho^b_{\rm YM} \Big\rangle_\ab 
\label{pal14}
\end{align} 
where the gluon colour charge $\rho_{\rm YM}^a(\vx) = f^{abc}\,A_k^b(\vx)\,\Pi_k^c(\vx)$
now contains a functional derivative, $\vec{\Pi}(\vx) = - i \delta / \delta \vec{A}(\vx)$. 
The expectation value in eq.~(\ref{pal14}) implies that the integrand
should be sandwiched between two trial wave functionals. It is then  convenient 
to functionally integrate by parts and let the derivative in the left factor 
$\rho_{\rm YM}$ act on the wave function to the left, and the right factor $\rho_{\rm YM}$
act to the right:
\[
-\frac{\delta}{\delta A_k^a(\vx)} \left[ \JJ_\ab[A]^{-\frac{1}{2}}\,\exp\left(- \frac{1}{2 g^2}
\int (A-\ab)\,\omega\,(A-\ab)\right)\right] = E^a_k(\vx) \cdot  
\JJ_\ab[A]^{-\frac{1}{2}}\,\exp\left(- \frac{1}{2 g^2} \int (A-\ab)\,\omega\,(A-\ab)\right)\,.
\]
The factor $E$ can be interpreted as an electric field and reads explicitly:
\begin{align}
E^a_i(\vx) = \frac{1}{g^2}\,\left[ \int dy\,\hat{\omega}^{ab}_{ik}(\vx,\vy)\, (A- \ab)^b_k(\vy) 
+ \frac{g^2}{2}\, \frac{\delta \ln \JJ_\ab[A]}{\delta A^a_i(\vx)}\right]\,.
\label{pal15}
\end{align}
The Coulomb energy can now be recast to
\begin{align}
\langle H_C^A \rangle_\ab = \frac{g^2}{2}\int d^3(x,y)\,\Big\langle
f^{acd} A_i^c(\vx)\,E_i^c(\vx)\cdot \hat{F}^{ab}(\vx,\vy)\cdot 
f^{bc'd'} A_k^{c'}(\vy) \,E_k^{c'}(\vy) 
\Big\rangle_\ab \,.
\label{pal16}	
\end{align}
To proceed, we use the curvature approximation for the Faddeev-Popov determinant \cite{Feuchter2005}
in the presence of a background field,
\begin{align}
\left \langle \frac{g^2}{2}\, \frac{\delta \ln \JJ_\ab[A]}{\delta A^a_i(\vx)} \cdots 
\right \rangle_\ab \approx \left \langle \int dy\,\hat{\chi}^{ab}_{ik}(\vx,\vy) \,  (A- \ab)^b_k(\vy) 
\cdots \right \rangle_\ab\,,
\end{align}
which is valid to the given order, but only under the expectation value. The electric
field now simplifies considerably,
\begin{align}
E^a_i(\vx) \approx \frac{1}{g^2} \int dy\,\big[\hat{\omega} - \hat{\chi}\big]^{ab}_{ik}(\vx,\vy)\,(A- \ab)^b_k(\vy)\,.
\label{pal18}
\end{align}
Furthermore, we can also factorize the expectation value in eq.~(\ref{pal16}) 
to the given order\,,
\begin{align}
	\langle H_C^A \rangle_\ab \approx \frac{1}{2}\int d^3(x,y)\,\Big\langle
	f^{acd} A_i^c(\vx)\,E_i^c(\vx)\cdot f^{bc'd'} A_k^{c'}(\vy) \,E_k^{c'}(\vy) 
	\Big\rangle_\ab\cdot \big\langle g^2 \hat{F}^{ab}(\vx,\vy)\big \rangle_\ab \,.
	\label{pal17}	
\end{align}
Here, the second expectation value is the non-Abelian Coulomb 
potential,\footnote{The first equality will be explained in eq.~(\ref{pal21}) below.}
\begin{align}
\big\langle g^2 \,\hat{F}^{ab}(\vx,\vy) \big \rangle_\ab \approx 
\big\langle g^2\,\hat{F}^{ab}(\vx,\vy) \big \rangle_0 = \delta^{ab}\,V_C(\vx-\vy)\,.
\label{pal19}
\end{align}
This potential has already been used in the Fermi sector. We can model its long-ranged 
part, as obtained from variational calculations in the Yang-Mills sector or the lattice,
by a linear rising potential, $V_C = - \sigma_C \,|\vx - \vy|$. In momentum space, this
amounts to $V_C(p) = 8 \pi \sigma_C / p^4 $, cf.~section \ref{sec:loop1:fermi}.
	
With eq.~(\ref{pal18}), the operator in the first expectation value in eq.~(\ref{pal17}) 
reduces to a monomial in the gauge field, which can be evaluated using Wick's theorem. 
The presence of the background field in our trial wave functional implies that the 
contractions are only colour diagonal when the Cartan basis is used
(cf.~appendix \ref{app:cartan}),
\[
\langle A^a_i(\vx) A_k^b(\vy)\rangle_\ab =  D^{ab}_{ik}(\vx,\vy) = 
\langle a \vert \sigma \rangle \,D^\sigma_{ik}(\vx,\vy) \,\langle \sigma \vert b \rangle\,.
\] 
We must also introduce finite temperature by compactifying the $3$-axis, Fourier transform 
and Poisson resum the Matsubara series. After a lengthy but straightforward calculation along 
the lines layed out earlier, we arrive at 
\begin{align}
\langle H_C^A \rangle_\ab =
\frac{g^2}{2}\,V_2 \beta \sum_{\sigma,\rho,\tau} | f_{\sigma,\rho,\tau}|^2 
\int \frac{d^3 p}{(2\pi)^3} \int \frac{d^3p}{(2\pi)^3}\, & F\big(\vq + \vp - \vec{e}_3 \,
\xvec{\ab}(\xvec{\sigma} + \xvec{\tau} + \xvec{\rho})\big) \, \widetilde{f}(\vp,\vq)\,
\times \nonumber
\\[2mm]
& \times 
\sum_{\nf \in \mathbb{Z}} \exp\big[i \nf \beta (p_z + \xvec{\ab}\cdot\xvec{\sigma})\big]
\sum_{\mf \in \mathbb{Z}} \exp\big[i \mf \beta (q_z + \xvec{\ab}\cdot\xvec{\tau})\big]\,.
\end{align} 
Here, the contractions give rise to the function 
\[
\widetilde{f}(\vp,\vq) = \big[\omega(\vp)\,D(\vp)\omega(\vp)]_{ij}\,D(\vq)_{ji} - 
\big[\omega(\vp)\,D(\vp)\big]_{ij} \,\big[\omega(\vq) D(\vq)\big]_{ji}
\]
in the integrand. Furthermore, the shift in the momentum argument of the Coulomb potential 
is due to the background field expectation value, cf.~eq.~(\ref{pal19}). For any function 
of the root vectors, $\phi(\xvec{\sigma})$, we have the relation (no sum over repated indices)
\begin{align}
| f_{\rho,\sigma,\tau}|^2\,\phi(\xvec{\rho} + \xvec{\sigma} + \xvec{\tau}) 
 = | f_{\rho,\sigma,\tau}|^2 \phi(\xvec{0})\,,
\label{pal21}
\end{align}
because the structure constants are only non-zero for such combination of roots for which 
the sum of the root vectors vanish. This entails that the momentum shift 
in the Coulomb potential can actually be omitted, which explains the first equality 
in eq.~(\ref{pal19}). Putting everything 
together and dividing by the space volume $\beta V_2$, the contribution of the 
Coulomb term to the energy density becomes
\begin{align}
e_C[\ab] = \frac{1}{8}\sum_{\rho,\sigma,\tau} | f_{\sigma,\rho,\tau}|^2 
\int \frac{d^3 p}{(2\pi)^3} \int \frac{d^3p}{(2\pi)^3}\, V_C(\vp + \vq)\,&
\big[1 + (\hat{\vp}\cdot \hat{\vq})^2\big] \,f(\vp,\vq)\,
\times \nonumber
\\[2mm]
& \times 
\sum_{\nf \in \mathbb{Z}} \exp\big[i \nf \beta (p_z + \xvec{\ab}\cdot\xvec{\sigma})\big]
\sum_{\mf \in \mathbb{Z}} \exp\big[i \mf \beta (q_z + \xvec{\ab}\cdot\xvec{\tau})\big]\,,
\label{pal20}
\end{align}
where the variation kernels appear only in the scalar function 
\begin{align}
f(\vp,\vq) = \frac{\Omega(\vp) \,\big[ \Omega(\vp) - \Omega(\vq)\big]}{\omega(\vp)\,\omega(\vq)}\,,
\qquad\qquad\quad\text{with}\quad \Omega(\vp) \equiv \omega(\vp) - \chi(\vp)\,.	
\end{align}
From the structure of eq.~(\ref{pal20}), it is clear that $f(\vp,\vq)$ can be symmetrized 
under the integral, 
\begin{align}
f(\vp,\vq) = \frac{1}{2}\,\frac{ \big[ \Omega(\vp) - \Omega(\vq)\big]^2}{\omega(\vp)\,\omega(\vq)}\,,
\end{align}
which is more convenient. The final form eq.~(\ref{pal20}) of the Coulomb contribution has 
the same mathematical structure as the non-Abelian magnetic energy in eq.~(\ref{pal6}). 
This allows us to use the same analysis and numerical technique in both cases. We will 
describe our method briefly in the next subsection. 

\subsection{General computation of 2-loop terms}
\label{sec:loop2N}
Both two-loop contributions (\ref{pal6}) and (\ref{pal20}) to the gluon energy density 
have the same general form in the presence of a background field and at finite temperature,
\begin{align}
e[\ab] = \int \frac{d^3 p}{(2\pi)^3}\int \frac{d^3 q}{(2\pi)^3}\,\sum_{\rho,\sigma,\tau}
| f_{\rho,\sigma,\tau}|^2\,\phi(\vp,\vq)\,
\sum_{\mf = - \infty}^\infty e^{i \mf \beta (p_z + \xvec{\ab}\cdot \xvec{\sigma})}
\sum_{\nf = - \infty}^\infty e^{i \nf \beta (q_z + \xvec{\ab}\cdot \xvec{\tau})}\,,
\label{pal30}
\end{align}  
where the symmetric function $\phi(\vp,\vq)$ contains the variation kernels. 
For simplicity, we will limit the following considerations to the colour group $G=SU(2)$ 
and also make use of some explicit properties of the $SU(2)$ Cartan base; 
the generalization to $G=SU(N)$ will be studied elsewhere.

We begin by collecting all factors in the integrand that depend on the background field $\ab$.
For $G=SU(2)$, the roots are scalar numbers from $\{-1,0,1\}$, and the background field is 
also a scalar $\xvec{\ab} = \ab$. Using $f_{\rho,\sigma,\tau} = \epsilon_{\rho\sigma\tau}$
and working out the colour trace yields
\begin{align}
	\sum_{\rho,\sigma,\tau} | f_{\rho,\sigma,\tau}|^2 \,e^{ i\beta \xvec{\ab}\cdot 
	(\mf \xvec{\sigma} + \nf \xvec{\tau})} 
= \sum_{\sigma = \pm 1} \Big[e^{i \beta \ab \sigma(\mf - \nf)} + e^{i \beta \sigma \ab \nf} + 
e^{i \beta \sigma \ab \mf}\Big]\,.
\label{pal32}
\end{align}
When inserting the last term on the rhs of eq.~(\ref{pal32}) back into  eq.~(\ref{pal30}),
the result can be put in the form
\begin{align}
\int \frac{d^3 p}{(2\pi)^3} 
\sum_{\mf = - \infty}^\infty \sum_{\sigma = \pm 1} e^{ i \mf \beta (p_z + \sigma \ab)}\cdot
g(\vp, \beta)\,,
\label{pal33}
\end{align}
where the function
\begin{align}
g(\vp, \beta) = \int \frac{d^3 q}{(2\pi)^3} \,\phi(\vp,\vq) \sum_{\nf = -\infty}^\infty 
e^{ i \nf \beta q_z}
\label{pal34}
\end{align}
depends on the temperature, but not on the background field $\ab$. By relabeling 
$\mf \leftrightarrow \nf$ and $\vp \leftrightarrow \vq$, it is easily seen that the 
second contribution on the rhs of eq.~(\ref{pal32}) yields the same result,
with the arguments in $\phi(\vp,\vq)$ reversed (which is irrelevant, as $\phi$ can
be assumed symmetric).

The first term on the rhs of eq.~(\ref{pal32}) is a bit more involved. Inserting in 
eq.~(\ref{pal30}) yields
\begin{align}
\int \frac{d^3 p}{(2\pi)^3}\int \frac{d^3 q}{(2\pi)^3} \,\phi(\vp,\vq)
\sum_{\mf = - \infty}^\infty \sum_{\nf = - \infty}^\infty \sum_{\sigma = \pm 1}
e^{ i \mf \beta p_z + i \nf \beta q_z + i \beta (\mf - \nf) \sigma \ab}\,.
\nonumber
\end{align}
If the Poisson series and the loop integration \emph{were} absolutely convergent, we could
shift the summation index $\mf \to \ell = \mf - \nf$ and the integration variable 
$\vq \to \vq' = \vq + \vp$ to coerce the contribution in the form eq.~(\ref{pal33}). 
After renaming again $ \ell \to \mf$ and $ \vq' \to \vq$, we would again find the form
eq.~(\ref{pal33}), with\footnote{Shifting instead $\vp$ and $\nf$ would give the same result, 
with the arguments in the (symmetric) function $\phi$ exchanged.}
\begin{align}
g(\vp,\beta) =   \int \frac{d^3 q}{(2\pi)^3} \,\phi(\vp,\vq-\vp) \sum_{\nf = -\infty}^\infty 
e^{ i \nf \beta q_z}\,.
\label{pal35}
\end{align}
Collecting all pieces gives the energy density in the form 
\begin{align}
e[\ab] &= \int \frac{d^3 p}{(2\pi)^3} \sum_{\mf = - \infty}^\infty \sum_{\sigma = \pm 1} 
         e^{ i \mf \beta (p_z + \sigma \ab)}\cdot g(\vp, \beta)
\label{pal40}
 \\[2mm]
 g(\vp, \beta) &= \int \frac{d^3 q}{(2\pi)^3} \sum_{\nf = -\infty}^\infty 
 e^{ i \nf \beta q_z} \,\left[\phi(\vp,\vq) + \phi(\vq,\vp) + \frac{1}{2}
 \phi(\vp, \vq - \vp) + \frac{1}{2} \phi(\vq - \vp, \vp) \right]\,.
\label{pal42}
\end{align}
For the Polyakov loop, we must compute the \emph{difference} of the energy density 
with and without the background field,
\begin{align}
 e[\ab] - e[0] =  \int \frac{d^3 p}{(2\pi)^3} \sum_{\mf = - \infty}^\infty
e^{ i \mf \beta p_z} \sum_{\sigma = \pm 1} \Big[ e^{i \mf \beta \sigma \ab} - 1 \Big]
\cdot g(\vp, \beta)\,.
\label{pal44}
\end{align}
The term with $\mf = 0$ does not contribute  and can be omitted. Furthermore, we can 
extend the $\sigma$-sum to all roots $\sigma \in \{-1,0,1\} $, since the term with 
$\sigma = 0$ vanishes identically. This allows us to write the final result in the 
same form as the one-loop contributions in eq.~(\ref{deriv3a}),
\begin{align}
 e[\ab] - e[0] =  \sum_{\sigma} \sum_{\mf \in \mathbb{Z} \atop  \mf \neq 0} \int \frac{d^3 p}{(2\pi)^3} 
e^{ i \mf \beta p_z}  \Big[ e^{i \mf \beta \sigma \ab} - 1 \Big]	\cdot g(\vp, \beta)\,.
\label{pal50}
\end{align}
The main difference to the one-loop term is that $g(\vp,\beta)$ from eq.~(\ref{pal42})
is now itself a temperature-dependent loop integral instead of just a simple algebraic 
function. Except for the $\nf = 0$ term, all contributions to $g(\vp,\beta) $ only have 
a reduced $O(2)$ symmetry, i.e. they have an angular dependency which requires the use of 
eq.~(\ref{deriv6}) instead of eq.~(\ref{hB2}) when evaluating eq.~(\ref{pal50}). 
Combined with the sum over $\nf$ and the triple momentum integration, the numerical effort 
to compute the finite temperature corrections for the two-loop contribution is easily three 
orders of magnitude larger than for the one-loop case. Below, we will therefore use the same 
approximation as for the 1-loop contributions, where the $T=0$ variation kernels and hence 
also the $T=0$ limit of the functions $g(p)$ were used. This amounts to taking only the $\nf=0$ 
contribution in eq.~(\ref{pal42}), which is $O(3)$ symmetric and independent of temperature. 
In the numerics section, we will briefly justify this approximation \emph{a posteriori}, 
i.e. we will show how to compute the first few $\nf \neq 0$ finite temperature corrections 
for selected momenta and assert that they are negligible as compared to the $T=0$ term, 
even for temperatures up to $T=2T^\ast$. 

\subsection{Renormalization of the 2-loop contribution}
\label{sec:loop2D}
Our consideration on the Fourier transform in the last chapter
indicate that eq.~(\ref{pal50}) is finite provided that $g(\vp,\beta)$ does not rise stronger 
than $g(\vp,\beta)\sim |\vp|$ at large momenta. This finiteness of the \emph{outer} loop 
integration is due to the subtraction of the $\ab=0$ background in eq.~(\ref{pal50}) and 
does not hold for other Green's functions. The possible divergences are thus all coerced 
in the \emph{inner} loop integration in eq.~(\ref{pal42}). The terms with $\nf\neq 0$ 
in that equation are finite as can be seen by the regulator method for the Fourier integral, 
or by appealing to appendix \ref{app:nonmag}. This restricts possible divergences to the 
$T=0$ contribution,
\begin{align}
	g_0(p)= \int \frac{d^3 q}{(2\pi)^3} \,\Big[2 \phi(\vp, \vq) + \phi(\vp, \vq - \vp)\Big]\,.	
	\label{palix0}
\end{align}
The leading UV divergence is thus the contribution
\begin{align}
	g_0(p) = 3 \int \frac{d^3 q}{(2\pi)^3} \,\phi(\vp, \vq) + \cdots\,,
	\label{palix1}
\end{align} 
where the dots contain subleading divergences and finite pieces. To check this assertion,
we can go back to the original expression (\ref{pal30}) and use the results of appendix 
\ref{app:nonmag}, where it was argued that only those terms in eq.~(\ref{pal30}) are possibly 
divergent in which one of the two Poisson indices vanishes. Picking only these contributions 
and using the symmetry of $\phi(\vp,\vq)$, we find the leading divergence with a different 
prefactor:\footnote{One prefactor of $2$ comes from setting $\mf$ or $\nf$ 
to zero, while the other factor of two comes from the colour trace.} 
\begin{align}
	g_0(p) = 4 \int \frac{d^3 q}{(2\pi)^3} \,\phi(\vp, \vq) + \cdots \,,
	\label{palix2}	
\end{align}
The catch is that we had to shift the summation index and the loop momentum in 
divergent expression in order to derive eq.~(\ref{palix1}). Such operations are known 
to change the UV divergence, which depends on the regularization scheme and the momentum
and Poisson routing. The ambiguity must eventually be removed as part of the 
renormalization procedure. At present, we do not have a fully consistent method to 
renormalize our two-loop contributions, by relating the necessary counter terms to 
physical observables. Instead, we take a pragmatic approach and \emph{cancel} the 
divergences in whatever momentum routing is numerically convenient.\footnote{We use 
eq.~(\ref{pal42}) due to its close analogy to the one-loop expressions.} 

Next we read off the core function $\phi(\vp, \vq)$ by comparing the general form 
eq.~(\ref{pal30}) with the bosonic 2-loop contributions,  eq.~(\ref{pal6}) and (\ref{pal20}):
\begin{align}
	\phi(\vp, \vq) &= \frac{g^2}{16}\,\frac{3 - (\hat{\vp}\cdot \hat{\vq})}{\omega(\vp) \omega(\vq)} 
	+	\frac{[1 + (\hat{\vp}\cdot \hat{\vq})^2]}{16}\cdot
	\frac{\big[\Omega(\vp)- \Omega(\vq)\big]^2}{\omega(\vp)\omega(\vq)} \cdot V_C(\vp + \vq)\,,
	\label{palix4}
\end{align}
where $\Omega(\vp) = \omega(\vp) - \chi(\vp)$.  With the $T=0$ kernels, this depends on the 
momenta only in the combination $p=|\vp|$, $q = |\vq|$ and $\xi = \hat{\vp}\cdot\hat{\vq}$, 
i.e. we can write 
\begin{align}
	\phi(\vp, \vq) \equiv \Phi(p,q,\xi) = \frac{g^2}{16}\frac{3 - \xi}{\omega(p)\omega(q)} + 
	\frac{1 + \xi^2}{16}\cdot \frac{[\Omega(p) - \Omega(q)]^2}{\omega(p)\omega(q)}\cdot 
	\frac{8 \pi \sigma_C}{[p^2 + q^2 + 2 \xi p q ]^2}\,.
\label{palix5}
\end{align}
The same dependency holds for the second term $\phi(\vp, \vq - \vp) \equiv \Phi(p,Q,\eta)$
in the square bracket in eq.~(\ref{palix0}), where 
\begin{align}
Q &= | \vq - \vp| = \sqrt{p^2 + q^2 - 2 \xi p q}\nonumber \\[2mm]
\eta &= \cos \sphericalangle(\vp, \vp-\vq) = \frac{q\xi - p}{Q}	\,.
\label{palix6}
\end{align}
As a consequence, the zero-temperature 2-loop contribution is $O(3)$ invariant, 
$g_0(\vp) = g_0(p)$.

To find the possible UV divergences explicitly, take the Gribov form eq.~(\ref{gribovx}) 
for the gluon propagator $\omega(\vp)$ and eq.~(\ref{gribovx2}) for the curvature $\chi(\vp)$, 
respectively, insert in 
eqs.~(\ref{palix4}) and (\ref{palix0}), and expand the integrand for large loop momenta. 
With a sharp momentum cutoff $\Lambda$, we obtain  
\begin{align}
	g_0(p) &= \frac{9 g^2}{64\pi^2}\cdot \frac{\Lambda^2}{\omega(p)} + 
	\frac{g^2\,\Lambda}{96\pi^2} \cdot\frac{p}{\omega(p)} + 
	\frac{\sigma_C}{\pi}\,\ln\big(\frac{\Lambda}{\mu_0}\big)\cdot\frac{1}{\omega(p)}
	+ \text{finite}\,,
	\label{palix7}
\end{align}
where $\mu_0$ is an arbitrary scale parametrizing the finite piece in the log divergence.
This function would then enter eq.~(\ref{hB2}) for the Fourier transform, and 
eventually eq.~(\ref{uBsu2}) to compute the effective potential for the Polyakov loop. 

\begin{figure}
	\centering
	\includegraphics[width=0.48\linewidth]{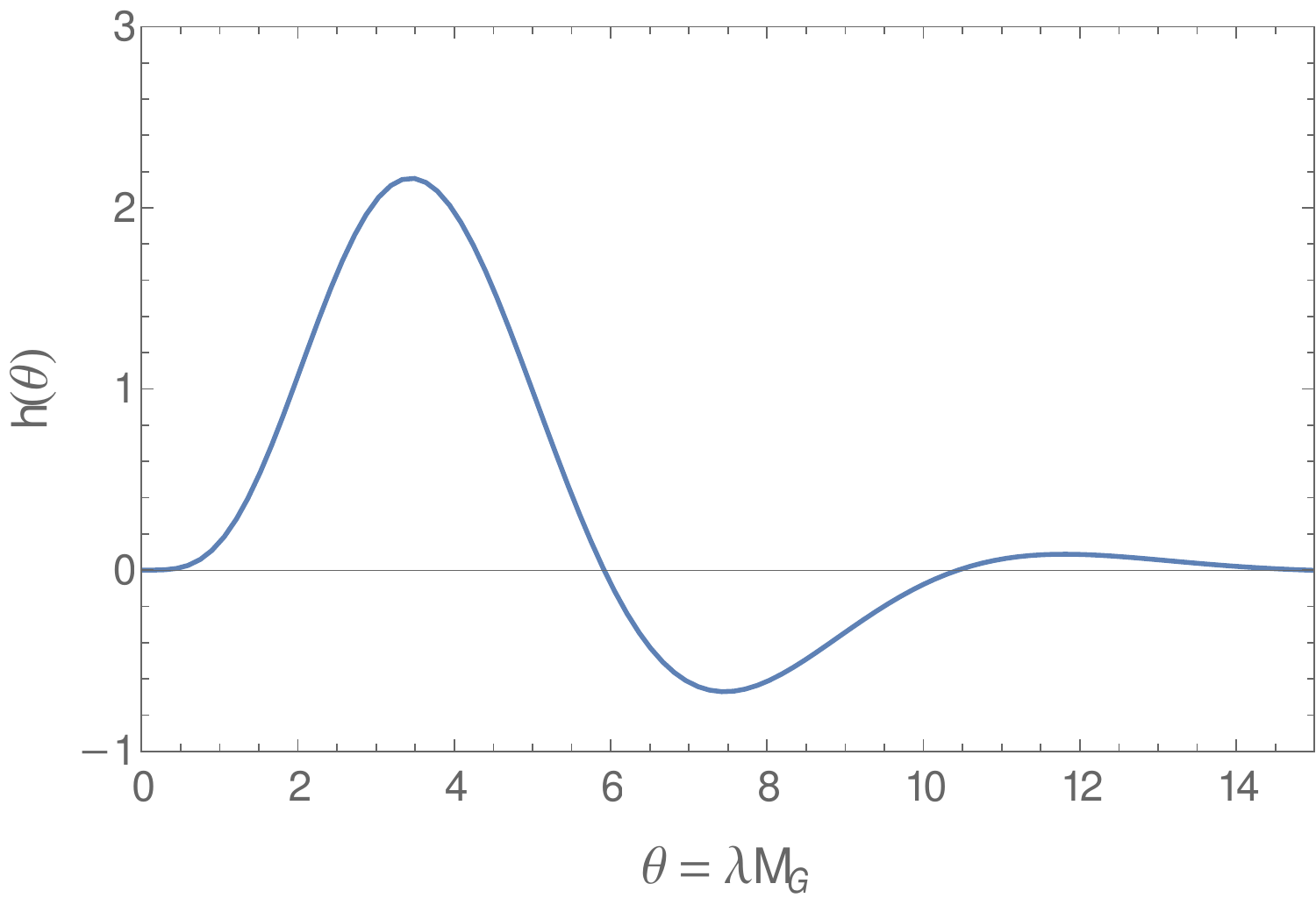}
	\hfill
	\includegraphics[width=0.48\linewidth]{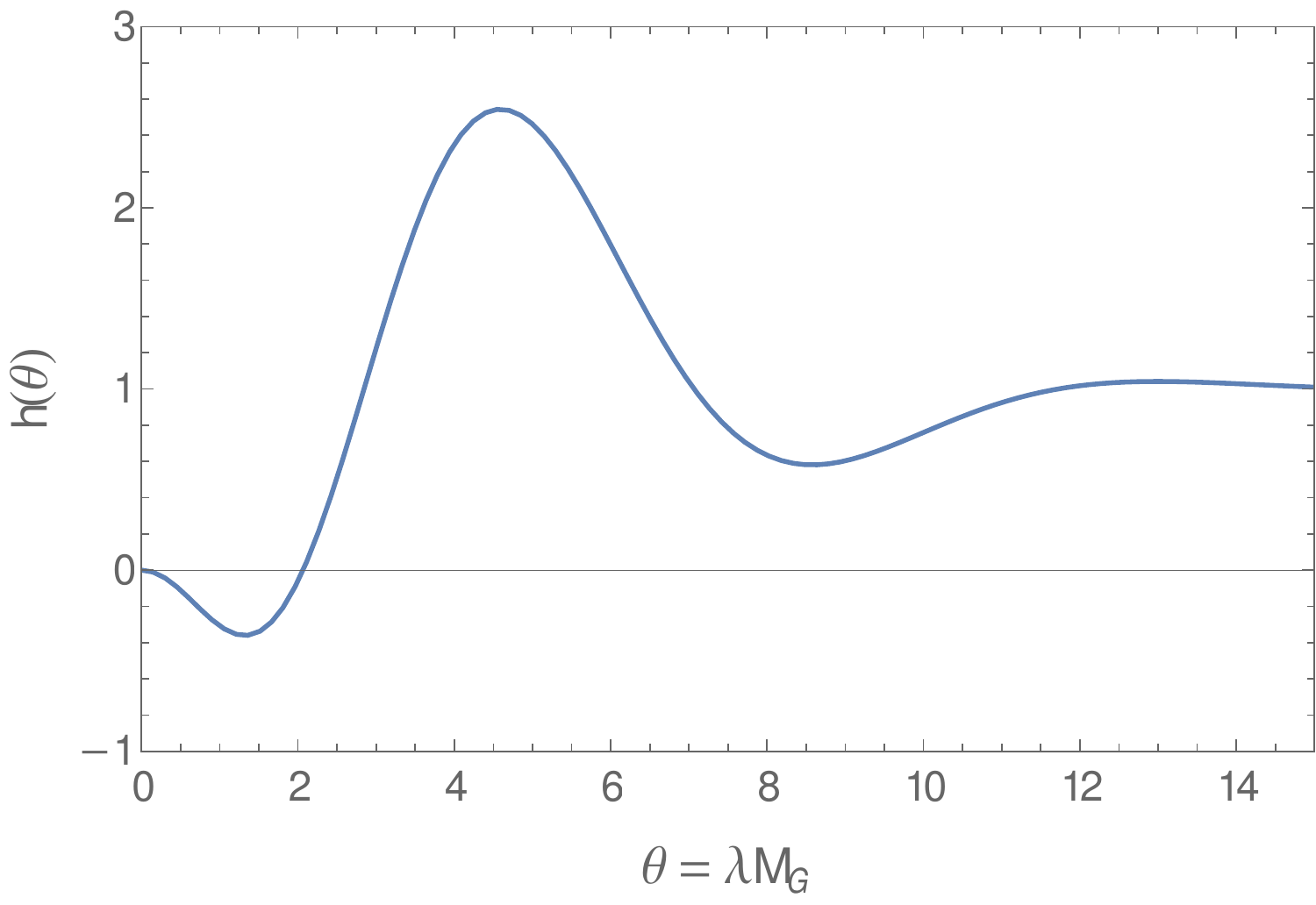}
	\caption{The Fourier transform $h(\lambda)$ from eq.~(\ref{hB2}) for 
		the momentum dependences $g(p)$ encountered in 2-loop divergences.
		\emph{Left}:  $g(p)\sim p/\omega(p)$ (spurious linear divergence). 
		\emph{Right:} $g(p) \sim 1/\omega(p)$ (quadratic and log divergence).}	
	\label{fig8}
\end{figure}

The linear divergence in eq.~(\ref{palix7}) is absent if we either take a gauge-invariant 
regularization scheme, or consider the free case $\omega(k) = k$. In general, divergences 
and counter terms should not depend on the specific form of the variational 
solution.\footnote{Furthermore, the momentum dependence $p/\omega(p)$ of the linear divergence 
is such that the corresponding Fourier transform $h(\lambda)$ vanishes both at $\lambda = 0$ 
and $\lambda \to \infty$, as can be seen from the left panel of Fig.~\ref{fig8}. Even if present, 
it would therefore affect neither the Stefan-Boltzmann law at large temperatures nor 
confinement at small temperatures.} We thus conclude that the linear divergence 
is spurious and should be canceled completely, if a hard momentum cutoff is used. 

The remaining quadratic and logarithmic divergences are universal, i.e. they do 
not depend on the form of the variation kernels and persist even in the free case, 
$\omega(q) = q$. Their momentum dependence is identical and leads to a Fourier transform
$h(\lambda)$  which vanishes at $\lambda = 0$, but goes to a non-trivial constant at 
$\lambda \to \infty$, cf.~Fig.~\ref{fig8}. As explained in the previous chapter, $h(0)$ counts the number 
of perturbative degrees of freedom, i.e. $h(0) = 0$ means that the Stefan-Boltzmann limit
of the 1-loop calculation is preserved. Furthermore, the 1-loop bosonic contributions had 
$h_B(\infty) = 0$ leading to a very delicate confinement that is easily overcome by fermions. 
A non-trivial limit $h(\infty) \neq 0$ at small temperatures could thus be very helpful.

The counter terms for the inner loop integration would remove the divergence but leave a 
finite piece (see eq.~(\ref{palix7})),
\begin{align}
\left[\frac{9 g^2}{64\pi^2}\cdot \Lambda^2 +
\frac{\sigma_C}{\pi}\,\ln\big(\frac{\Lambda}{\mu_0}\big)\right]
\cdot \frac{1}{\omega(p)} \to c_2\cdot \frac{\sigma_C}{\omega(p)}
\label{palix8}
\end{align}
In principle, the dimensionless coefficient $c_2$ should be fixed by relating 
it to another physical input observable. This quantity would have to be computed 
to the same 2-loop order, using the same regularization technique and momentum routing. 
Renormalization would then remove the divergence and trade the coupling $g$ for 
the (dimensionfull) input quantity. This is beyond the scope of the present paper. 
Instead, we take a more pragmatic approach 
and treat $c_2$ and $g$ as free parameters, just as the corresponding coefficient 
$c_0$ in the 1-loop case. This allows us to study e.g.~how different 
values of $c_2$ could affect the physical outcome.

With these arrangements, the renormalized inner loop integral (\ref{pal42}) becomes 
(using a sharp momentum cutoff),
\begin{align}
	g_0(p) = \frac{1}{4 \pi^2} \int_0^\infty dq\,\Bigg[q^2\int_{-1}^1 d\xi \,\Big(
	2 \Phi(p,q,\xi) + \Phi(p,Q,\eta)\Big) - \Phi_\infty(p,q)\Bigg]+ c_2\,\frac{\sigma_C}{\omega(p)}\,,
\label{palix10}
\end{align}
with the subtraction
\begin{align}
	\Phi_\infty(p,q) = \left[ \frac{9g^2}{8} q\, + \frac{g^2}{24}\,p + \frac{4 \pi \sigma_C}{q}
	\,\tanh(q / \sqrt{\sigma_C})\right]\frac{1}{\omega(p)}\,.
\end{align}
The $\tanh$ is introduced to avoid the infrared singularity arising from the 
subtraction of the logarithmic UV divergence. This could be replaced by any other 
regulator function $f(q)$ with the limits $f(\infty)=1$ and $f(q) \sim q$ at $q \to 0$.
A different regulator $f(q)$ would amount to a slightly different subtraction of the form 
of a finite numerical constant times $1/\omega(p)$. This can always be absorbed by a slight 
change of the renormalization parameter $c_2$.

\subsection{Numerical results}
\label{sec:loop2R}

\subsubsection{Core function at zero temperature}
\begin{figure}
	\centering
	\includegraphics[width=0.48\linewidth]{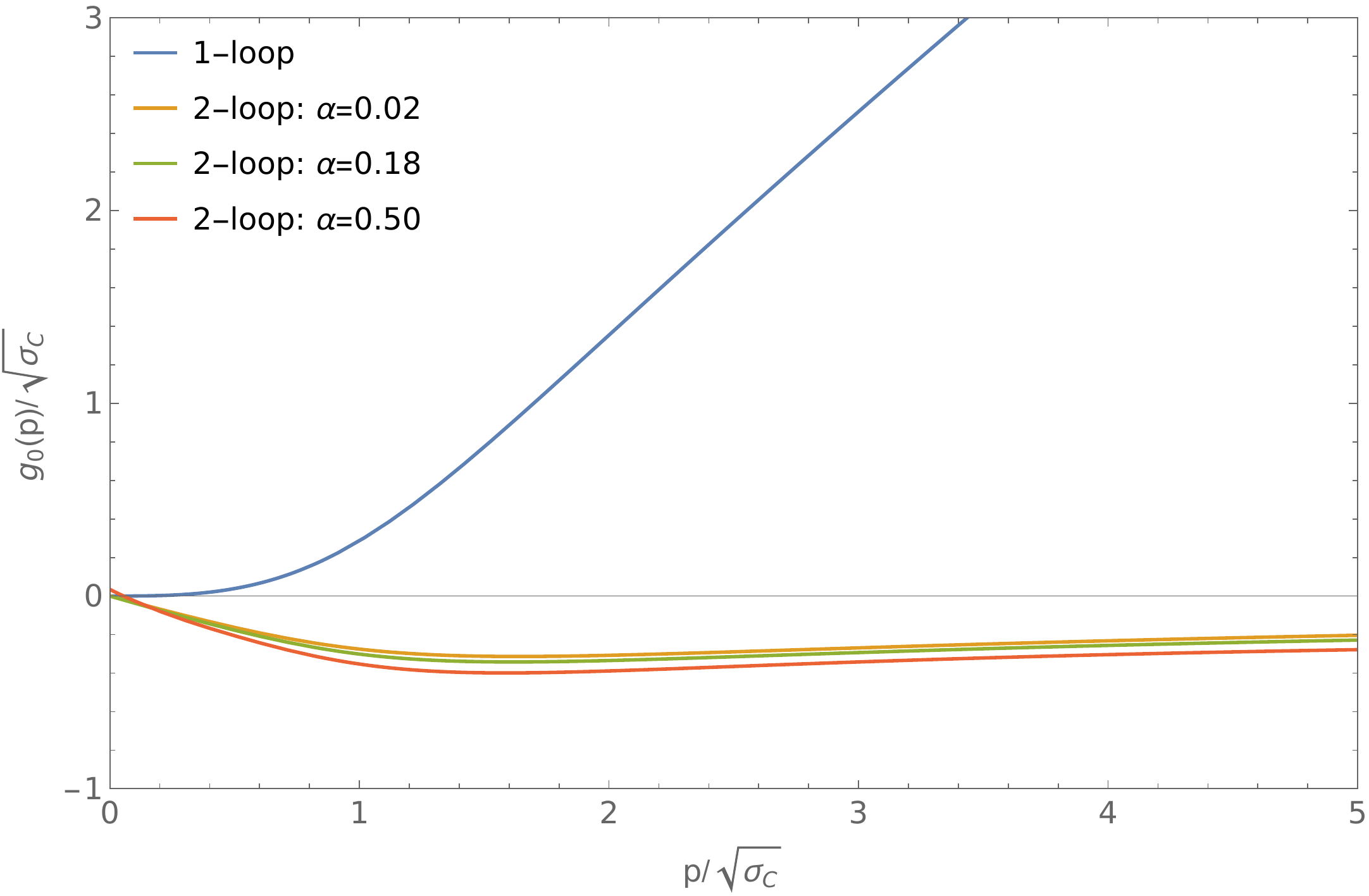}
	\hfill
	\includegraphics[width=0.48\linewidth]{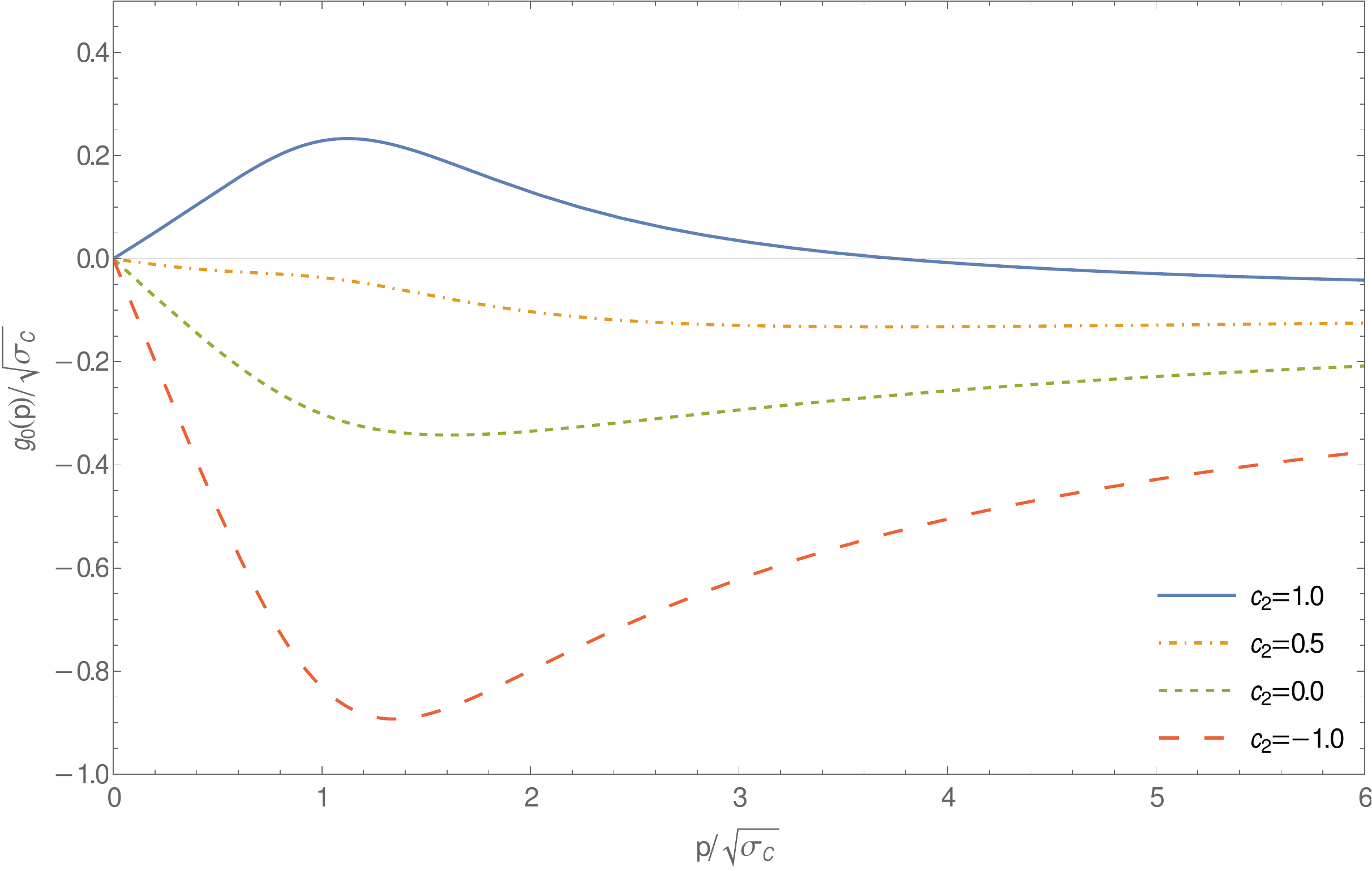}
	\caption{\emph{Left}: The inner loop integral (core function) $g_0(p)$ for the 
		renormalization parameter $c_2 = 0$ at various couplings $\alpha = g^2 / 4 \pi$.
		\emph{Right:} The core function at the preferred coupling $\alpha = 0.18$ 
		for various values of $c_2$.}	
	\label{fig9}
\end{figure}

For our numerical code, we measure all dimensionfull quantities in units of the 
Coulomb string tension, i.e.~we use the mass scale $\sqrt{\sigma_C}$ from here on. 
First, we compute the inner loop integration (core function) $g_0(p)$ from eq.~(\ref{palix10}).
In the left panel of Fig.~\ref{fig9}, we show $g_0(p)$ for $c_2 = 0$ and various 
values of the coupling $\alpha = g^2 / (4 \pi)$. To put this in perspective, we 
have also included the 1-loop  contribution $g_B(p) = \omega(p) - \chi(p)$.
From the plot, it is clear that the 2-loop corrections are small in 
magnitude (even at couplings of order unity) for most momenta, but they 
dominate at small momenta because of the cancellation in the one-loop result 
mentioned earlier. This will help to make the bosonic confinement more robust.

We have varied the coupling $\alpha$ in the expected region by a factor $25$ and 
found only a mild effect on the 2-loop result. Furthermore, the paramters $g$ and 
$c_2$ are related, i.e.~changes in $g$ can largely be compensated by changes in $c_2$. 
In the following, we will therefore fix $g$ to a reasonable value $g=1.5$ corresponding 
to $\alpha = g^2 / 4 \pi =0.18$, and vary only $c_2$.

In the right panel of Fig.~\ref{fig9}, we plot $g_0(p)$ at the preferred $g$ and various 
values of $c_2$. Negative values increase the confining effect by making $g_0(p)$ more 
negative at small momenta. By contrast, positive values have the opposite effect and 
may reduce (or even destroy) the confining property. Note that we also expect that a 
strong confinement leads to a higher transition temperature, as more thermal fluctuations
are necessary to overcome the confining order.

\subsubsection{Finite Temperature corrections}
Next, we want to corroborate that the finite temperature corrections to $g_0(p)$ 
are indeed negligible. We take the $\nf \neq 0$ terms from eq.~(\ref{pal42})
and combine the mirror pairs $\pm \nf$. Then we introduce spherical coordinates and 
consider each Poisson index $\nf \ge 1$ separately. Since we no longer have $O(3)$ symmetry,
the external momentum $\vp$ cannot be rotated to the $z$-direction, but only to the
$xy$-plane. As usual, we replace the polar and azimuthal angles by their cosine, 
$\xi_p = \cos \vartheta_q$ or $\gamma_q = \cos \varphi_q$ etc. The finite temperature 
corrections to $g_0(p)$ then read
\begin{align}
	g_\nf(\vp, \beta) = g_\nf(p,\xi_p,\beta) = 
	\frac{1}{\pi^3} \int_0^\infty dq\,q^2\int_{-1}^{1} d\xi_q \int_{-1}^{+1} \frac{d\gamma_q}
	{\sqrt{1-\gamma_q^2}}\,\cos(\nf \beta q \xi_q)\,\big[ 2 \phi(\vp, \vq) +  \phi(\vp, \vq - \vp) \big]\,. 
\end{align}
Since this is not $O(3)$ symmetric, we also have to average over the angle $\xi_p$ 
using eq.~(\ref{ave}). The parameter in that average is $\lambda = \mf \beta$. Larger 
values of $\mf \ge 1$ are strongly suppressed by the factor $1/m^4$ in the effective 
action of the Polyakov loop eq.~(\ref{uBsu2}), and we expect the largest finite 
temperature corrections from $\mf = 1$, i.e.~$\lambda = \beta$. In total,
\begin{align}
	g_\nf(p, \beta) \simeq \frac{1}{2 \pi^3}\frac{\beta p}{\sin(\beta p)}\int_0^\infty dq\,q^2
	\int_{-1}^1 d\xi_p\int_{-1}^1 d\xi_q\int_{-1}^{1} \frac{d\gamma_q}{\sqrt{1-\gamma_q^2}}\,
	\cos(\beta p \xi_p) \cos(\nf \beta q \xi_q)\,\Big[2 \Phi(p,q,\xi) + \Phi(p,Q,\eta) \Big]\,.
	\label{palix12}
\end{align}
In this equation, $Q$ and $\eta$ are defined as in eq.~(\ref{palix6}), with the angle 
$\xi$ now computed from
\[
\xi \equiv \hat{\vp}\cdot \hat{\vq} = \xi_p \xi_q + \gamma \sqrt{(1-\xi_p^2)(1- \xi_q)^2}\,. 
\]
As can be expected, the 4-fold integral (\ref{palix12}) is numerically quite challenging, 
although the 3 angular integrations can usually be done efficiently using Gauss-Legendre 
techniques. We have computed the correction eq.~(\ref{palix12}) for $\nf \le 5$,
spot values of the momentum and selected temperatures between $100\,\mathrm{MeV}$ and 
$450\mathrm{MeV}$, and found that the result is generally of the order or even smaller 
than the numerical accuracy in $g_0(p)$. This confirms that we can make the same 
approximation as in the 1-loop case, viz.~replace the core function entering 
the calculation of the Polyakov loop by its $T \to 0$ limit.
%
%

\subsubsection{Fourier transform of the 2-loop contribution} 

\begin{figure}
	\centering
	\includegraphics[width=0.48\linewidth]{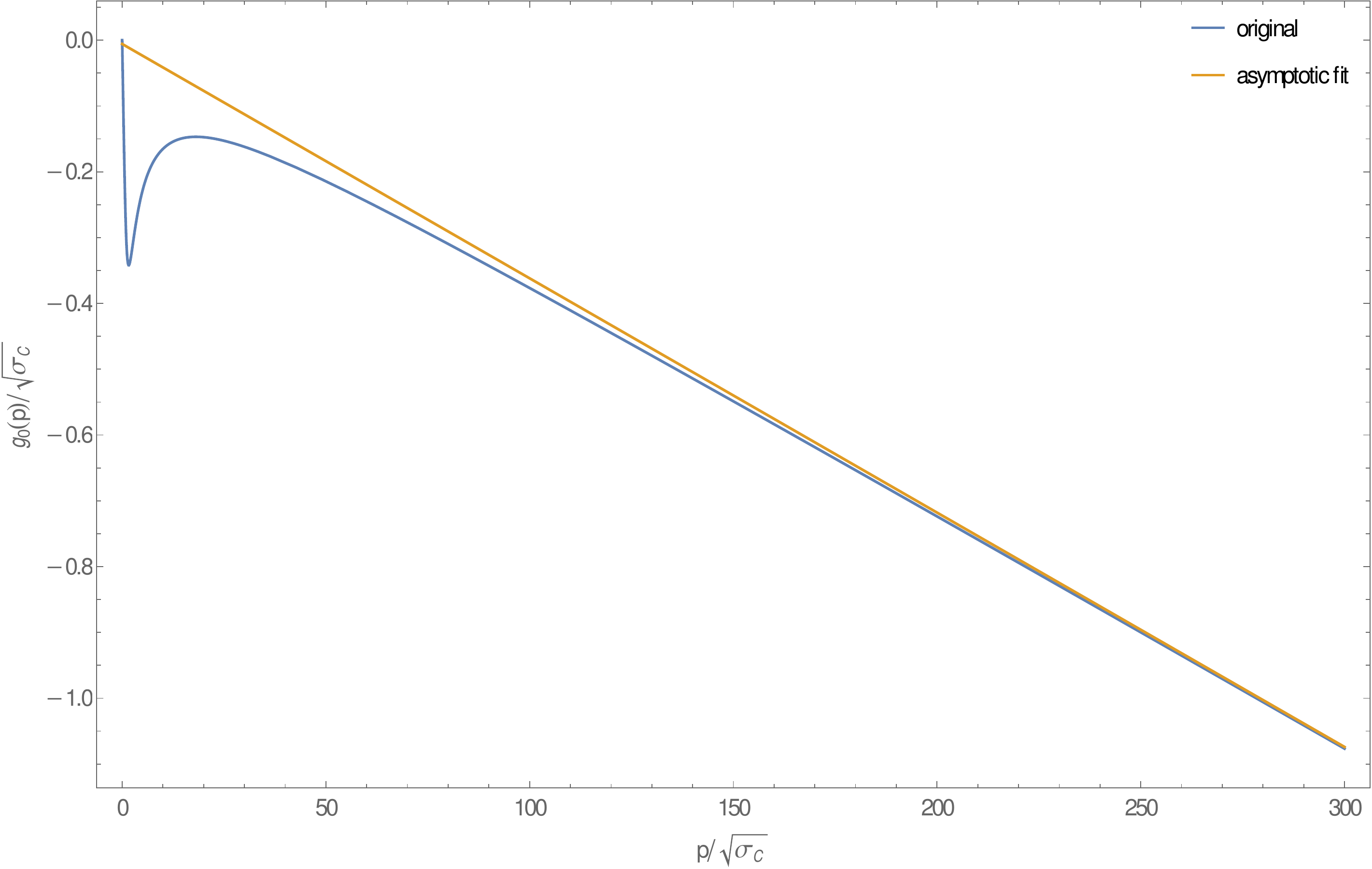}
	\hfill
	\includegraphics[width=0.48\linewidth]{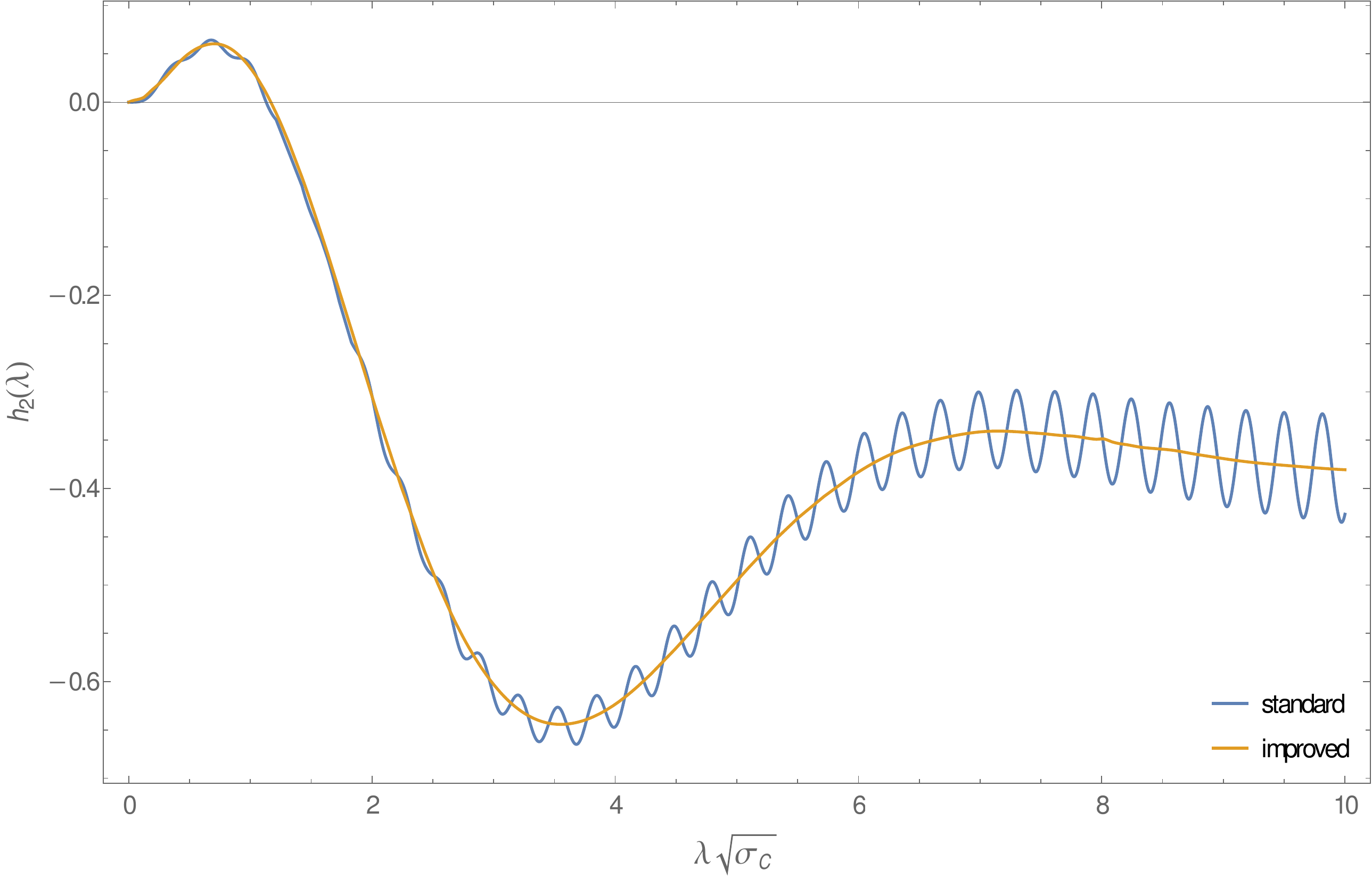}
	\caption{\emph{Left}: The 2-loop core function $g_0(p)$ with its 
		asymptotics at large momenta $p \gg 1$. 
		\emph{Right}: The 2-loop Fourier transform $h_2(\lambda)$ computed with 
		and without the improved regularization technique eq.~(\ref{palix20}).}	
	\label{fig10}
\end{figure}

The next step is to Fourier transform the core function $g_0(p)$
according to eq.~(\ref{hB2}), which yields the temperature-dependent
2-loop amplitude $h_2(\lambda= \mf \beta)$. The regulator method shows
numerical instabilites (oscillations) that worsen at large
$\lambda$, cf.~Fig.~\ref{fig10}. The cause 
of the problem can be traced to the fact that our 2-loop core function 
$g_0(p)$ does not vanish at large momenta, which makes the computation of the 
Fourier transform rather delicate. Since we cannot compute $g_0(p)$ for 
arbitrarily large $p$, we would have to cut off the Fourier integral at some 
upper limit $\Lambda$. This leads to typical oscillations of the type   
\begin{align}  
g_0(\Lambda)\,\frac{\lambda \Lambda \cos(\lambda \Lambda) - \sin(\lambda \Lambda)}{\lambda^2}
\nonumber\,.
\end{align}  
The point here is that we must not introduce boundary 
terms by a hard cutoff to the Fourier integral, since boundaries at large but finite momenta 
spoil the regulator method. Numerically, $g_0(p)$ actually becomes \emph{linear} at large 
momenta, albeit with a very small slope, cf.~the left panel of Fig.~\ref{fig10}. 
We determine the coefficients 
$a$ and $b$ from a linear regression $g_0(p) \simeq a + b\,p$ at large 
$p \ge \Lambda \simeq 200$ and find the small values $a = -0.0057$ and $b = -0.00356$.
After subtracting the asymptotics, the Fourier integration can be done, and the subtracted
linear function can be transformed analytically\footnote{According to eq.~(\ref{qn}), the 
regulator method  gives $\displaystyle\lim_{\mu \to 0} \frac{-\lambda^3}{2}\int_0^\infty
 dp\,p\,\sin(\lambda p) e^{-\mu p} (a + b\, p) = b$.} and added back in:
\begin{align}
	-\frac{\lambda^3}{2}\lim_{\mu \to 0}\int_0^\infty dp\,p\,\sin(\lambda p) 
	e^{-\mu p} \Big[g_0(p)- (a + b\cdot p)\Big] +b\,.
	\label{palix20}
\end{align}
The result of this procedure is a much smoother Fourier integration without the numerical
artifacts, cf.~again the right panel of Fig.~\ref{fig10}. 

In the next Fig.~\ref{fig11} we first present the results of the regulator method.
The Fourier transform converges nicely when $\mu \to 0$, but it requires 
quite small values to reach the limit. In practice, we have decreased $\mu$ progressively
in up to 20 steps and used Richardson extrapolation to the limit $\mu \to 0$. 

The right panel of Fig.~\ref{fig11} shows the Fourier amplitude $h_2(\lambda$) for the 
preferred coupling $\alpha = 0.18$ and several values of the renormalization constant $c_2$. 
All functions $h_2(\lambda)$ vanish at the origin, which means that the mode count 
at high temperatures and the Stefan-Boltzmann law from 1-loop is preserved.
Negative values for $c_2$ increase the negative constant 
$h_2(\infty)$ and hence the strength of the bosonic confinement, while also 
increasing the deconfinement temperature. Conversely, positive values for $c_2$
have the opposite effect of weakening or even destroying confinement, if taken 
too large. However, such large values of $|c_2|$ over-emphasize the two-loop contribution,
which should remain a subleading effect. Coefficients $|c_2| \le 0.2$ 
are natural and seem to exhibit the correct qualitative behaviour.   

\begin{figure}
	\centering
	\includegraphics[width=0.48\linewidth]{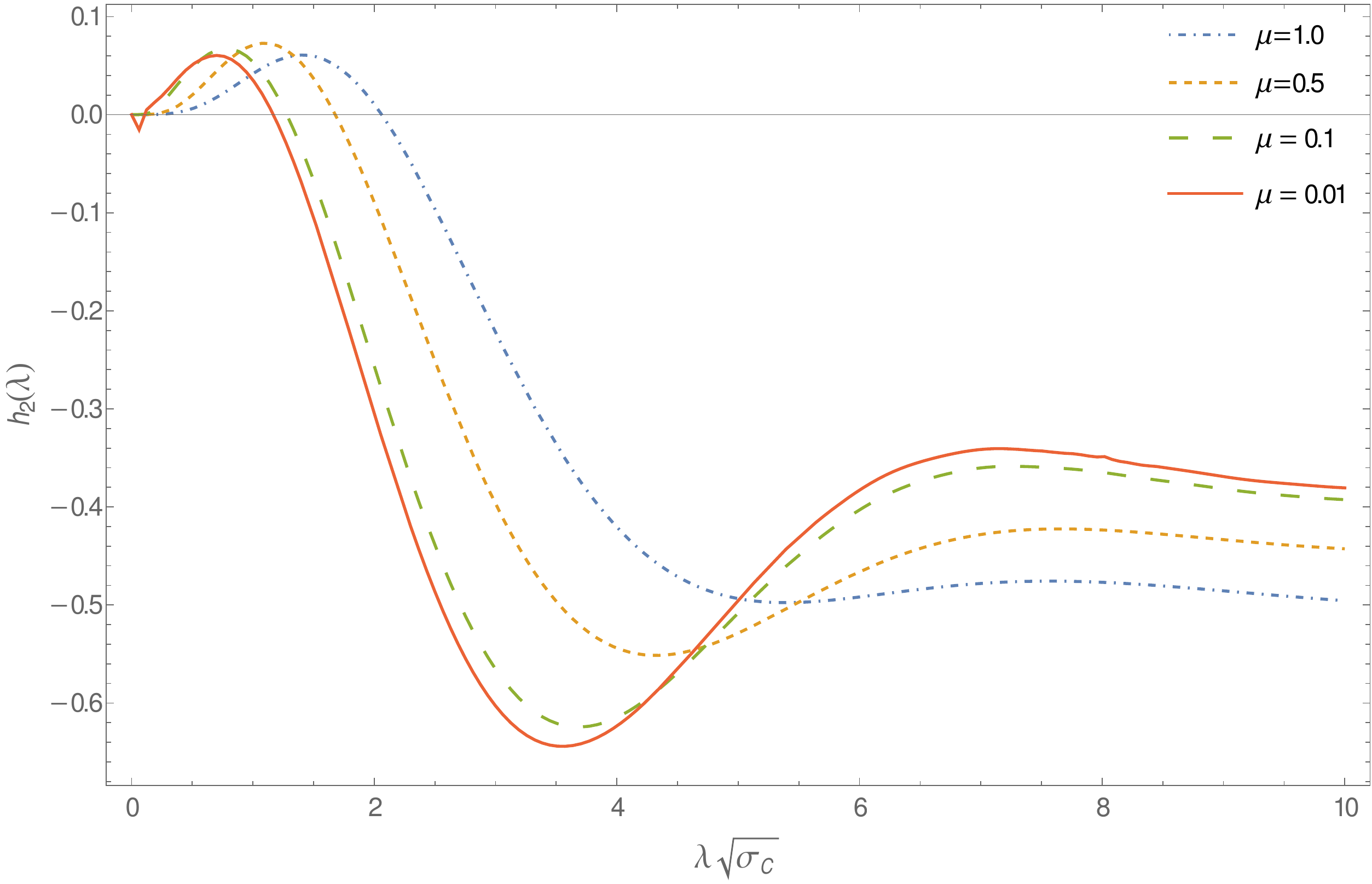}
	\hfill
	\includegraphics[width=0.48\linewidth]{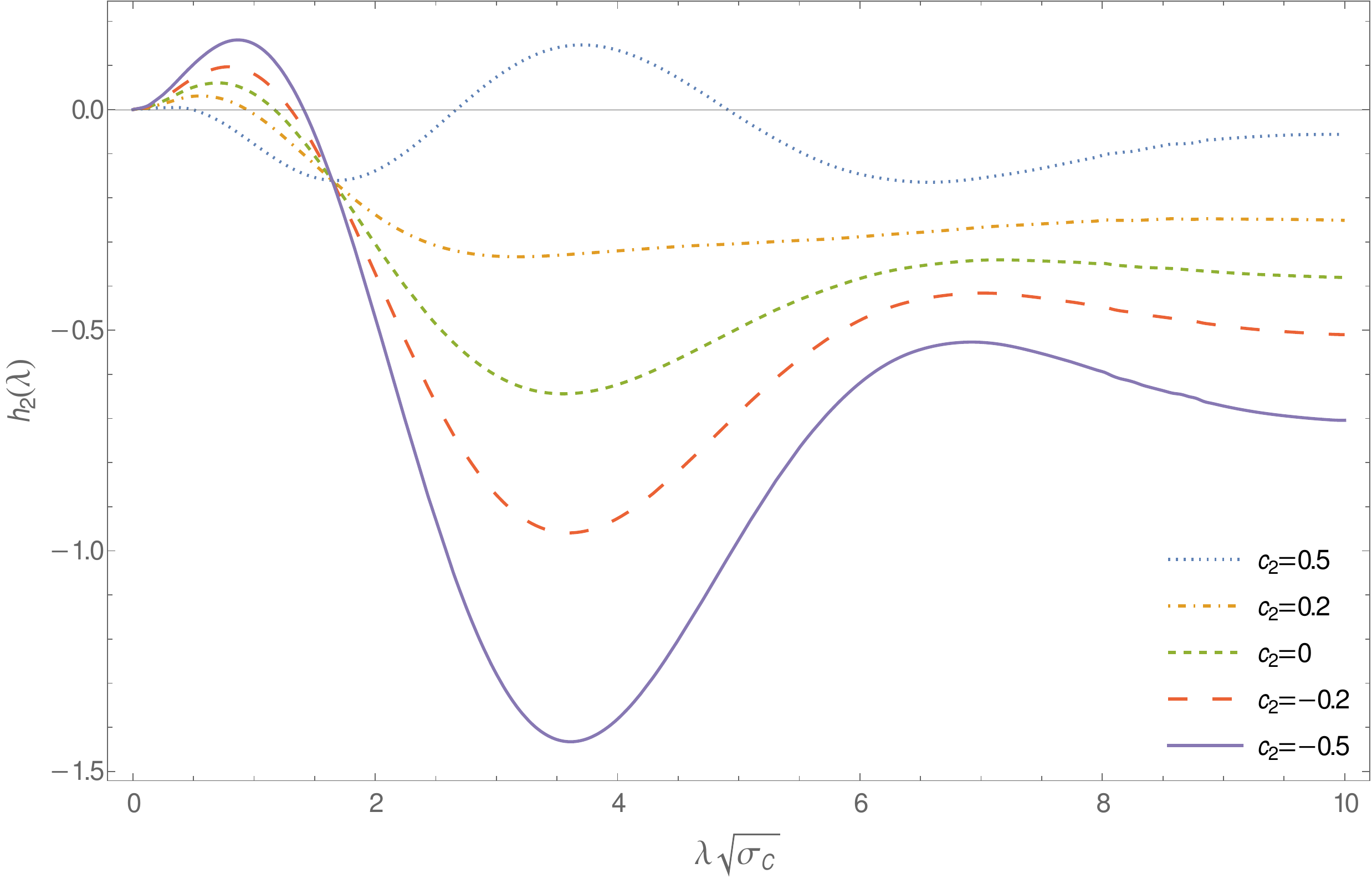}
	\caption{\emph{Left}: The approach to the limiting function $h_2(\lambda)$ as 
		the regulator is removed.
		\emph{Right}:The 2-loop Fourier amplitude $h_2(\lambda)$ computed 
		for several values of the renormalization parameter $c_2$. }	
	\label{fig11}
\end{figure}    
 
\subsubsection{The Polyakov loop} 

\begin{figure}
	\centering
	\includegraphics[width=0.48\linewidth]{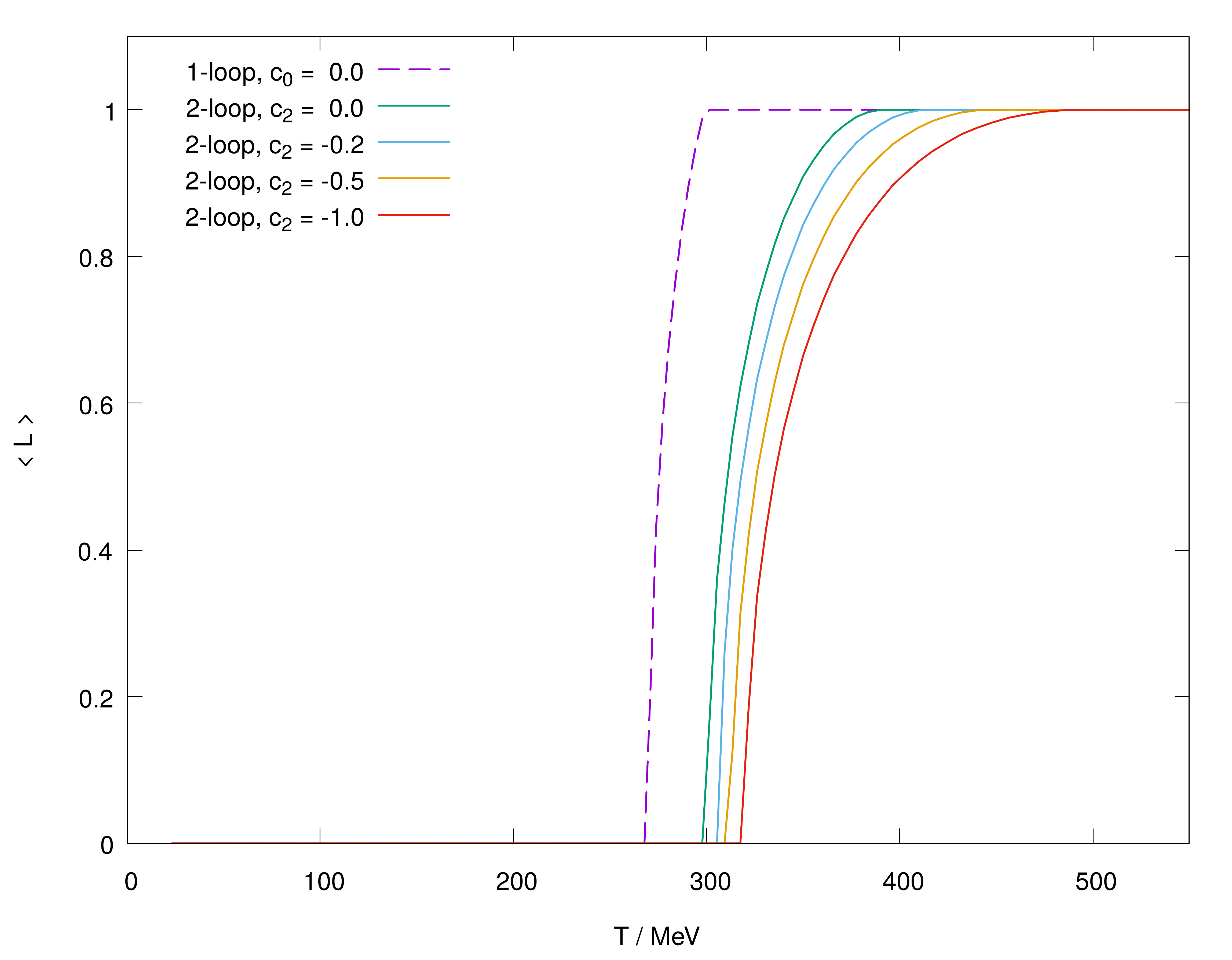}
	\hfill
	\includegraphics[width=0.48\linewidth]{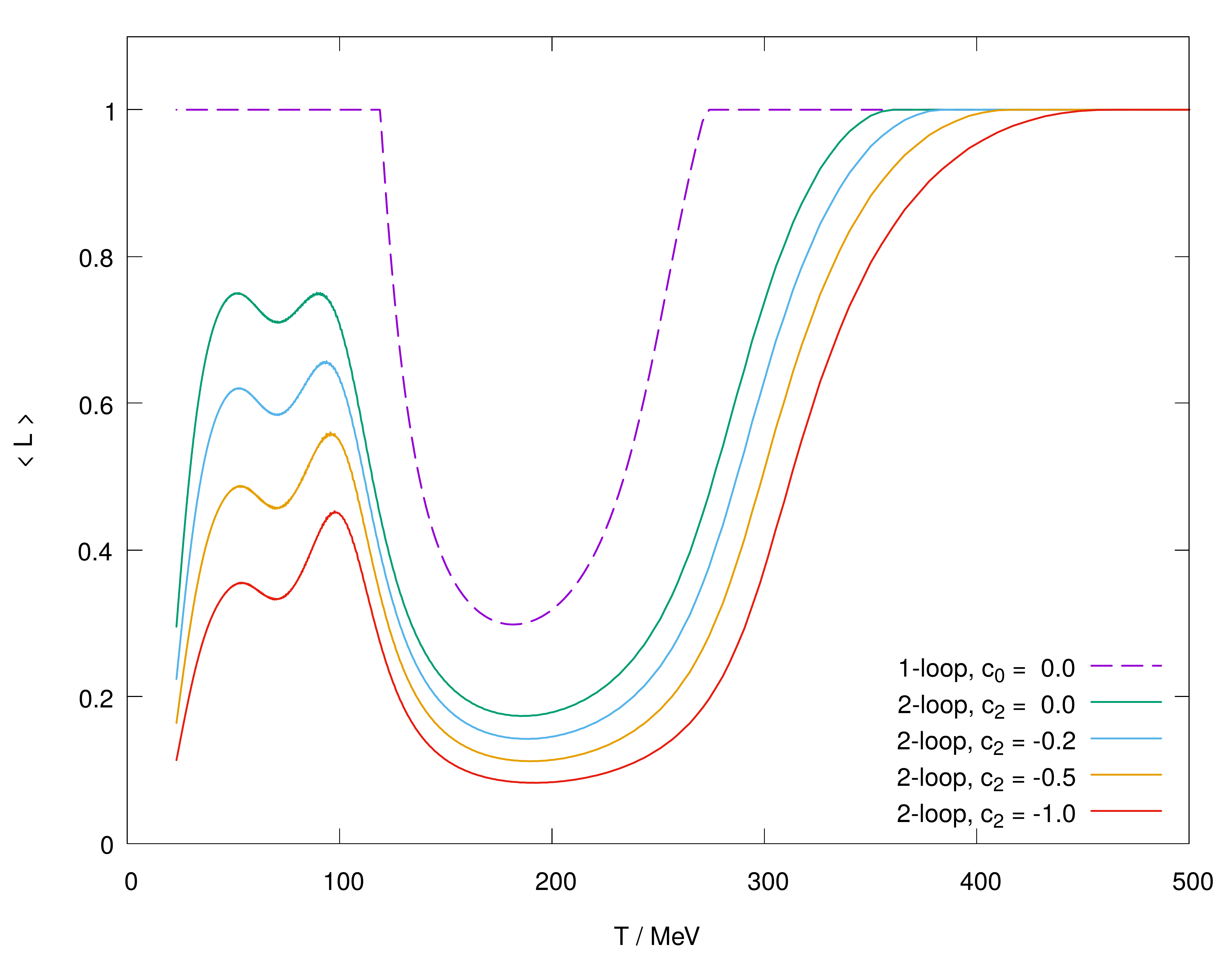}
	\caption{The Polyakov loop as a function of temperature for various values of the 
		2-loop renormalization constant $c_2$. For reference, the 1-loop findings is also 
		included. The \emph{left} panel shows the pure Yang-Mills case, while the \emph{right} 
		panel includes $N_f=2$ flavours of (light) quarks. All plots are made with the 
	    preferred value $c_0 = 0$ for the mass counterterm. }	
	\label{fig12}
\end{figure}

Let us now collect all the pieces and study the effect of the 2-loop corrections
on the Polyakov loop. 
Fig.~\ref{fig12} shows our cumulative results for the expectation value of the 
Polyakov loop as a function of temperature. The renormalization constant for the 
one-loop terms is fixed to the preferred value $c_0 = 0$, and the $2$-loop contribution
is varied in the range $c_2 \in [0, -1.0]$ as discussed earlier. 

In the left panel, we show only the gluon contribution at one and two-loop level.
As expected, the inclusion of the two-loop terms make for a stronger gluon confinement,
so that the critical temperature rises. The effect is not dramatic, as critical 
temperatures of $T^\ast \approx 300\,\mathrm{MeV}$ are still in agreement with 
lattice calculations, in particular since the Coulomb string tension determining 
the absolute scale is not known to high precision. The onset of confinement when 
cooling the system seems to be somewhat softer at two-loop order, but the overall 
qualitative features and the order of the phase transition remain unchanged. 

After including quarks, the picture changes qualitatively, as can be seen from 
the right panel of Fig.~\ref{fig12}. We haven taken $N_f = 2$ flavours and 
included the one-loop result for reference.  At one-loop level, the confinement 
is incomplete and reverts to a fully deconfined phase as we cool the system 
below $T \simeq 100\,\mathrm{MeV}$. As we switch on the 2-loop contribution,
this unphysical phase becomes suppressed by the stronger gluon confinement, 
although the original tendency to increase the Polyakov loop in this region 
remains clearly visible. We could say that the unphysical deconfined phase 
is replaced by a region of \emph{incomplete} confinement, with the separation
energy for two static colour sources being large, but not infinite. 
At \emph{very} low temperatures, the Polyakov loop always vanishes 
$\langle \LL \rangle = 0$, irrespective of the renormalization parameter.
This is different from the one-loop results, where the unphysical phase 
seems to persists up to $T = 0$.

\begin{figure}
	\centering
	\includegraphics[width=0.48\linewidth]{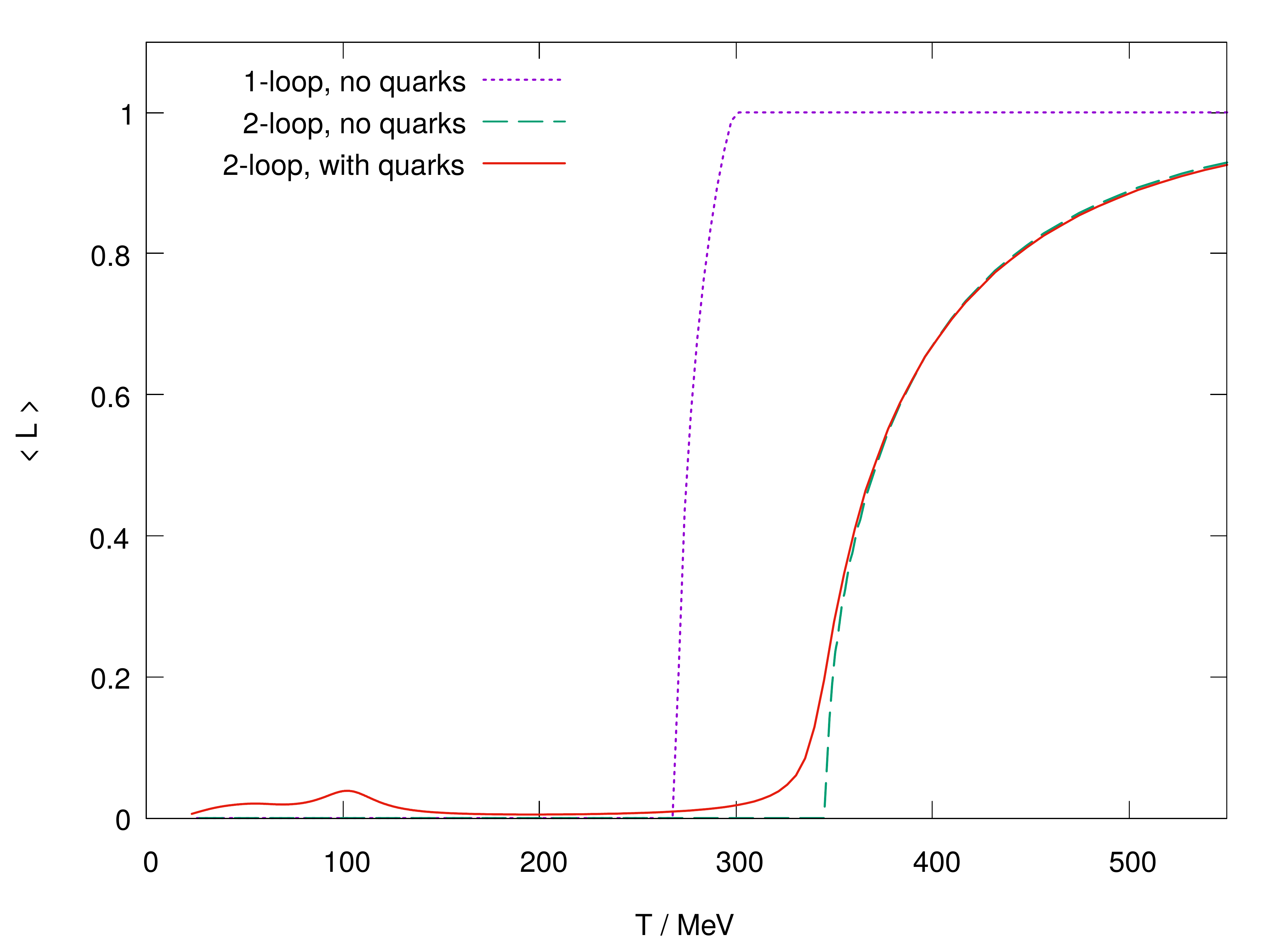}
	\hfill
	\includegraphics[width=0.48\linewidth]{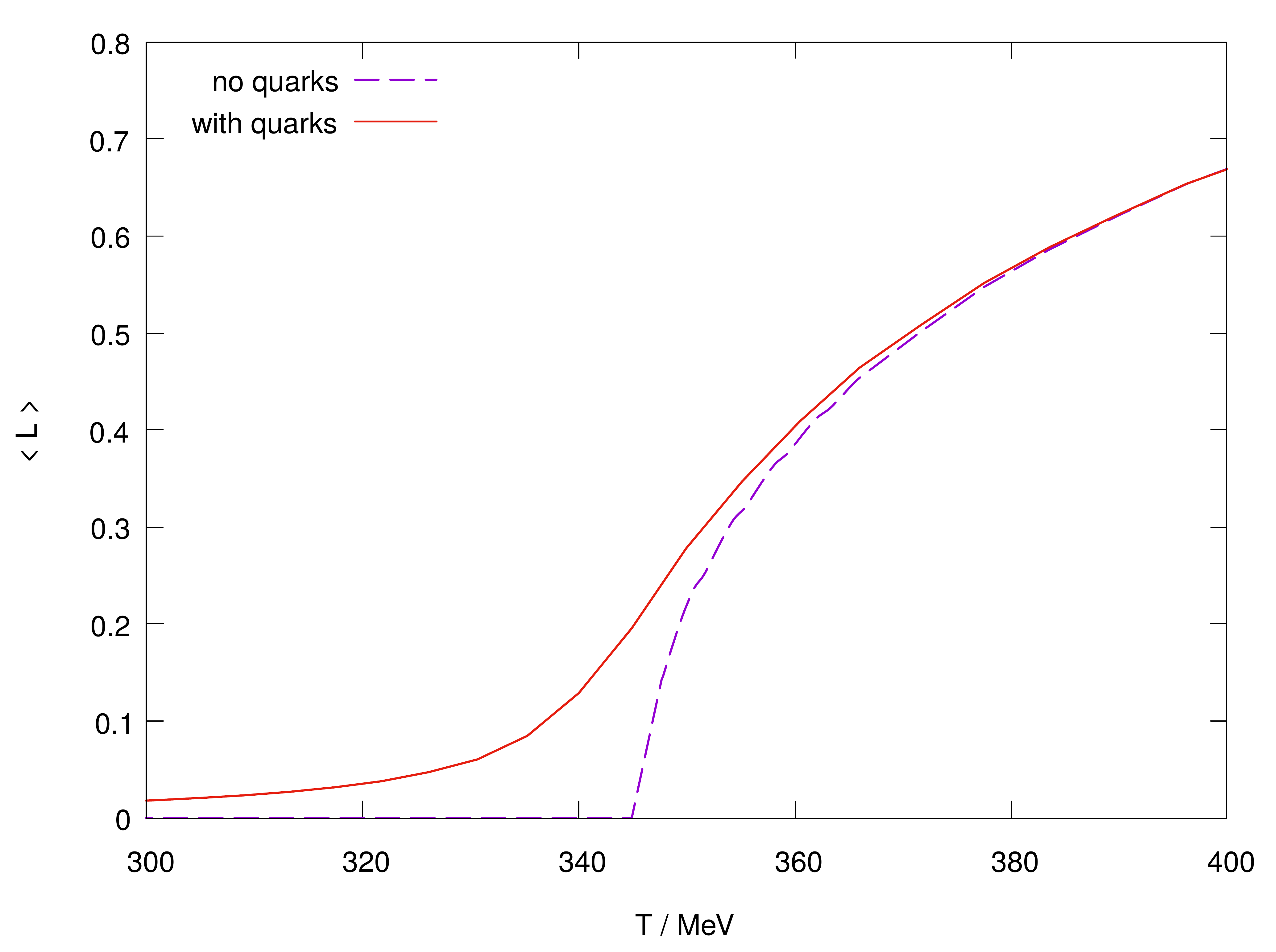}
	\caption{The Polyakov loop for an extreme value of the 
		renormalization parameter $c_2 = -20$, which (over-)emphasizes
		the 2-loop contribution. The \emph{left} panel shows the entire 
		temperature range, while the \emph{right} panel highlights 
		the region near the phase transition.}	
	\label{fig13}
\end{figure}

As we have mentioned above, larger values of $-c_2$ tend to strengthen the 
gluon confinement by emphasize the bosonic 2-loop contribution. At the same 
time, this also increases the (pseudo-)critical temperature as more thermal
fluctuations are necessary to overcome the confining order. For curiosity,
we have explored the extreme case of $c_2 = -20$ 
in Fig.~\ref{fig13}. The unphysical effects in the confined phase are now 
completely suppressed except for a tiny bump near $T = 100\,\mathrm{MeV}$.
At the same time, the critical temperature in the Yang-Mills case increases
to about $T^\ast = 344\,\mathrm{MeV}$, due to the strong gluon confinement.

In the right panel of Fig.~\ref{fig13}, we highlight the temperature region
near the phase transition. It is clearly seen that the quarks have the 
tendency to weaken the transition, which changes from a strong 2nd order
transition in the pure Yang-Mills case to a \emph{cross-over}. The 
pseudo-critical temperature for this cross-over is, however,  not lowered 
substantially compared to the pure Yang-Mills $T^\ast$. The quarks have 
practically no effect in the deconfined region as well, and only affect 
the physics close to the phase transition.  
 
\section{Summary and conclusion} \label{sec:summary}

In this paper, we have studied the Polyakov loop and the deconfinement phase 
transition within the Hamiltonian approach to QCD. We found excellent agreement with 
lattice data for the pure Yang-Mills case, but an unphysical deconfined phase at 
low temperatures as we introduce one or two flavours of light quarks. This phenomenon
could be traced to the weakness of the gluon confinement in the Hamiltonian approach, 
since the effective potential for the Polyakov loop becomes very small at low temperatures
and the center symmetric phase is thus only slightly lower (in free energy) than its 
center-broken counter part. Even $N_f=1$ flavour of light quarks can thus easily overcome
the bosonic order and deconfine the system at low temperatures. 

Since the shortcoming of the Hamiltonian approach at one-loop level is clearly in the
gluon sector, we have studied the effect of including the neglected 2-loop gluon 
contributions. (The quark 2-loop terms will not make a qualitative difference.) As 
in the one-loop case, we treat the renormalization constant $c_2$ arising at this order
as a free parameter, which can be tuned to control the strength of the 2-loop effects.
Though the numerical effort is several orders of magnitude larger, we were able to 
compute the 2-loop corrections with high accuracy and study a wide range of parameters. 
We find that the 2-loop terms indeed strengthen the gluon confinement and partially 
suppress the unphysical effects in the confined phase. This suppression becomes stronger 
as we increase the renormalization constant $c_2$ to emphasize the 2-loop terms. At the same 
time, the stronger confinement increases the critical temperature in the Yang-Mills case 
to about $T^\ast \approx 300\,\mathrm{MeV}$, which is still supported by lattice results.

To completely suppress the unphysical effects in the confined phase, we had to take 
the 2-loop renormalization parameter to rather extreme values $c_2 \approx -20$. 
This is a rather unnatural setup, because the 2-loop terms start to 
dominate and it general remains unclear how such a large value for $|c_2|$ should emerge naturally.
Nonetheless, it is interesting to see that the quarks in this scenario affect only the 
region near the transition, turning the 2nd order transition to a cross-over. The (pseudo-)critical 
temperature in the extreme scenario rises to about $344\,\mathrm{MeV}$.

There are several ways in which our findings could be further improved. First, a full 
renormalization should be carried out that fixes the constants $c_0$ and $c_2$ at 
one- and two-loop level by relating it to a physical input quantity. This should, 
in particular, clarify if large values of $c_2$ as in the previous scenario
are physically sensible. Furthermore, we could include the fermion 2-loop terms 
to have a complete description at two loops, as well as the finite temperature 
corrections to the kernels and core functions. We do not expect these terms 
to contribute significantly, while the computational effort would increase considerably. 
A better strategy might be to incorporate the effects of higher loops and finite 
temperature in the gluon variational kernels already at one-loop. This might require 
to deviate substantially from the Gaussian ansatz, which we plan to investigate 
in a future study. 


\appendix
\section{Root decomposition of $SU(N)$}
\label{app:cartan}
The semi-simple Lie algebra $SU(N)$ has rank $r=(N-1)$ and there are hence 
$r$ mutually commuting generators $H_k$ which span the Cartan subalgebra
of $SU(N)$. As explained in the main text, the background field 
$\ab$ must be chosen in the Cartan subalgebra,
\[
\ab = \sum_{k=1}^r \ab^k \,H_k\,. 
\]
Since the $H_k$ are anti-hermitean and mutually commuting, they can be 
simultaneously diagonalized with purely imaginary eigenvalues $(-i \mu_k)$.
The real numbers $\mu_k$ are called the weights of $H_k$, and the collection
of one eigenvalue from each $H_k$ forms a weight vector 
$\xvec{\mu} = (\mu_1,\ldots,\mu_r)$. The number of such vectors, i.e.~the number 
of the eigenvalues of $H_k$ depends on the representation. In the fundamental
representation, for instance, $H_k$ is $(N \times N)$ and there are hence 
$N$ weight vectors.

In the present paper, we are mainly concerned with the background field in
the \emph{adjoint} representation, $\hat{\ab}^{ab} = - f^{abc}\,\ab^c$. The weights 
$\sigma_k$ in the adjoint representation are called the roots,
\begin{align}
	\hat{H}_k\,|\sigma\rangle = - i \sigma_k \,\vert\,\sigma\rangle
	\label{HH}
\end{align}
and the real numbers $\sigma_k$ from all Cartan generators are collected in 
root vectors $\xvec{\sigma} = (\sigma_1,\ldots,\sigma_r)$. The corresponding 
eigenvector $\vert \sigma \rangle$ diagonalizes all generators $\hat{H}_k$ 
simultaneously. Since $\hat{H}_k$ is $(N^2-1) \times (N^2-1)$ dimensional, 
the eigenvector $\vert \sigma \rangle$ is an adjoint colour vector with 
$N^2-1$ components $\angle a \vert \sigma \rangle$, and there can be at most
$(N^2-1)$ such eigenvectors and hence $(N^2-1)$ root vectors.
Of these roots, $r=(N-1)$ must vanish and the entire root system of $SU(N)$ 
thus contains $N(N-1)$ non-vanishing root vectors. They can be given a partial 
ordering by the first element, i.e.~the eigenvalues of $\hat{H}_1$. Then the 
non-vanishing roots $\xvec{\sigma}$ come in pairs $\pm \sigma$ and half of 
them are positive, half of them are negative. From eq.~(\ref{HH}), the 
adjoint background field is diagonal in the basis $|\sigma\rangle$, 
\[
\hat{\ab}\,| \sigma \rangle = - i (\mathsf{a}\cdot\xvec{\sigma})\,| \sigma \rangle
= -i \left(\sum_{k=1}^r \ab_k\,\sigma_k\right)\,\vert \sigma \rangle\,.
\]
The colour group $G=SU(2)$ has rank $r=1$ and both the root and weight vectors 
are pure numbers. There are two weights $\pm \frac{1}{2}$ and three
roots $\{-1,0,1\}$, of which only two are non-vanishing. The corresponding 
eigenvectors in the adjoint are the well-known cyclic basis,
\[
\vert \sigma=1\rangle  = - \frac{1}{\sqrt{2}} \begin{pmatrix}1 \\ i \\ 0 \end{pmatrix}
\,,\qquad\qquad
\vert \sigma=0\rangle  = \begin{pmatrix}0 \\ 0 \\ 1 \end{pmatrix}
\,,\qquad\qquad
\vert \sigma=-1\rangle = \frac{1}{\sqrt{2}} \begin{pmatrix}1 \\ -i \\ 0 \end{pmatrix}\,.
\]
This structure easily generalizes to $G=SU(3)$, which has rank $r=2$. The two 
Cartan generators are usually taken as $H_1 = T_3 = \lambda_3 / (2i)$ and 
$H_2 = T_8 = \lambda_8 / (2i)$ in terms of Gell-Mann matrices. The root and 
weight vectors are both two-dimensional. Explicitly, the weights read
\[
\xvec{\mu}\,:\qquad \Big( 0 \,,\, \frac{1}{\sqrt{3}} \Big)\,,
\qquad\quad \Big(\frac{1}{2} \,,\, \frac{1}{2 \sqrt{3}}\Big)\,,
\qquad\quad \Big( \frac{1}{2} \,,\, - \frac{1}{2\sqrt{3}}\Big)\,.
\]
More important are the $N^2-1 = 8$ root vectors, of which $N(N-1)=6$ are 
non-vanishing. As they come in pairs with opposite sign of $\sigma_1$, 
there are three non-vanishing positive roots
\begin{align}
	\xvec{\sigma}\,:\qquad \Big( 1 \,,\, 0 \Big)\,,
	\qquad\quad \Big(\frac{1}{2} \,,\, \frac{1}{2} \sqrt{3}\Big)\,,
	\qquad\quad \Big( \frac{1}{2} \,,\, - \frac{1}{2}\sqrt{3}\Big)\,.
	\label{ro}
\end{align}
From these roots, it is clear that any $SU(3)$ background field 
in the Cartan algebra, $\ab = \ab^3\,T^3 + \ab^8 T^8$ 
can conveniently be described by the rescaled components
\begin{align}
	x = \frac{\beta \,\ab^3}{2\pi}\,,\qquad \qquad 
	y = \frac{\beta \,\ab^8}{2 \pi}\,.
\end{align}
The fundamental domain (Weyl alcove) in these variables is given by 
\begin{align}
	x \in \big[0,\,1\big]\,,\qquad\qquad 
	y \in \big[0,\, \frac{2}{\sqrt{3}}\big]\,.
	\label{alk}
\end{align}
Finally, the momentum shift $p_\sigma$ for the three positive roots is
\begin{align}
	\xvec{\sigma}  = \big(0\,,\,1\big)\,&:\qquad \qquad 
	p_\sigma = \frac{2\pi}{\beta}\,x \nonumber \\[2mm]
	\xvec{\sigma}  = \big(\frac{1}{2}\,,\,\frac{1}{2}\,\sqrt{3}\big)\,&:\qquad \qquad 
	p_\sigma = \frac{\pi}{\beta}\,\left(x + \sqrt{3}\,y\right) \nonumber \\[2mm]
	\xvec{\sigma}  = \big(\frac{1}{2}\,,\,-\frac{1}{2}\,\sqrt{3}\big)\,&:\qquad \qquad 
	p_\sigma = \frac{\pi}{\beta}\,\left(x - \sqrt{3}\,y\right)\,.
\end{align}
The structure constants $f^{abc}$ of the $su(N)$ algebra can also be
transformed to the Cartan basis, 
\[
f_{\rho,\sigma,\tau} = f^{abc}\,\langle a \vert \rho \rangle 
\langle b \vert \sigma \rangle \langle c \vert \tau \rangle\,.
\]
They are still antisymmetric, and most relations for the original structure 
copnstants have simple counter parts in the Cartan basis. For instance,
from $ f^{abc}\,f^{a'bc} = N \delta^{aa'}$ (sum over repated indices) we have
\begin{align} 
\sum_{\rho,\sigma,\tau} | f_{\rho,\sigma,\tau}|^2\, u(\xvec{\rho}\cdot \xvec{a})
= N\,\sum_\rho u(\xvec{\rho}\cdot \xvec{a})\,,
\qquad \qquad \qquad
\sum_{a,b,c} | f^{abc}|^2 = \sum_{\rho,\sigma,\tau} | f_{\rho,\sigma,\tau}|^2 
= N (N^2-1)\,. \label{schnabel}
\end{align}

\section{Self-consistent two-loop energy}
\label{app:sc}
In the main text, it has been stressed repeatedly that the gap equation mixes
loop orders, and the self-consistent one-loop energy already contains parts of 
the two-loop terms. This raises the question whether the self-consistent 
two-loop energy must be corrected to avoid double counting.

To clarify this point, it is sufficient to work at $T=0$ without a background 
field; the modifications for the general case are straightforward. We write the 
complete energy in the gluon sector (\ref{EBOS})  in the form
\begin{align}
	E_B = \langle H_B \rangle = \frac{1}{2}\, V_3\, (N^2-1)\,\int \frac{d^3 q}{(2\pi)^3}\,
	\frac{\Omega(q)^2 + q^2}{\omega(q)} + E_2\,,
	\label{fra1}
\end{align}
where $\Omega(q) \equiv \omega(q) - \chi(q) $, and $E_2$ denotes all 2-loop 
contributions. Note that this expression is valid for \emph{any} 
kernel $\omega(q)$, irrespective of whether it satisfies the gap equation or not. 
To minimize the energy, we now take the variation with respect to the propagator 
$\omega(k)^{-1}$, set it to zero and cancel the common factor
\begin{align}
	c \equiv \frac{2\, (2\pi)^3}{V_3 (N^2-1)}. 	
	\label{fracx}
\end{align}
Further evaluation, using $\delta \omega(q) / \delta \omega^{-1}(k),
= - \omega(q)^2 \,\delta(q-k) $ yields the gap equation in the form
\begin{align}
	\omega(k)^2 = k^2 + \chi(k)^2 - 2 \nu(k) + c\,\frac{\delta E_2}{\delta \omega^{-1}(k)}\,.
	\label{gappy}
\end{align}
Here, we have obtained an additional 2-loop contribution through the implicit dependency
of the curvature on the gluon kernel,
\[
\nu(k) \equiv \int d^3 q \,\frac{\Omega(q)}{\omega(q)}\,\frac{\delta \chi(q)}{\delta \omega^{-1}(k)}\,.
\]
This term is usually neglected in the gap equation, and we will also
do so in the present paper. For the moment, however, we keep this term and 
insert the gap equation back into the full energy eq.~(\ref{fra1}) to obtain the 
total self-consistent energy. After a straightfoward calculation, 
\begin{align}
	E_B^{\rm sc} = 
	V_3\,(N^2-1) \int \frac{d^3q}{(2\pi)^3}\,\big[ \Omega(q) + \epsilon(q)\big] + E_2^{\rm sc}\,,
	\label{frax}
\end{align}
where the (self-consistent) two-loop contributions read explicitly 
\begin{align}
	\epsilon(q) & = \int d^3 p\,\frac{\Omega(p)}{\omega(p)}\,\omega(q)^{-1}\,
	\frac{\delta \chi(p)}{\delta \omega^{-1}(q)} \\[2mm]
	E_2^{\rm sc} & = E_2 - \int d^3 q \,\omega(q)^{-1}\,\frac{\delta E_2}{\delta \omega^{-1}(q)}\,.	\label{fra2}
\end{align}
These formulas for the total boson energy are only valid for solutions of the gap equation
(\ref{gappy}). 
If we follow the standard approach and neglect the implicit dependency of the curvature 
on the gluon kernel, $\epsilon(q) \approx 0$, it becomes evident that the first term 
in eq.~(\ref{frax}) is precisely the self-consistent one-loop energy, which was used 
back in eq.~(\ref{eb1}) in the main text. The second term, $E_2^{\rm sc}$, is therefore 
the \emph{self-consistent two-loop energy}. As can be seen from eq.~(\ref{fra2}), 
$E_2^{\rm sc}$ \emph{differs} from the original two-loop term $E_2$ by a subtraction -- 
this is precisely the subtraction necessary to compensates for the two-loop terms 
which have been moved into the self-consistent one-loop contribution via the gap 
equation. 

The subtraction in $E_2^{\rm sc}$ can be substantial: if the functional 
$E_2$ is e.g. quadratic in the propagator $\omega(k)^{-1}$, it is 
easily seen that $E_2^{\rm sc} = - E_2$, i.e. the two-loop contribution 
flips the sign. Fortuantely, only the $T=0$ tadpole contribution from the 
non-Abelian magnetic field is affected by this subtlety, and this term drops 
out when computing the Polyakov loop. 

\section{Finiteness of the non-Abelian magnetic contribution}
\label{app:nonmag}
In the following, we show that all contributions to the non-Abelian magnetic energy, 
eq.~(\ref{pal6}), with both Poisson indices non-vanishing, are UV \emph{finite} ---
provided that the same regularization procedure for the momentum integrals 
as in the one-loop terms is employed, cf.~eq.~(\ref{qn}). For the proof it is sufficient
to consider the most singular terms in eq.~(\ref{pal6}) only, i.e. we can ignore the 
angular dependency $(\hat{p}\cdot \hat{q})$  in the numerator against the constant $3$
and, furthermore, replace the full gluon energy $\omega(p)$ by its perturbative form $p$. 
Then the terms of eq.~(\ref{pal6}) with no zero Poisson indices $ \mf \neq 0 \neq \nf$
become
\begin{align}
	\langle \widetilde{H^{\rm NA}_{\rm YM}}\rangle_0 = 
	\frac{3 g^2}{16}\,V_2 \beta \,\frac{1}{(2\pi)^4}
	\int_0^\infty dp\,p \int_0^\infty dq\,q 
	\sum_{\rho,\sigma,\tau} | f_{\rho,\sigma,\tau}|^2 
	\sum_{\mf = - \infty \atop  \mf \neq 0}^\infty \int_{-1}^1 dz\,
	e^{i \mf \beta (p z + \xvec{\sigma}\cdot \xvec{\ab})}
	\sum_{\nf = - \infty \atop \nf \neq 0}^\infty
	\int_{-1}^1 dy\, e^{i \nf \beta (yq + \xvec{\tau}\cdot \xvec{\ab})}\,.
	\label{nonmag6}
\end{align}
Performing the $z$- and $y$-integrals gives 
\begin{align}
\langle \widetilde{H^{\rm NA}_{\rm YM}}\rangle_0 = \frac{3 g^2}{(2 \pi)^4}\,V_2\beta\,
\sum_{\rho,\sigma,\tau} | f_{\rho,\sigma,\tau}|^2 \, I(\xvec{\sigma}\cdot \xvec{\ab})
\, 	I(\xvec{\tau}\cdot \xvec{\ab})\,,
\label{na1}
\end{align}
where we have introduced the abbreviation
\begin{align}
I(x) \equiv \int_0^\infty dp\,p\sum_{\mf = 1}^{\infty}\,\frac{\sin(\mf \beta p)}{\mf \beta p}\,
\cos(\mf \beta x)\,.	
\end{align}
At this point, we can use eq.~(\ref{qn}) with $\alpha = -1$ to carry out the $p$-integral,
\begin{align}
	I(x) = \sum_{\mf = 1}^\infty \frac{\cos(\mf \beta x)}{(\mf \beta)^2}= 
	\frac{1}{\beta^2}\,\Big[ \frac{\pi^2}{6} - \frac{\pi}{2}\,\beta x + \frac{1}{4}\,(\beta x)^2 \Big]\,,
	\qquad \qquad \quad x \in [0, 2\pi]\,.
	\label{na17}
\end{align}
After inserting this result in eq.~(\ref{na1}), we must carry out the remaining summation 
over the roots. Since we are only interested in the finiteness of the loop integrals, 
it is sufficient to consider the colour group $G=SU(2)$, for which the $N^2-1 = 3 $ roots
are given by $\sigma \in \{ 0, \pm 1\}$ and 
\begin{align}
\sum_{\rho,\sigma,\tau} | f_{\rho,\sigma,\tau}|^2 \, I(\xvec{\sigma}\cdot \xvec{\ab})
\, 	I(\xvec{\tau}\cdot \xvec{\ab}) = 2 \,I(a)^2 + 4\, I(a) \,I(0)\,.	
\label{na21}
\end{align}
Inserting finally eq.~(\ref{na21}) into eq.~(\ref{na1}) and subtracting the zero-background field 
contribution, we obtain eventually
\begin{align}
\langle \widetilde{H^{\rm NA}_{\rm YM}}\rangle_0	 - 
\langle \widetilde{H^{\rm NA}_{\rm YM}}\rangle_0 \big\vert_{\ab = 0} = 
\frac{3 g^2}{8 \pi^4}\,\beta V_2\,\big[I(\ab)- I(0)\big]\cdot\big[ I(\ab) + 3 \,I(0) \big] \,.
\end{align}
By eq.~(\ref{na17}), this is indeed finite for all background fields $ \ab $\,.

\bibliography{polyham}

\end{document}